\documentclass[12pt]{article}
\usepackage[utf8]{inputenc}
\usepackage[round]{natbib}
\usepackage[normalem]{ulem}
\usepackage[colorinlistoftodos,textsize=tiny]{todonotes}
\usepackage{amsmath,amsfonts,accents,url,comment,float,multirow,multicol,adjustbox,makecell}
\numberwithin{equation}{section}
\usepackage[left=2.5cm, right=2.5cm, top=2.5cm]{geometry}
\def\independenT#1#2{\mathrel{\rlap{$#1#2$}\mkern2mu{#1#2}}}
\newcommand\independent{\protect\mathpalette{\protect\independenT}{\perp}}
\allowdisplaybreaks

\newtheorem{assumption}{Assumption}
\newtheorem{theorem}{Theorem}[section]
\newtheorem{lemma}{Lemma}[section]

\DeclareMathOperator*{\argmin}{arg\,min}
\usepackage[colorlinks = true,
            linkcolor = red,
            urlcolor  = blue,
            citecolor = blue,
            anchorcolor = blue]{hyperref}
\providecommand{\keywords}[1]
{
  \small	
  \textbf{\textit{Keywords---}} #1
}
\def\R{{\mathbb{R}}}
\def\P{{\mathbb{P}}}
\def\spacingset#1{\renewcommand{\baselinestretch}%
{#1}\small\normalsize} \spacingset{1}

\title{Instrumental variable estimation of the proportional hazards model by presmoothing}

\author{
{\large Lorenzo T\textsc{edesco}}
\footnote{e-mail: lorenzo.tedesco@kuleuven.be}\\\texttt{\small ORSTAT, KU Leuven}
\and
\addtocounter{footnote}{2}
{\large Jad B\textsc{eyhum}}
\footnote{e-mail: jad.beyhum@gmail.com}
\\\texttt{\small Department of Economics, KU Leuven}
\and {\large Ingrid V\textsc{{an} K{eilegom}}}
\footnote{e-mail: ingrid.vankeilegom@kuleuven.be}
\\\texttt{\small ORSTAT, KU Leuven}
}
\date{}

\begin{document}
\newif\ifband
\newif\ifcorr
\bandfalse
\corrfalse
\maketitle
\spacingset{1.2}

\begin{abstract}
We consider instrumental variable estimation of the proportional hazards model of \cite{Cox1972}. The instrument and the endogenous variable are discrete but there can be (possibly continuous) exogenous covariables. By making a rank invariance assumption, we can reformulate the proportional hazards model into a semiparametric version of the instrumental variable quantile regression model of \cite{chernozhukov2005iv}.  A naïve estimation approach based on conditional moment conditions generated by the model would lead to a highly nonconvex and nonsmooth objective function. To overcome this problem, we propose a new presmoothing methodology. First, we estimate the model nonparametrically - and show that this nonparametric estimator has a closed-form solution in the leading case of interest of randomized experiments with one-sided noncompliance. Second, we use the nonparametric estimator to generate ``proxy'' observations for which exogeneity holds. Third, we apply the usual partial likelihood estimator to the ``proxy'' data.  While the paper focuses on the proportional hazards model, our presmoothing approach could be applied to estimate other semiparametric formulations of the instrumental variable quantile regression model. Our estimation procedure allows for random right-censoring. We show asymptotic normality of the resulting estimator. The approach is illustrated via simulation studies and an empirical application to the Illinois Unemployment Incentive Experiment.
\end{abstract}

\newpage
\spacingset{1.4}
\section{Introduction}
The proportional hazards (henceforth, PH) model, first introduced in \cite{Cox1972}'s seminal paper, has become a popular tool for analyzing time-to-event data, such as the unemployment duration of job seekers or the duration until the bankruptcy of some firms. It is a semiparametric model of the hazard rate, which has the convenient property that the nonparametric component of the model - called the baseline hazard - can be profiled out in the likelihood. The resulting partial likelihood allows for a simple and stable estimation procedure of the parametric part of the model.

In practice, observational data may be affected by unmeasured confounding that influences both the treatment and the outcome, creating an endogeneity issue. In econometrics, a common solution to the endogeneity problem is the use of an instrumental variable (henceforth, IV) approach. There exists several different IV models, which differ in terms of the unobserved heterogeneity that they allow. 

In this paper, we address the problem of endogeneity in the PH model by embedding the problem in the instrumental variable quantile regression (henceforth, IVQR) model introduced in \cite{chernozhukov2005iv}. Indeed, by making a rank invariance assumption, we reformulate the proportional hazards model into a semiparametric version of the instrumental variable quantile regression model of \cite{chernozhukov2005iv}. This allows us to use some of the tools from the literature on the IVQR model. The IVQR model and its associated estimation procedures have become popular in econometrics research, see Chapter 9 in \cite{koenker2017handbook} for a recent review. By constraining the treatment effects to only vary by quantile, this approach allows estimating quantile treatment effects over the full population. This is in contrast to the local average treatment effects framework of \cite{angrist1996identification}, which does not restrict the heterogeneity of treatment effects but requires a monotonicity assumption to hold and only allows to estimate treatment effects on subpopulations. See \cite{wuthrich2020comparison} for a detailed comparison of the two approaches.

One of the main difficulties generated by the IVQR model is estimation. The causal regression function of interest can be shown to be the solution of a system of nonlinear integral equations. A natural approach to estimation is to solve an empirical version of this system. However, as noted in \cite{chernozhukov2006instrumental}, when the causal  regression function  follows a (semi)parametric model this leads to a severely nonsmooth and nonconvex objective function. For the linear-in-parameters semiparametric quantile regression model, some ingenuous solutions to this problem have been developed in \cite{chernozhukov2006instrumental}, \cite{kaplan2017smoothed} and \cite{kaido2021decentralization} among others. However, to the best of our knowledge, there does not exist a general computationally advantageous approach to the estimation of semiparametric IVQR models.

In this paper, we focus on the case where the endogenous variable and the instrument are both discrete and there are (possibly continuous) exogenous covariables. To simplify the exposition, we assume that the endogenous variable and the instrument have the same number of modalities. This setting is very relevant in practice since it includes randomized experiments with noncompliance (where both the instrument and the endogenous variable are binary), which is the main focus of \cite{chernozhukov2005iv}, or, more recently, \cite{wuthrich2019closed}. In the remainder of the paper, we often call ``treatment'' the discrete endogenous variable because of this link. As common with time-to-event data, we assume that the outcome variable is randomly right-censored, where the censoring time is conditionally independent of the outcome.

We propose a new solution to the problem of estimating IVQR models with a semiparametric quantile regression function. We apply this new methodology to the PH model but stress that it could be useful in other semiparametric models, such as the proportional odds model or distribution regression. This approach is based on a \textit{presmoothing} strategy. Our procedure can be summarized in three steps.
\begin{enumerate}
\item First, we estimate nonparametrically the causal quantile regression function.
\item Second, we use this nonparametrically estimated causal quantile regression function to generate ``proxy'' observations for which the treatment is exogenous.
\item Finally, we apply the celebrated partial likelihood estimator of \cite{Cox1972} to the ``proxy'' observations.
\end{enumerate}
Step 1 is simplified by the fact that the treatment and instrument are discrete, making the inverse problem well-posed. Moreover, as we argue in the paper, in the practically relevant case of randomized experiments with \textit{one-sided} noncompliance,  where both the treatment and the instrument are binary and there is full compliance in the control group, our nonparametric estimator possesses a closed-form expression. For Step 1, we use \cite{beran1981nonparametric}'s estimator of the conditional survival function to construct an empirical version of the system of equations generated by the model and then solve the estimated system to obtain the causal regression function of interest. This allows us to deal with random right-censoring, a common feature of duration data. The fact that we can use the partial likelihood estimator in Step 3 is very convenient since the latter has a convex objective function and is implemented in traditional statistical software. We stress again that the proposed procedure could be applied to other models (such as the proportional odds model or distribution regression) by replacing Step 3 with standard estimators of these models (under exogeneity). The main advantage of our procedure is to bring us back (in Step 2) to the case of exogenous data, which is well-studied and for which most semiparametric models were originally motivated.

We show that the resulting estimator is asymptotically normal. We also develop a theory for the nonparametric estimator of Step 1 which is new to the literature. Simulations demonstrate that our estimator possesses good finite sample properties. Finally, the procedure is applied to evaluate the causal effect of reemployment bonuses using data from the Illinois Unemployment Incentive Experiment.

We call our methodology \textit{presmoothing} because it is related to literature in (semi)parametric statistics advocating for estimation procedures that first use nonparametric estimators before projecting the nonparametric estimators on (semi)parametric spaces in a second step, see \cite{cristobal1987class,akritas1996use} or, more recently, \cite{musta2022presmoothing}. The main takeaway from this statistical literature is that, asymptotically, the two-step estimator recovers the property of the ``oracle'' second-step estimator (which assumes that the nonparametric object estimated in the first step is known). \textit{Presmoothing} is particularly useful in cases where estimating \textit{directly} the semiparametric model is complicated while nonparametric estimation is simpler. One of the contributions of this article is to bring this \textit{presmoothing} tool to econometrics.

Finally, before outlining the different sections of the paper, we note that there already exist quite a few papers studying instrumental variable methods for duration models. To the best of our knowledge, no other paper considers the PH model embedded in the IVQR model. Please note, that for reasons of space, we do not cite all relevant works. \cite{bijwaard2005correcting} consider a mixed proportional hazards model in which the unobserved heterogeneity enters multiplicatively, that is differently from the IVQR model. \cite{wang2022instrumental} considers a PH model with endogeneity but they make an additive (rather than quantile) homogeneity of treatment effects assumption. As argued in \cite{chernozhukov2005iv}, the restrictions on unobserved heterogeneity in the IVQR model can be seen as weaker. \cite{beyhum2022nonparametric} studies the nonparametric IVQR model with random right censoring with discrete treatment and instrument but no covariates. The present paper generalizes \cite{beyhum2022nonparametric} by allowing for continuous covariables in nonparametric estimation and studying the three-step estimation of the PH model. \cite{hong2003inference,chen2018sequential,wang2021moment,beyhum2023instrumental} study the linear-in-parameters IVQR model under random right-censoring but do not cover the PH model. The way we reformulate the PH duration model into an IVQR model is related to the reformulation of the dynamic duration model into a dynamic IVQR model in \cite{beyhum2023variable}.

The rest of the paper is organized as follows. In Section \ref{sec:model}, we specify the model. Identification results are derived in Section \ref{sec:identification}. Section \ref{sec:estimation} is devoted to the estimation theory. 
In Section \ref{sec:asymptotic_theory} we discuss the asymptotic theory of the proposed estimator.
In Section \ref{sec:simulations}, the finite sample performance of the proposed method is investigated through simulations. In Section \ref{sec:empirical_application}, we illustrate the method by means of an empirical application of the method. All the technical details are deferred to the Appendix. 
\section{Model}\label{sec:model}
\subsection{The duration model}
We are interested in the effect of an endogenous treatment $Z$ on a non-negative outcome $T$. We consider a treatment $Z=(Z_1,\dots,Z_{d_Z})^\top$ expressed as a vector of $d_Z$ binary variables and taking $L$ different values $\{z_1,\dots,z_L\} = \mathcal{Z}$. In our data example e.g., we have a categorical treatment with three levels that can be defined by means of two binary (dummy) variables, so $d_Z=2$.   Since there are three levels, $L=3$ in this case and $z_1=(0,0)^\top, z_2=(1,0)^\top$and $z_3=(0,1)^\top$. There are some exogenous covariables $X$ with compact support $\mathcal{X}\subset\mathbb{R}^{d_X}$.  Let $T(z,x)$ be the potential outcome duration under treatment equal to $z\in\mathcal{Z}$ and covariables equal to $x\in\mathcal{X}$. Let us assume that $T(z,x)$ is a continuous random variable and also define by $$\lambda(z,x,t):= \lim_{dt \to 0}\frac{\P(T(z,x)\in[t,t+dt]|T(z,x)\ge t)}{dt} $$
the structural hazard of $T(z,x)$ at time $t\in\R_+$. We impose the consistency condition $T=T(Z,X)$. We also assume that $T(z,x)$ follows a proportional hazards model, that is there exists a baseline hazard $\lambda_0:\R_+\xrightarrow{}\R_+$ and a vector of regression coefficients $(\beta_{0,z},\beta_{0,x})\in\mathbb{R}^{d_Z+d_X}$ such that 
\begin{align}\label{eq:model_cox}
    \lambda(z,x,t) = \lambda_0(t)\exp((z,x)^\top\beta_0),\ \text{for all } (z,x,t)\in\mathcal{Z}\times \mathcal{X}\times \R_+ .
\end{align}
Remark that this proportional hazards assumption implies that the support of $T(z,x)$ does not depend on $z,x$ and is equal to that of $\lambda_0$, or equivalently of $T$. We denote by $\mathcal{T}$ this support of $\lambda_0$. Let us now define the structural cumulative hazard $$\Lambda(z,x,t):=\int_0^t \lambda(z,x,s)ds=\Lambda_0(t)\exp((z,x)^\top\beta_0),$$
where $\Lambda_0(t):=\int_0^t \lambda_0(s)ds$ is the cumulative baseline hazard. Let us also introduce $Q(z,x):= \Lambda(z,x,T(z,x))$, which is the structural hazard of $T(z,x)$ evaluated at $T(z,x)$. The random variables $\{Q(z,x)\}_{(z,x)\in\mathcal{Z}\times \mathcal{X}}$ can be thought of as the unobserved heterogeneity of the model. We make the following Assumption.
\begin{assumption}\label{RI} The following holds
\begin{itemize}
\item[(i)] There exists a random variable $Q$ such that $Q(z,x)=Q$ for all $(z,x)\in\mathcal{Z}\times \mathcal{X}$; 
\item[(ii)] The baseline hazard $\lambda_0$ is continuous and its support $\mathcal{T}$ is a bounded interval.
\end{itemize}
\end{assumption}
By Assumption \ref{RI} $(ii)$, the structural cumulative hazard $\Lambda(z,x,\cdot)$ is strictly increasing on $\mathcal{T}$. Hence, we can define the inverse of $\Lambda(z,x,\cdot)$ on $\mathcal{T}$, which we denote by $\Lambda(z,x,\cdot)^{-1}$. This means that $\Lambda(z,x,\cdot)^{-1}(s)$ is the unique element $t$ of $\mathcal{T}$ such that $\Lambda(z,x,t)=s$ (by definition the image of the cumulative hazard on $\mathcal{T}$ is $\R_+$ so that $\Lambda(z,x,\cdot)^{-1}(s)$ is defined on all $\R_+$). Because of this, we have $T(z,x)= \Lambda(z,x,\cdot)^{-1}(Q(z,x))$. Assumption \ref{RI} $(i)$ then implies that for two subjects $i$ and $j$ , $T_i(z,x)>T_j(z,x)$ implies $Q_i(z,x) = Q_i>Q_j = Q_j(z,x)$, which leads to $T_i(z',x')>T_j(z',x')$, for all $(z,x),(z',x')\in\mathcal{Z}\times \mathcal{X}$. Hence, under Assumption \ref{RI} $(i)$ the rank in the outcome of any two subjects is the same across all potential outcomes. Because of this, Assumption \ref{RI} $(i)$ is a rank invariance assumption as in \cite{chernozhukov2005iv}. This assumption restricts the heterogeneity of the treatment effects on the duration: the treatment can change the quantiles of the distribution of the potential outcomes but it cannot change the rank that a subject has in this distribution. Moreover, note that the rank invariance assumption does not restrict the possible values of the structural hazard $\lambda(z,x,t)$ beyond the continuity condition in Assumption \ref{RI} $(ii)$ and therefore, in this sense, rank invariance is not a constraint on the marginal distribution of $T(z,x)$. It only imposes limits on the joint distribution of potential outcomes, that is the distribution of $(T(z,x))_{(z,x)\in \mathcal{Z}\times \mathcal{X}}$. This rank invariance assumption could be relaxed into a rank similarity assumption as in \cite{chernozhukov2005iv} while keeping all results valid.  
Remark that Assumption \ref{RI} allows us to write \begin{equation}\label{model2704} \Lambda_0(T)\exp((Z,X)^\top\beta_0)=Q.\end{equation}

We assume to observe a categorical instrumental variable $W$. To simplify the theoretical arguments, we also suppose that the instrument has the same number of modalities as the treatment. Therefore, $W$ has support $\mathcal{W} = \{w_1,\dots,w_L\}\subset \mathbb{R}^{d_W}$. Differently from the variable $Z$, the instrument $W$ does not need to consist of binary variables.  Importantly, assuming that the cardinalities of the support of $Z$ and $W$ are equal does not limit our discussion to a specific case. First, it would be possible, but burdensome, to extend both identification and estimation results to the case where the number of modalities of $W$ is greater or equal to the number of modalities of $Z$. Second, it is always possible to obtain the same number of modalities between $Z$ and $W$ by aggregation of modalities of $W$. We impose the following condition. 
\begin{assumption}\label{IVind}
$(W,X)\independent Q$,
\end{assumption}
where `$\independent$' stands for statistical independence.
This assumption formalizes the exogeneity of the instrument $W$ and covariables $X$.
The duration $T$ is randomly right censored by a random variable $C$ with support in $\R_+$ so that we do not observe $T$ but $Y=\min(T,C)$. The observables are $(Y, \delta, Z, X, W)$, where $\delta =I(T\le C)$. We impose the following standard independent censoring assumption:
\begin{assumption}\label{Cind}
$T\independent C|Z,X,W$.
\end{assumption}
\subsection{Reformulation as a semiparametric IVQR model}\label{sec:reformulation}
In this section, thanks to Assumption \ref{RI}, we reformulate the duration model into a semiparametric IVQR model. 
Let us define, for $u\in[0,1]$,  \begin{equation}\label{semiparametric2704}\varphi(z,x,u)=\Lambda_0^{-1}\left(\frac{-\log(1-u)}{\exp((z,x)^\top\beta_0)}\right).\end{equation}
and $U=1-\exp(-Q)$. By standard arguments in duration analysis, $Q$ follows a unit exponential distribution, so that $U$ follows a standard uniform distribution. Then, inverting equation \eqref{model2704}, we obtain that
\begin{equation} \label{IVQR} T=\varphi(Z,X,U),\ U|W,X\sim\mathcal{U}[0,1],  \end{equation}
where $\mathcal{U}[0,1]$ stands for the uniform distribution on $[0,1]$. Remark that by Assumption \ref{RI} $(ii)$ and the inverse function theorem, $\varphi(z,x,\cdot)$ is strictly increasing and differentiable with continuous derivative on $[0,1]$, for all $(z,x)\in\mathcal{Z}\times \mathcal{X}$. As claimed in the introduction, this is an IVQR model as in \cite{chernozhukov2005iv}, where the causal quantile regression function $\varphi$ follows the semiparametric model given in \eqref{semiparametric2704}, which is generated by the proportional hazards assumption on the structural hazard. From now on, this paper will analyze directly this semiparametric IVQR model.
 \\

\noindent \textbf{Notation. } Finally, we introduce some additional notation.  Define $V = (Z,X) \in \mathcal{V} = \mathcal{Z}\times \mathcal{X}\subset\mathbb{R}^{d_V}$, where $d_V=d_Z+d_X$. For $v = (z,x)$, we write $\varphi(v,\cdot)$ for $\varphi(z,x,\cdot)$. For $l = 1,\dots,L$, we also write $\varphi_l^x(\cdot)$ for $\varphi(z_l,x,\cdot)$ and $\varphi^x(\cdot)$ for  $(\varphi_l^x(\cdot))_{l=1}^L $. 

\section{Identification} \label{sec:identification}
 In this section, we provide an identification result of the parameter of interest $\beta_0$. The identification can be obtained in two steps. First, we analyze the identification of $\varphi$. Second, we study the identification of $\beta_0$ given the first step. Note that we keep the analysis of identification brief since it does not constitute the main contribution of the paper (which is the estimation procedure).
 \subsection{Identification of $\varphi$}
Let us first recall the standard characterization of $\varphi$ as the solution to a system of nonlinear equations, that is, for every $x\in\mathcal{X}$, $\varphi^x(u)$ is a solution of the following system of equations in $(\theta_l)_{l=1}^L\in\mathbb{R}^{L}_{+}$:
\begin{align}\label{eq:system}
    \sum_{l=1}^L F(\theta_l,z_l|x,w_k)= u \quad\text{for}\, k=1,\dots,L,\,u\in(0,1),
\end{align}
where $ F(t,z|x,w) = P(T\le t, Z =z|X = x,W=w)$.
 Indeed, by the fact that $\varphi(z,x,\cdot)$ is strictly increasing and the independence between $U$ and $(W,X)$, for all $k\in\{1,\dots,L\}$, it follows that: 
\begin{align*}
    \sum_{l=1}^L F(\varphi_{l}^x(u),z_l|x,w_k) &=  \sum_{l=1}^L  P(T\le \varphi_{l}^x(u), Z =z_l|X=x,W = w_k) \\
    &=\sum_{l=1}^L  P(\varphi_{l}^x(U)\le \varphi_{l}^x(u), Z =z_l|X=x,W = w_k)\\
        &=\sum_{l=1}^L  P(U\le u, Z =z_l|X=x,W = w_k)\\
            &=P(U\le u|X=x,W = w_k) = u.
\end{align*}
\cite{chernozhukov2005iv} give conditions under which there is a unique solution to \eqref{eq:system}. In the context of the PH model, there is an additional layer of complexity due to the fact that $F$ is not identified everywhere because of censoring. For $z\in\mathcal{Z}, x\in\mathcal{X}$ and $w\in\mathcal{W}$, let us denote by $c_{z,x,w}$ the upper bound of the support of the distribution of $C$ given $Z=z,X=x, W=w$. Also, given two real numbers $c,d$, we denote by $c\wedge d$ the value $\min(c,d)$.  Moreover, let $\bar t$ be the upper bound of $\mathcal{T}$. Thanks to Assumption \ref{Cind} and by standard ``Kaplan-Meier type'' arguments from the duration analysis literature, $F(\cdot, z|x,w)$ is identified on $[0,\bar t\wedge c_{z,x,w}]$, and $\bar t\wedge c_{z,x,w}$ is itself identified.
Let us now introduce $$\bar u^x= \inf_{z\in\mathcal{Z}, w\in\mathcal{W}} \varphi(z,x,\cdot)^{-1}(\bar t\wedge c_{z,x,w}),$$ 
where $\varphi(z,x,\cdot)^{-1}(t)=\inf\{u\in\mathbb{R}_+:\varphi(z,x,u)\ge t\} $ is the pseudo-inverse of $\varphi(z,x,u)$. For $u\le \bar u^x$, the left-hand-side of \eqref{eq:system} will be identified at $(\theta_\ell)_{\ell=1}^L =\varphi^x(u)$ for all $x\in\mathcal{X}$, so that identification can follow along the lines of \cite{chernozhukov2005iv}. We provide below an identification result assuming that the solutions to the system \eqref{eq:system} are unique, a property for which sufficient conditions can be found in \cite{chernozhukov2005iv}.

\begin{theorem} \label{GID} Let Assumptions \ref{RI}, \ref{IVind} and \ref{Cind} hold and assume that \eqref{eq:system} has a unique solution in $\R^L_{+}$ for all $u\in[0,1]$. Then, $\bar u^x$ is identified and $\varphi(z,x,u)$ is identified for all $z\in\mathcal{Z},x\in\mathcal{X},u\in[0,\bar{u}^x]$.
\end{theorem}

 \subsection{Identification of $\beta_0$}
The proof of identification of $\beta_0$ is constructive, meaning that we can use its rationale to estimate $\beta_0$ in the subsequent section. The proof uses the knowledge of $\varphi$ to solve the problem of identifying $\beta_0$ under exogeneity.

Formally, let $\bar u:=\inf_{x\in\mathcal{X}} \bar u^x$. From Theorem \ref{GID}, $\varphi$ is identified for $(z,x,u)\in \mathcal{Z}\times\mathcal{X}\times[0,\bar u]$. As $\bar u$ is not known precisely in practice, we choose a point $\bar U$ such that $0<\bar U<\bar u$. We then define two random variables $U_g$ and $U_g^c$ such that $(U^g,U_g^c)\independent(Z,X)$. The subscript `g' highlights that these random variables are generated. The variable $U_g$ is drawn from a uniform distribution $\mathcal{U}[0,1]$, is independent of $U^c_g$, and serves as the generative process. $U_g^c$ acts as a censoring variable for $U_g$ with support in $[0,\bar U]$, ensuring $\min(U_g,U_g^c)$ does not exceed $\bar U$, thus respecting the identification constraints. Hence, any continuous distribution on $[0,\bar U]$ for $U_g^c$ is acceptable.

We then define the random variables $T_g=\varphi(Z,X,U_g)$, $C_g =\varphi(Z,X,U_g^c)$, $Y_g= T_g\wedge C_g$ and $\Delta=I(T_g\le C_g)$. By Theorem \ref{GID}, the distribution of $(Y_g,\Delta, Z,X)$ is identified. As shown in Section \ref{sec:model}, the random variable $T_g$ follows a PH model with parameter $\beta_0$, where exogeneity holds as $U_g$ is independent of $(Z,X)$. Furthermore, the censoring mechanism is conditionally independent as $T_g\independent C_g|Z,X$ by construction. Hence, $\beta_0$ is identifiable from the distribution of $(Y_g,\Delta, Z,X)$ under standard assumptions for the identification of the PH model (e.g., see \cite{tsiatis1981large}).

This argument leads us to the following theorem.

\begin{theorem} \label{BID} Let the assumptions of Theorem \ref{GID} hold and assume that $\bar u>0$ and $E[VV^\top]$ has full rank, then $\beta_0$ is identified.
\end{theorem}

For later theoretical discussions, it will be beneficial to set $\tilde U = \min(U_g,U_g^c)$. We will also employ the equalities $Y_g = \varphi(Z,X,\tilde U)$ and $\Delta = I(U_g\le U_g^c)$.

\section{Estimation}\label{sec:estimation}
In this section, we discuss the estimation of the parameter vector $\beta_0$, given an independent and identically distributed (iid) sample $\{(Y_i,\delta_i,Z_i,X_i,W_i)\}_{i=1}^n$ of observables. We first outline a naïve, but ultimately unsuitable approach to estimation before presenting the three steps of the estimation procedure mentioned in the introduction. For sake of simplicity, we restrict ourselves to the case $d_X=1$, so that the already involved conditions we state next do not depend on the parameter $d_X$.

\subsection{Naïve approach} A naïve approach to estimation would proceed as follows. By substituting equation \eqref{semiparametric2704} into \eqref{eq:system}, we obtain
$$   \sum_{l=1}^L F\left(\left.\Lambda_0^{-1}\left(\frac{-\log(1-u)}{\exp((z_l^\top,x^\top)\beta_0)}\right),z_l\right|x,w_k\right)= u \quad\text{for}\, k=1,\dots,L,\,u\in[0,\bar u).$$
This leads to 
\begin{equation}\label{naive}  E\left[\int_0^{\bar u} \sum_{k=1}^L\left[\sum_{l=1}^L F\left(\left.\Lambda_0^{-1}\left(\frac{-\log(1-u)}{\exp((z_l^\top,X^\top)\beta_0)}\right),z_l\right|X,w_k\right)- u\right]^2du\right]=0.\end{equation}
One could estimate $F$ by a conditional  \cite{beran1981nonparametric}'s estimator (see Section \ref{sec:beran_definition} for its definition) and then obtain an empirical analog to equation \eqref{naive}. This empirical criterion could be minimized over $\Lambda_0$ and $\beta_0$. As noted in the introduction, this approach would result in a highly nonsmooth and nonconvex problem.
\subsection{Our estimation procedure}

\subsubsection{Step 1: Nonparametric estimator of $\varphi$}\label{sec:beran_definition}
We first build a nonparametric estimator of $\varphi$. The estimator is based on \eqref{eq:system}, where $F$ is first estimated via a conditional Beran's estimator  \\

\noindent \textbf{Step 1.1: Nonparametric estimator of $F$ under random right censoring}

\noindent Remark that $F(t,z|x,w) = F(t|z,x,w)p_{z,x,w}$, where $F(t|z,x,w) = P(T\le t|Z=z,X=x,W=w)$ and $p_{z,x,w}= P(Z=z|X=x,W=w)$. 
 We propose to estimate $F(t|z,x,w)$ non-parametrically using the \cite{beran1981nonparametric} estimator, given by
\begin{align} \label{eq:estimator_KM}
    \tilde F(t|z,x,w) = 1- \prod_{j=1}^n\Big[1- \frac{B_{h}^{z,w}(x-X_j,Z_j,W_j)}{\sum_{k=1}^nI(Y_k\ge Y_j)B_{h}^{z,w}(x-X_k,Z_k,W_k)}\Big]^{\eta_j(t) } ,
\end{align}
where $\eta_j(t) = I(Y_j\le t, \delta_j = 1)$ and $B_{h}^{z,w}(x-X_k,Z_k,W_k)$ is a sequence of non-negative weights adding up to 1. 
In our case, we adopt the Nadaraya-Watson type weights, which are specified as follows:
\begin{align}\label{eq:ND_weigths}
    B_{h}^{z,w}(x-X_k,Z_k,W_k)= \frac{K(\frac{x - X_k}{h})I(Z_k=z,W_k=w)}{\sum_{i=1}^nK(\frac{x - X_i}{h})I(Z_i=z,W_i=w)},
\end{align}
where $K(\cdot)$ is a univariate kernel function, and $h=h_n$ is a bandwidth depending on $n$ and converging to zero as $n\xrightarrow{}\infty$.
To ensure that the final criterion function is smooth, we smooth $\tilde F$ with respect to time. Consider a further kernel $\tilde K$ and define $H(t) = \int_{-\infty}^t \tilde K(u)\text{d}u$, with a bandwidth $\epsilon = \epsilon_n$ depending on $n$ and converging to zero as $n\xrightarrow{} \infty$. We  obtain
\begin{align}\label{eq:F_smoothing}
    \hat F(t|z,x,w) = \int  H\Big(\frac{t-u}{\epsilon}\Big)\text{d}\tilde F(u|z,x,w).
\end{align}
Consider now the following estimator for the quantity 
$p_{z,x,w}$: 
\begin{align*}
    \hat p_{z,x,w} = \frac{\sum_{i=1}^n I(Z_i=z,W_i=w)K(\frac{x-X_i}{h})}{\sum_{i=1}^n I(W_i=w)K(\frac{x-X_i}{h})}.
\end{align*}
The final estimator for $F(t,z|x,w) $ is given by
\begin{align}\label{eq:estimator_F}
     \hat F(t,z|x,w) = \hat  F(t|z,x,w)\hat p_{z,x,w}.
\end{align}
\noindent \textbf{Step 1.2: Final nonparametric estimator of $\varphi$}\\
Let $\bar T<\infty$  be an upper bound on $\max_{v\in\mathcal{V}}\varphi(v,\bar U)$, let 
 $\mathcal{F}_{Z}^{\bar U, \bar T} = \{f:\mathcal{Z}\times [0,\bar U]\xrightarrow{} [0,\bar T]\}$ and let $\mathcal{F}_{\uparrow}^{Z,W}$ be the set of maps from $\mathbb{R}_+\times \mathcal{Z}\times \mathcal{W}$ to $\mathbb{R}_+$ that are continuous and increasing in the first argument. Let $x\in\mathcal{X}$ be fixed.
We denote by $F^x(t,z|w)$ the function $F(t,z|x,w)$.
 We have shown that for each $x\in\mathcal{X}$, $\varphi^x$ belongs to the set of solutions to the equation:
\begin{align}\label{eq:A_equal_zero}
A(\varphi^x,F^x)\equiv0,
\end{align}
where $A$ is an operator that associates to an element $(\tilde\varphi^x,\tilde F^x)\in \mathcal{F}_{Z}^{\bar U, \bar T}\times \mathcal{F}_{\uparrow}^{Z,W}$, a map from $[0,\bar U]$ to $\mathbb{R}^L$, denoted by $A(\tilde\varphi^x,\tilde F^x)$ and defined as
$$
A(\tilde\varphi^x,\tilde F^x)(u) = \Big(\sum_{l=1}^L \tilde F^x(\tilde \varphi^x_{l}(u), z_l|w_k) - u\Big)_{k=1}^{L},
$$
where we wrote $\tilde\varphi^x_l(u)$ for $\tilde\varphi^x(z_l,u)$.  
Thus, we define  the estimator of $\varphi^x$ by 
\begin{align}\label{eq:estimator_phi}
    \hat\varphi^x \in \argmin_{\theta\in \mathcal{F}_{Z}^{\bar U, \bar T}}\|A(\theta,\hat F^x)\|,
\end{align}
where $\|\cdot\|$ denotes the Euclidean norm.


\subsubsection{Step 2: ``Proxy" generation process}
We now show how to use the nonparametric estimator obtained in Step 1 to generate ``proxy'' observations for which the treatment is exogenous. For each $i=1,\dots,n$, we generate a random observation of $(U_g,U_g^c)$, namely $(U_{g,i},U_{g,i}^c)$, from which we obtain $\tilde U_i = \min(U_{g,i},U_{g,i}^c)$ and $\Delta_i = I(U_{g,i}\le U_{g,i}^c)$. Then, the proxy observations correspond to $\{(\hat Y_{g,i}, \Delta_i,V_i)\}_{i=1}^n$, where $\hat Y_{g,i}=\hat \varphi(V_i,\tilde U_{i}) = \hat\varphi^{X_i}(Z_i,\tilde U_i)$.

\subsubsection{Step 3: Partial likelihood estimator based on the ``proxy'' observations.}
The last step involves obtaining an estimator $\hat\beta$ for $\beta_0$ using the standard partial likelihood estimator of \cite{Cox1972} based on the ``proxy'' observations obtained in Step 2. Denote by $\mathcal{B}$ a compact set that contains $\beta_0$ as an internal point. Thus, $\hat \beta$ is obtained as the minimizer of the standard score function for a PH model, and so 
\begin{align}\label{eq:estimator_phi_partial}
	\hat \beta \in \argmin_{\beta\in\mathcal{B}} \Big\| n^{-1}\sum_{i:\Delta_{i}=1}V_i^\top\beta -n^{-1}\sum_{i=1}^n\log\big(n^{-1}\sum_{j:\hat Y_{g,j}\ge \hat Y_{g,i}}\exp(V_j^\top\beta)\big)\Big\|.
\end{align}
\subsection{Computation}
We now discuss the optimization program \eqref{eq:estimator_phi}, noticing that the objective function $\|A(\theta,\hat F^x)\|$ might have multiple local minima. Therefore, the computation of the estimator $\hat \varphi^x$ can require starting the optimization algorithm at different initialization points. However, when the system of equations provided in \eqref{eq:system} reduces to a triangular system, the estimator is easier to compute. This means that, after possibly relabelling the points in $\mathcal{Z}$ and $\mathcal{W}$, for a fixed value $u\in[0,\bar U]$ the system takes the form:
\begin{align}\label{eq:system_triangular}
\sum_{l=1}^k F(\theta_l,z_l|x,w_k)= u \quad\text{for}\,k=1,\dots,L.
\end{align}

This case occurs, for instance, in the presence of one-sided noncompliance, that is, $Z$ and $W$ are binary with support $\mathcal{Z}=\mathcal{W}=\{0,1\}$, and $P(Z=1|W=0,X) = 0$. In fact, the corresponding system takes the form: 
\begin{align*}
    \begin{cases}
        F(\theta_1,0|x,0)= u\\
         F(\theta_1,0|x,1) + F(\theta_2,1|x,1) = u.
    \end{cases}
\end{align*}
Another example, in which $Z$ and $W$ are not binary, corresponds to the empirical application discussed in Section \ref{sec:empirical_application}.

When the system is of the form provided in \eqref{eq:system_triangular}, the estimator has a closed-form expression based on the following inductive algorithm:
\begin{align}\label{eq:triangular_solution}
\begin{split}
\hat\varphi(z_1,x,u) &= \hat F\big(\cdot,z_1|x,w_1\big)^{-1}(u),\quad\text{and}\\
\hat\varphi(z_k,x,u) &= \hat F\big(\cdot,z_k|x,w_k\big)^{-1}\Big(u-\sum_{l=1}^{k-1}\hat F\big(\hat \varphi(z_l,x,u),z_l|x,w_k\big)\Big),\quad \text{for } k\ge 2,
\end{split}
\end{align}
where $\hat F(\cdot,z|x,w)^{-1}(u) = \inf\{t\in\mathbb{R}_+:\hat F(t,z|x,w)\ge u\} $ denotes the pseudo-inverse of the function $\hat F(\cdot,z|x,w)$. Note that, based on the assumptions on the model, $F(\cdot,z|x,w)$ is strictly monotone increasing on $[0,\bar T]$, for each $z\in\mathcal{Z},x\in\mathcal{X},w\in\mathcal{W}$. In addition, by its definition, $\hat F(t,z|x,w)$ is monotone increasing in $t\in[0,\bar T]$, and this implies the inversion procedure in \eqref{eq:triangular_solution} is well-posed. 

In conclusion, the computation of $\hat \varphi^x$ only requires the inversion of some functions, which is easily available in any programming software. For instance, in \texttt{R}, the inversion of a function can be performed using \texttt{uniroot}. 

We summarize the algorithm for the estimation of $\hat \beta$ as follows:
\begin{enumerate}
    \item For $i=1,\dots,n$ generate $\tilde U_i$ and $\Delta_i$. 
    \item For $i=1,\dots,n$ estimate $\hat \varphi(Z_i,X_i,\tilde U_i)$, possibly using the closed-form expression given in \eqref{eq:triangular_solution}.
    \item Obtain $\hat \beta$ using the standard partial likelihood estimator based on the proxy observations $\{(\hat\varphi(Z_i,X_i,\tilde U_i),\Delta_i,V_i)\}_{i=1}^n$.
\end{enumerate}
\section{Asymptotic theory}\label{sec:asymptotic_theory}
In this section, we will discuss the asymptotic results of the estimator $\hat \beta$. First, we will consider its consistency. Then, we will discuss its asymptotic normality.
\subsection{Consistency}\label{sec:consistency}
Define the following quantities:
\begin{align*}
        G(t|z,x,w) &= P(C\le t|Z=z,X=x,W = w);\\
	F_{Y|Z,X,W}(y|z,x,w) &= P(Y \leq y|Z=z,X=x,W = w);\\
	F_{Y,1|Z,X,W}(y|z,x,w) &= P(Y \leq y,\delta = 1|Z=z,X=x,W = w);\\
	f_{Z,X,W}(z,x,w) &= \frac{\partial}{\partial x}P(Z = z,X\le x,W = w);\\
 	f_{X,W}(x,w) &= \frac{\partial}{\partial x}P(X\le x,W = w).
\end{align*}
Define also 
\begin{align*}
    f(t,z|w,x) &= \frac{\partial}{\partial t} F(t,z|x,w);\\
    f(t|z,w,x) &= \frac{\partial}{\partial t} F(t|z,x,w);
\end{align*}
and 
\begin{align*}
    \hat f(t,z|w,x) &= \frac{\partial}{\partial t} \hat F(t,z|x,w);\\
    \hat f(t|z,w,x) &= \frac{\partial}{\partial t} \hat F(t|z,x,w).\\
\end{align*}
Consider the following regularity assumption, where $\alpha\in(0,1)$ corresponds to the order of Lipschitz continuity needed for later theoretical results. A further specification of the value of $\alpha$ is given in Appendix \ref{appendix:asymptotic_normality_proof}.
\begin{assumption}\label{assumtpion:K} The following holds:
	\begin{itemize}
		\item[(i)] For each $(z,w)\in\mathcal{Z}\times\mathcal{W}$, the conditional distribution functions $F(t|z,x,w)$ and $G(t|z,x,w)$ have continuous derivatives in $t\in[0,\bar T]$ and $x\in\mathcal{X}$ up to the third order and bounded fourth-order (included mixed derivatives).
		\item[(ii)] The univariate kernel functions $K(\cdot)$ and $\tilde K(\cdot)$ are compactly supported. The kernel $K(\cdot)$ is a continuously differentiable function of order $\nu$ satisfying $\int K(u)du=1$, $\int K^2(u)du<\infty$, and $\int u^j K(u)du=0$ for $j<\nu$, where $\nu\geq 4$ is an integer.
		 The kernel $\tilde K(\cdot)$ is a continuously differentiable function of order $\pi$ satisfying $\int \tilde K(u)du=1$, $\int \tilde K^2(u)du<\infty$, and $\int u^j \tilde K(u)du=0$ for $j<\pi$, where $\pi\geq 3$ is an integer. In addition, the support of $\tilde K$ is [-1,1]. Lastly,  
    the kernel $\tilde K$ and its derivatives up to order $\pi-1$ are equal to zero at the border of the support. 
		\item[(iii)]For some finite constant $C>0$ the bandwidths $h$ and $\epsilon$ satisfy   $h = Cn^{-u}$ and $\epsilon=Cn^{-w}$ 
  with
  \begin{align*}
\frac{1}{4\nu}<u<\min(\frac{1}{2},\frac{1}{5+2\alpha})
\end{align*}
and
\begin{align*}
	\max(\frac{u(1+\alpha)}{\pi}, u,\frac{1}{4\pi})<w<\min(\frac{1}{3+\alpha},2\nu u-1,\frac{u\nu}{\alpha},\frac{u(\nu+1)}{3+\alpha},\frac{1-u}{5+2\alpha})
\end{align*}
  \item[(iv)] The functions $f_{Z,X,W}(z,x,w)$, $f_{X,W}(x,w)$ are bounded away from zero on their relative support and $\inf_{t\in[0,\bar T], z\in\mathcal{Z},x\in\mathcal{X}, w\in\mathcal{W}}f(t|z,x,w)>0$.
	\end{itemize}
\end{assumption}
A valid example of the values for $(u,w,\alpha,\nu,\pi)$ is, for instance, $(0.143,0.144,0.001,4,3)$. With this choice, we can also construct a valid example for $K(x)$ and $\tilde K(x)$. Consider the function $\bar K(x)= I(|x|\le 1)\frac{693}{512}(1-x^2)^5$. Then, it is easy to check that $K(x) = \tilde K(x) = \frac{3}{2}\bar K(x) + \frac{1}{2}x\bar K'(x)$ satisfies condition $(ii)$.

We obtain the following result, which concerns the consistency and the rate of convergence of $\hat F(t,z|x,w)$ to $F(t,z|x,w)$. In addition, it states the consistency of $\hat f(t,z|x,w)$ in $t$ and $x$ up to order $2+\alpha$. These results will be used to ensure the regularity and the rate of convergence of the estimator $\hat \varphi$, which is crucial for the rate of convergence of $\hat\beta$.  Lastly, we show that the estimators $\hat F(t,z|x,w)$ and $\hat f(t,z|x,w)$ admit an iid representation that will be employed to obtain an analogous representation for $\hat \varphi$, which will lead to the asymptotic normality of $\hat{\beta}$.
\begin{theorem}\label{theo:F_results}
	Under Assumptions \ref{Cind} and \ref{assumtpion:K}, the following results hold:
	\begin{itemize}
		\item[(i)]$
		\sup\limits_{t\in[0,\bar T],z\in\mathcal{Z},x\in\mathcal{X},w\in\mathcal{W}}|\hat F(t,z|x,w) - F(t,z|x,w)|=  O_p((\log n/(nh))^{1/2} + h^\nu + \epsilon^\pi ).$
		\item[(ii)] $
		\sup\limits_{t\in[0,\bar T],z\in\mathcal{Z},x\in\mathcal{X},w\in\mathcal{W}}|\hat f(t,z|x,w) - f(t,z|x,w)| =  O_p((\log n/(nh\epsilon))^{1/2} + h^\nu + \epsilon^\pi ).$
		\item[(iii)]$\sup\limits_{t\in[0,\bar T],z\in\mathcal{Z},x\in\mathcal{X},w\in\mathcal{W}} |\frac{\partial}{\partial t}\hat f(t,z|x,w) -\frac{\partial}{\partial t}f(t,z|x,w) | = o_p(1).$
			\item[(iv)] $\sup\limits_{t\in[0,\bar T],z\in\mathcal{Z},x\in\mathcal{X},w\in\mathcal{W}} |\frac{\partial^2}{\partial t^2}\hat f(t,z|x,w) -\frac{\partial^2}{\partial t^2}f(t,z|x,w) | = o_p(1).$
		\item[(v)] $\sup\limits_{t_1,t_2\in[0,\bar T],z\in\mathcal{Z},x\in\mathcal{X},w\in\mathcal{W}} \frac{|\frac{\partial^2}{\partial t^2}\hat f(t_1,z|x,w) -\frac{\partial^2}{\partial t^2}f(t_2,z|x,w) |}{|t_1-t_2|^\alpha} = o_p(1).$
		\item[(vi)]$\sup\limits_{t\in[0,\bar T],z\in\mathcal{Z},x\in\mathcal{X},w\in\mathcal{W}} |\frac{\partial}{\partial x}\hat f(t,z|x,w) -\frac{\partial}{\partial x}f(t,z|x,w) | = o_p(1).$
		\item[(vii)] $\sup\limits_{t\in[0,\bar T],z\in\mathcal{Z},x\in\mathcal{X},w\in\mathcal{W}} |\frac{\partial^2}{\partial x^2}\hat f(t,z|x,w) -\frac{\partial^2}{\partial x^2}f(t,z|x,w) | = o_p(1).$
		\item[(viii)] $\sup\limits_{t\in[0,\bar T],z\in\mathcal{Z},x_1,x_2\in\mathcal{X},w\in\mathcal{W}} \frac{|\frac{\partial^2}{\partial x^2}\hat f(t,z|x_1,w) -\frac{\partial^2}{\partial x^2}f(t,z|x_2,w) |}{|x_1-x_2|^\alpha} = o_p(1).$
		\item[(ix)] For $(t,z,x,w)\in[0,\bar T]\times \mathcal{Z}\times\mathcal{X}\times\mathcal{W}$, the quantity $\hat F(t,z|x,w) - F(t,z|x,w)$  can be expressed as 
		\begin{align*}
   \hat F(t,z|x,w) - F(t,z|x,w)
    &= (nh)^{-1} \sum_{i=1}^n K\left(\frac{x-X_i}{h}\right)\eta^F(Y_i, \delta_i, Z_i,  W_i, t, z, x, w) \\
    &\quad + R_n(t,z,x,w);
		\end{align*}
		where
		\begin{align*}
  	\eta^F&(Y_i, \delta_i, Z_i,  W_i, t, z, x, w) \\ &=F(t|z,x,w) I(W_i=w)\frac{I(Z_i=z)-p_{z,x,w}}{f_{X,W}(x,w)} + p_{z,x,w}\xi^F(Y_i, \delta_i, Z_i, W_i, t, z, x, w);\\
			\xi^F&(Y_i, \delta_i, Z_i,  W_i, t, z, x, w) \\
			&=   
   (1 - F(t|z,x,w))\Big[ \int_0^{\min(Y_i,t)} \frac{-\text{d}F_{Y,1|Z,X,W}(y|z,x,w)}{\big(1 - F_{Y|Z,X,W}(y|z,x,w)\big)^2} \\
			&\quad\quad + \frac{\delta_iI(Y_i\le t)}{1-F_{Y|Z,X,W}(Y_i|z,x,w)} \Big];
		\end{align*}
  and \begin{align}
      \sup_{t\in [0,\bar T],z\in\mathcal{Z},x\in\mathcal{X},w\in\mathcal{W}}|R_n(t,z,x,w)| =  O_p\left((\log n/(nh))^{3/4} + h^\nu + \epsilon^\pi \right).
  \end{align}
		\item[(x)] For $(t,z,x,w)\in[0,\bar T]\times \mathcal{Z}\times\mathcal{X}\times\mathcal{W}$, the quantity  $\hat f(t,z|x,w) - f(t,z|x,w)$ can be expressed as 
		\begin{align*}
			\hat f(t,z|x,w) -&f(t,z|x,w)=(nh)^{-1}\sum_{i=1}^n K\left(\frac{x - X_i}{h}\right) I(W_i=w)\frac{I(Z_i=z)-p_{z,x,w}}{f_{X,W}(x,w)} f(t|z,x,w) \\
			&\quad+ (nh\epsilon)^{-1}\sum_{i=1}^n p_{z,x,w} K\left(\frac{x - X_i}{h}\right) \int \tilde K(u)\xi^F(Y_i, \delta_i, Z_i,  W_i, t-u\epsilon, z, x, w)\text{d}u\\
			&\quad+ r_n(t,z,x,w)
		\end{align*}
  and 
  \begin{align*}
       \sup_{ t\in [0,\bar T],z\in\mathcal{Z},x\in\mathcal{X},w\in\mathcal{W}}|r_n(t,z,x,w)| = O_p\big((\log n/(nh\epsilon))^{3/4} + h^\nu + \epsilon^\pi \big).
  \end{align*}
	\end{itemize}
\end{theorem}
To proceed, denote by $\|\cdot\|_\infty$ the infinity norm. The argument over which the supremum is taken may vary throughout the paper, and it is specified whenever any doubt might arise.  Introduce the following assumption:
\begin{assumption}\label{assumption:C}
	For all $\varepsilon>0$, there exists $\varsigma>0$ such that for all $x\in\mathcal{X}$, we have 	
	$\inf_{\theta\in\mathcal{F}_{Z}^{\bar U, \bar T}: \|\theta-\varphi^x\|_{\infty}\ge \varsigma} \|A(\theta,F^x)\|_{\infty}\ge \varepsilon$.
\end{assumption}
Assumption \ref{assumption:C} is related to the shape of the objective function $A$ and ensures that there is a unique solution to the system of equations \eqref{eq:system} in $[0,\bar U]$ and, hence, that there is a unique minimum to the program \eqref{eq:estimator_phi} when $\hat F$ estimates $F$ well enough. 
\begin{theorem}\label{theo:consistency_beta} Let Assumption \ref{assumption:C} and the conditions of Theorem \ref{BID} hold. Then, $\hat\beta-\beta_0 = o_p(1)$.
\end{theorem}

\subsection{Asymptotic distribution}\label{sec:asymptotic_normality}
Denote the Fréchet derivative of $A$ in its first argument at the point $(\tilde\varphi^x,\tilde F^x)\in \mathcal{F}_{Z}^{\bar U, \bar T}\times \mathcal{F}_{\uparrow}^{Z,W}$ by 
\begin{align*}
	\big(\Gamma(\tilde\varphi^x,\tilde F^x)(u)\big)_{l,k} = - \tilde f^x(\tilde \varphi^x_{l}(u),z_l|w_k),
\end{align*}
where the subscript $l,k$ indicates the $(l,k)$th entry of the matrix for $l,k=1,\dots,L$, and $\tilde f^x$ is the derivative in the first argument of $\tilde F^x$. Consider the following assumption.
\begin{assumption}\label{assumption:N}
 There exists $\varsigma>0$ such that the lowest eigenvalue of $\Gamma(\varphi^x,F^x)(u)$ is greater than $\varsigma$ for all $u\in[0,\bar U]$, $x\in\mathcal{X}$.
\end{assumption}

\begin{theorem}\label{theo:asymptotic_normality}  Let Assumptions \ref{assumtpion:K}, \ref{assumption:C}, \ref{assumption:N} and the conditions of Theorem \ref{BID} hold. Then $\sqrt{n}(\hat\beta-\beta_0)$ converges weakly to a mean zero Normal random variable. 
\end{theorem}
Note that Theorem \ref{theo:asymptotic_normality} does not specify the asymptotic covariance matrix of $\sqrt{n}(\hat\beta-\beta_0)$ since the proof is based on showing that the estimator can be written as an empirical process over a Donsker class of functions. This result ensures asymptotic normality, but to obtain an explicit expression of the covariance matrix, one would need a complete representation of the empirical process. This is possible, but, as can be easily seen from the proof, the expression would be so involved that it would lack meaningful insights into the distribution and be difficult to use for practical inference purposes.

\section{Simulations}\label{sec:simulations}
In this section, we evaluate the finite sample performance of our proposed method using Monte Carlo simulation. Our main objective is to compare the performance of the partial likelihood estimator proposed by Cox, an estimator that does not take endogeneity into account, with our new estimator. We will consider several designs, of which the specifications are given in Table \ref{table:simulation_design}. Since our method allows for the inclusion of both discrete and continuous exogenous variables $X$, we will consider several univariate distributions for $X$: the continuous Beta design, the continuous uniform design, and the discrete Bernoulli design.

For all designs, we set the structural baseline cumulative hazard function $\Lambda_0$ to that of an exponential variable with mean 1, i.e. $\Lambda_0(s) = s$, and fix the value of $\beta_0 =  (\beta_{0,z},\beta_{0,x})^\top$. This results in the potential outcome $T(z,x)$ following an exponential distribution with a rate (reciprocal of the mean) equal to $\exp(x\beta_{0,x})$ if $z = 0$ and  $\exp(\beta_{0,z}+x\beta_{0,x})$ if $z=1$. The specifications of the design are given in Table \ref{table:simulation_design}. It is worth noting that the variable $Z$ is endogenous as it depends on the unobserved heterogeneity term $U$, and that $W$ is a proper instrumental variable.

In each design, the probability of $Z = 1$ given a certain value of the variable $W$ is as follows: $P(Z = 1|W = 0) = 0$, and $P(Z = 1|W = 1)$ is around 0.54, mimicking the empirical application considered in the next section. Moreover, in each design, $X$ and $W$ are independent. 
\begin{table}[H]
\centering
\begin{adjustbox}{width=1\textwidth}
\begin{tabular}{|c|c|c|c|c|}
\hline
Design     & $W$    & $X$         & $Z$                                         & $(\beta_{0,z},\beta_{0,x})$ \\ \hline
\makecell{continuous\\Beta} & B(0.5) & Beta(2,5) & $I(W=1)I(0.5U-W+0.45+X+0.5\epsilon\ge0)$     & $(0.7,0.3)$                 \\ \hline
\makecell{continuous\\uniform} & B(0.5) & $\mathcal{U}$(-0.5,0.5) & $I(W=1)I(0.5U-W+0.8+X+0.5\epsilon\ge0)$     & $(0.7,0.3)$                 \\ \hline
\makecell{discrete\\Bernoulli }   & B(0.5) & B(0.5)      & $I(W=1)I(0.5U-W+0.65+0.3X+0.5\epsilon\ge0)$ & $(0.7,0.7)$                 \\ \hline
\end{tabular}
\end{adjustbox}
\caption{\label{table:simulation_design} Distribution and coefficient specifications for each design. Note that B($p$) stands for the Bernoulli distribution of parameter $p$, $\mathcal{U}$($a,b$) represents the uniform distribution on the interval [$a,b$], Beta($\alpha$,$\beta$) represents the Beta distribution with shape parameters $\alpha$ and $\beta$, and $\epsilon$ is a standard normal variable. The random variable $U$ is the unobserved heterogeneity of the model, which follows a standard uniform distribution $U\sim\mathcal{U}(0,1)$. }
\end{table}
We set the censoring variable, $C$, to be distributed as an exponential variable with rate (reciprocal of the mean) $\lambda$. We then consider different levels of censoring, specifically 20\% and 40\%, by setting $\lambda = 0.43$ and $\lambda= 1.15$, respectively, for the discrete Bernoulli design, $\lambda = 0.30$ and $\lambda = 0.82$, respectively, for the continuous uniform design, and $\lambda = 0.33$ and $\lambda = 0.87$, respectively, for the continuous Beta design. Finally, with $Y=\min(T,C)$ and $\delta = I(T\le C)$ we generate an iid sample of size $n = 500,1000$ observations having the same distribution as $(Y,\delta,Z,X,W)$.

We employ the Epanechnikov kernel to smooth the variable $X$ when estimating $F^x$ and, for each $X=x$, we select the bandwidth with a data-driven direct plug-in estimate of the Integrated Mean Squared Error-optimal bandwidth. This corresponds to the default bandwidth selection proposed in the \texttt{R} package \texttt{nprobust}, for which we refer to \cite{calonico2019nprobust}. Our simulation studies showed that smoothing in the argument $t$ during the estimation procedure for $F^x$ leads to a little deterioration of the result and significantly increases the estimation time. Therefore, we consider smoothing in the argument $t$ to be a theoretical requirement that permits to show the asymptotic normality of the estimator, but that can be safely avoided during the method's application. Similarly, we chose to use the Epanechnikov kernel for our simulations, which is a function of order $\nu =2$, even though Assumption \ref{assumtpion:K} specifies an order of $\nu \ge 4$ for the kernel, due to its widespread acceptance in the field.


The upper bound of $C$ is infinite, allowing for any value of $\bar U\in(0,1)$ to be theoretically chosen, and so we set $\bar U = 0.9$. As we explain in Appendix \ref{appendix:technicalities}, we can choose $U_g^c$ having a degenerate distribution with a unique point mass at $\bar U$. 

The results given in Table \ref{table:simulation_results} are based on $N=500$ replications. We present the average bias, standard deviation, and Mean Squared Error (MSE) for each component of $\beta = (\beta_z,\beta_x)^\top$. Additionally, we report the coverage of $95\%$ bootstrap confidence intervals. The intervals are constructed by estimating the standard deviation of the proposed estimator based on a naive bootstrap resampling procedure with replacement, and then plugging in the standard deviation in a normal approximation of the bounds of confidence intervals. To accelerate the simulations, we employed the method proposed in \cite{giacomini2013warp}.
To evaluate the estimator on a component-wise basis, we also include the Root Mean Squared Error (RMSE) as a measure to assess the overall power of the estimator. The value is computed following the formula $RMSE = \sqrt{N^{-1}\sum_{j=1}^N \|\hat\beta^{(j)}-\beta_0\|^2}$, where $\hat\beta^{(j)}$ is the estimation of simulation $j$. For comparison purposes, the same quantities are also given for the partial likelihood estimator.

The results show that the proposed estimator has low bias. Additionally, as the sample size increases, the bias, and coverage of the estimator approach their theoretical values of 0 and 0.95 respectively. The performance of the estimator improves when the proportion of censored observations is lower. It can be argued that the estimator performs better under the discrete Bernoulli design, which may be attributed to the simpler structure of this design, as the exogenous component is discrete rather than continuous.  As expected, the standard partial likelihood estimator is biased.

\begin{table}[H]
\centering
\begin{adjustbox}{width=1\textwidth}
\begin{tabular}{cccc|rrrrr|rrrr|}
\cline{5-13}
                                                                                           &                                            &                                            &           & \multicolumn{5}{c|}{proposed estimator}                                                                                             & \multicolumn{4}{c|}{partial likelihood estimator}                                                       \\ \hline
\multicolumn{1}{|c|}{Design}                                                               & \multicolumn{1}{c|}{Cens}                  & \multicolumn{1}{c|}{$n$}                     & Comp      & \multicolumn{1}{c}{Bias} & \multicolumn{1}{c}{Sd} & \multicolumn{1}{c}{MSE} & \multicolumn{1}{c}{RMSE} & \multicolumn{1}{l|}{CP95} & \multicolumn{1}{c}{Bias} & \multicolumn{1}{c}{Sd} & \multicolumn{1}{c}{MSE} & \multicolumn{1}{c|}{RMSE} \\ \hline
\multicolumn{1}{|c|}{\multirow{8}{*}{\makecell{continuous\\Beta}}}    & \multicolumn{1}{c|}{\multirow{4}{*}{20\%}} & \multicolumn{1}{c|}{\multirow{2}{*}{500}}  & $\beta_z$ & -0.084                   & 0.283                  & 0.087                   & \multirow{2}{*}{0.229}   & 0.946                      & -0.262                   & 0.163                  & 0.095                   & \multirow{2}{*}{0.285}    \\
\multicolumn{1}{|c|}{}                                                                     & \multicolumn{1}{c|}{}                      & \multicolumn{1}{c|}{}                      & $\beta_x$ & -0.023                   & 0.586                  & 0.344                   &                          & 0.956                      & 0.093                    & 0.468                  & 0.228                   &                           \\ \cline{3-13} 
\multicolumn{1}{|c|}{}                                                                     & \multicolumn{1}{c|}{}                      & \multicolumn{1}{c|}{\multirow{2}{*}{1000}} & $\beta_z$ & -0.047                   & 0.215                  & 0.049                   & \multirow{2}{*}{0.159}   & 0.934                      & -0.265                   & 0.106                  & 0.081                   & \multirow{2}{*}{0.275}    \\
\multicolumn{1}{|c|}{}                                                                     & \multicolumn{1}{c|}{}                      & \multicolumn{1}{c|}{}                      & $\beta_x$ & -0.040                   & 0.408                  & 0.168                   &                          & 0.964                      & 0.083                    & 0.319                  & 0.108                   &                           \\ \cline{2-13} 
\multicolumn{1}{|c|}{}                                                                     & \multicolumn{1}{c|}{\multirow{4}{*}{40\%}} & \multicolumn{1}{c|}{\multirow{2}{*}{500}}  & $\beta_z$ & -0.050                   & 0.303                  & 0.094                   & \multirow{2}{*}{0.237}   & 0.964                      & -0.291                   & 0.184                  & 0.119                   & \multirow{2}{*}{0.319}    \\
\multicolumn{1}{|c|}{}                                                                     & \multicolumn{1}{c|}{}                      & \multicolumn{1}{c|}{}                      & $\beta_x$ & -0.004                   & 0.754                  & 0.569                   &                          & 0.974                      & 0.112                    & 0.532                  & 0.295                   &                           \\ \cline{3-13} 
\multicolumn{1}{|c|}{}                                                                     & \multicolumn{1}{c|}{}                      & \multicolumn{1}{c|}{\multirow{2}{*}{1000}} & $\beta_z$ & -0.042                   & 0.237                  & 0.058                   & \multirow{2}{*}{0.179}   & 0.966                      & -0.298                   & 0.125                  & 0.105                   & \multirow{2}{*}{0.311}    \\
\multicolumn{1}{|c|}{}                                                                     & \multicolumn{1}{c|}{}                      & \multicolumn{1}{c|}{}                      & $\beta_x$ & 0.002                    & 0.492                  & 0.242                   &                          & 0.986                      & 0.112                    & 0.359                  & 0.141                   &                           \\ \hline
\multicolumn{1}{|c|}{\multirow{8}{*}{\makecell{continuous\\uniform}}} & \multicolumn{1}{c|}{\multirow{4}{*}{20\%}} & \multicolumn{1}{c|}{\multirow{2}{*}{500}}  & $\beta_z$ & -0.049                   & 0.287                  & 0.085                   & \multirow{2}{*}{0.206}   & 0.960                      & -0.230                   & 0.162                  & 0.079                   & \multirow{2}{*}{0.256}    \\
\multicolumn{1}{|c|}{}                                                                     & \multicolumn{1}{c|}{}                      & \multicolumn{1}{c|}{}                      & $\beta_x$ & -0.040                   & 0.331                  & 0.111                   &                          & 0.964                      & 0.084                    & 0.256                  & 0.073                   &                           \\ \cline{3-13} 
\multicolumn{1}{|c|}{}                                                                     & \multicolumn{1}{c|}{}                      & \multicolumn{1}{c|}{\multirow{2}{*}{1000}} & $\beta_z$ & -0.038                   & 0.191                  & 0.038                   & \multirow{2}{*}{0.144}   & 0.960                      & -0.231                   & 0.110                  & 0.066                   & \multirow{2}{*}{0.244}    \\
\multicolumn{1}{|c|}{}                                                                     & \multicolumn{1}{c|}{}                      & \multicolumn{1}{c|}{}                      & $\beta_x$ & -0.020                   & 0.236                  & 0.056                   &                          & 0.950                      & 0.081                    & 0.181                  & 0.039                   &                           \\ \cline{2-13} 
\multicolumn{1}{|c|}{}                                                                     & \multicolumn{1}{c|}{\multirow{4}{*}{40\%}} & \multicolumn{1}{c|}{\multirow{2}{*}{500}}  & $\beta_z$ & -0.023                   & 0.321                  & 0.103                   & \multirow{2}{*}{0.226}   & 0.966                      & -0.257                   & 0.185                  & 0.100                   & \multirow{2}{*}{0.287}    \\
\multicolumn{1}{|c|}{}                                                                     & \multicolumn{1}{c|}{}                      & \multicolumn{1}{c|}{}                      & $\beta_x$ & -0.098                   & 0.481                  & 0.241                   &                          & 0.926                      & 0.102                    & 0.297                  & 0.099                   &                           \\ \cline{3-13} 
\multicolumn{1}{|c|}{}                                                                     & \multicolumn{1}{c|}{}                      & \multicolumn{1}{c|}{\multirow{2}{*}{1000}} & $\beta_z$ & -0.030                   & 0.210                  & 0.045                   & \multirow{2}{*}{0.153}   & 0.964                      & -0.262                   & 0.125                  & 0.084                   & \multirow{2}{*}{0.276}    \\
\multicolumn{1}{|c|}{}                                                                     & \multicolumn{1}{c|}{}                      & \multicolumn{1}{c|}{}                      & $\beta_x$ & -0.064                   & 0.351                  & 0.127                   &                          & 0.972                      & 0.103                    & 0.204                  & 0.052                   &                           \\ \hline
\multicolumn{1}{|c|}{\multirow{8}{*}{\makecell{discrete\\Bernoulli}}} & \multicolumn{1}{c|}{\multirow{4}{*}{20\%}} & \multicolumn{1}{c|}{\multirow{2}{*}{500}}  & $\beta_z$ & -0.005                   & 0.291                  & 0.085                   & \multirow{2}{*}{0.205}   & 0.962                      & -0.245                   & 0.159                  & 0.086                   & \multirow{2}{*}{0.269}    \\
\multicolumn{1}{|c|}{}                                                                     & \multicolumn{1}{c|}{}                      & \multicolumn{1}{c|}{}                      & $\beta_x$ & 0.013                    & 0.232                  & 0.054                   &                          & 0.950                      & 0.036                    & 0.155                  & 0.025                   &                           \\ \cline{3-13} 
\multicolumn{1}{|c|}{}                                                                     & \multicolumn{1}{c|}{}                      & \multicolumn{1}{c|}{\multirow{2}{*}{1000}} & $\beta_z$ & -0.011                   & 0.188                  & 0.035                   & \multirow{2}{*}{0.140}   & 0.960                      & -0.247                   & 0.108                  & 0.072                   & \multirow{2}{*}{0.259}    \\
\multicolumn{1}{|c|}{}                                                                     & \multicolumn{1}{c|}{}                      & \multicolumn{1}{c|}{}                      & $\beta_x$ & 0.019                    & 0.149                  & 0.023                   &                          & 0.972                      & 0.033                    & 0.105                  & 0.012                   &                           \\ \cline{2-13} 
\multicolumn{1}{|c|}{}                                                                     & \multicolumn{1}{c|}{\multirow{4}{*}{40\%}} & \multicolumn{1}{c|}{\multirow{2}{*}{500}}  & $\beta_z$ & 0.037                    & 0.319                  & 0.103                   & \multirow{2}{*}{0.220}   & 0.976                      & -0.270                   & 0.184                  & 0.107                   & \multirow{2}{*}{0.299}    \\
\multicolumn{1}{|c|}{}                                                                     & \multicolumn{1}{c|}{}                      & \multicolumn{1}{c|}{}                      & $\beta_x$ & -0.007                   & 0.271                  & 0.074                   &                          & 0.964                      & 0.045                    & 0.178                  & 0.034                   &                           \\ \cline{3-13} 
\multicolumn{1}{|c|}{}                                                                     & \multicolumn{1}{c|}{}                      & \multicolumn{1}{c|}{\multirow{2}{*}{1000}} & $\beta_z$ & 0.022                    & 0.203                  & 0.042                   & \multirow{2}{*}{0.162}   & 0.960                      & -0.276                   & 0.125                  & 0.092                   & \multirow{2}{*}{0.290}    \\
\multicolumn{1}{|c|}{}                                                                     & \multicolumn{1}{c|}{}                      & \multicolumn{1}{c|}{}                      & $\beta_x$ & 0.010                    & 0.163                  & 0.027                   &                          & 0.972                      & 0.045                    & 0.119                  & 0.016                   &                           \\ \hline
\end{tabular}
\end{adjustbox}
\caption{\label{table:simulation_results}Simulation results for each design. We include for the proposed estimator and the standard partial likelihood estimator, for each level of censoring (Cens) and for different sample sizes $(n)$, the average of the bias (Bias),  the standard deviation (Sd) and the Mean Squared Error (MSE) for each component (Comp) of $\beta = (\beta_z,\beta_x)^\top$. For each case, we also include the Root Mean Squared Error value (RMSE). For the proposed estimator, we also provide the coverage of the 95\% confidence intervals (CP95) for each component of $\beta$.  }
\end{table}

\section{Empirical application}\label{sec:empirical_application}
The Illinois Unemployment Incentive Experiment was a controlled social experiment conducted by the Illinois Department of Employment Security in 1984 to evaluate if  cash bonuses reduce the duration of unemployment. New claimants for Unemployment Insurance (UI) were randomly assigned to one of three groups: the Job Search Incentive Experiment group (JSIE), the Hiring Incentive Experiment group (HIE), or the control group. The JSIE group was eligible for a \$500 bonus if they found a job of at least 30 hours per week within 11 weeks of the start of their unemployment period and held the job for 4 months. The Hiring Incentive Experiment group had the same eligibility requirements, but the \$500 bonus was given to the hiring company instead. A detailed description of the experiment can be found in \cite{woodbury1987bonuses}. 

It is worth noting that to be part of one of the three aforementioned groups, claimants must be between 20 and 55 years old and have a valid unemployment insurance claim. The duration of unemployment $T$ was recorded as the number of weeks in which participants received unemployment benefits, resulting in a discrete data set. However, the true unemployment duration is continuous but only observed by intervals. This creates a partial identification issue, not within the scope of this paper. The data is treated as continuous. It is important to note that unemployment insurance is granted for 26 weeks, so observations can only be made up until the end of the UI claim and are therefore right censored at 26 weeks. Approximately 40\% of the observations are censored.

To clarify the connection between our model and the experiment, we examine the impact of the cash bonus in the JSIE (or HIE) experiment. Although group assignment is randomly determined, participant agreement to participate in the experiment is not independent. This is because, to participate in the program, job seekers must also read the experiment description and sign an agreement form at the start of their follow-up, and this can be influenced by the participant’s motivation to find work or by other personal attributes, such as specific skills. A framework with two treatments, as indicated by the following variable $Z=(Z_1,Z_2)^\top$, will be considered:
\begin{align*}
    Z_1&= \begin{cases}
      1 & \text{if the individual is in the JSIE group and agrees to participate}\\
       0 & \text{otherwise}\\
       \end{cases}\\
       Z_2 &=  \begin{cases}
      1 & \text{if the individual is in the HIE group and agrees to participate}\\
       0 & \text{otherwise.}\\
       \end{cases}
\end{align*}
Therefore, $Z=(0,0)^\top$ corresponds to the control group, and $Z=(1,0)^\top$, $Z=(0,1)^\top$, correspond to the JSIE and HIE group, respectively. To address selection bias we utilize the group assignment variable $W$ as instrumental variable:
\begin{align*}
    W= \begin{cases}
      0 & \text{if the individual is assigned to the control group }\\
      1 & \text{if the individual is assigned to the JSIE group}\\
      2 & \text{if the individual is assigned to the HIE group.}
    \end{cases}
\end{align*}
We consider for this analysis a subset of 1,543 non-white women out of the 12,101 available individuals and we report in Table \ref{table:ZWtable} the sample sizes for each $(W, Z)$ combination. JSIE and HIE experiments have refusal rates of 20\% and 35\%, indicating significant selection biases. 
\begin{table}[H]
\centering
\begin{tabular}{|c|c|c|c|}
\hline
   & $W=0$ & $W=1$ & $W=2$ \\ \hline
$Z=(0,0)^\top$ & 527  & 104   & 176   \\ \hline
$Z=(1,0)^\top$ & 0     & 408  & 0     \\ \hline
$Z=(0,1)^\top$ & 0     & 0     &328  \\ \hline
\end{tabular}
\caption{\label{table:ZWtable} Number of observations for each combination of levels of the variable $W$ and $Z$. }
\end{table}
We use the proposed method to estimate the vector of proportional hazards coefficients. Similar to Section \ref{sec:simulations}, we estimate the conditional distribution function using the Epanechnikov kernel, and the same selection method for the bandwidth using data-driven direct plug-in estimates of the Integrated Mean Squared Error-optimal bandwidth. To avoid numerical problems, we set $\bar U = 0.5$ and $\bar T = 26$. Table \ref{table:ApplicationResults} displays the estimation results of the proposed estimator compared to the partial likelihood estimator, where the standard deviation is computed using 500 bootstrap resamples with replications, and the confidence intervals are constructed using the estimated standard deviation and a normal approximation.

\begin{table}[H]
\centering
\begin{adjustbox}{width=1\textwidth}
\begin{tabular}{c|rrrr|rrrr|}
\cline{2-9}
                                    & \multicolumn{4}{c|}{proposed estimator}                                                                                 & \multicolumn{4}{c|}{partial likelihood estimator}                                                                        \\ \hline
\multicolumn{1}{|c|}{Comp}          & \multicolumn{1}{c|}{Est}   & \multicolumn{1}{c|}{Sd}    & \multicolumn{1}{c|}{CI 0.025} & \multicolumn{1}{c|}{CI 0.975} & \multicolumn{1}{c|}{Est}    & \multicolumn{1}{c|}{Sd}    & \multicolumn{1}{c|}{CI 0.025} & \multicolumn{1}{c|}{CI 0.975}                         \\ \hline
\multicolumn{1}{|c|}{$\beta_{z,1}$} & \multicolumn{1}{r|}{0.910} & \multicolumn{1}{r|}{0.327} & \multicolumn{1}{r|}{0.270}    & 1.550                         & \multicolumn{1}{r|}{0.079}  & \multicolumn{1}{r|}{0.090} & \multicolumn{1}{r|}{-0.098}   & 0.256                         \\ \hline
\multicolumn{1}{|c|}{$\beta_{z,2}$} & \multicolumn{1}{r|}{0.929} & \multicolumn{1}{r|}{0.339} & \multicolumn{1}{r|}{0.264}    & 1.594                         & \multicolumn{1}{r|}{0.075}  & \multicolumn{1}{r|}{0.097} & \multicolumn{1}{r|}{-0.115}   & 0.264 \\
\hline
\multicolumn{1}{|c|}{$\beta_{x}$}   & \multicolumn{1}{r|}{0.058} & \multicolumn{1}{r|}{0.098} & \multicolumn{1}{r|}{-0.135}   & 0.250                         & \multicolumn{1}{r|}{-0.136} & \multicolumn{1}{r|}{0.040} & \multicolumn{1}{r|}{-0.215}   & -0.058\\ \hline
\end{tabular}
\end{adjustbox}
\caption{\label{table:ApplicationResults} Estimation results for the proposed estimator compared to the partial likelihood estimator. For each component (Comp)  of $\beta = (\beta_{z,1},\beta_{z,2},\beta_{x})^\top$, we give the estimated value (Est), the standard deviation (Sd) and the confidence intervals at the 95\% level (CI 0.025 and CI 0.975).  }
\end{table}

The results in Table \ref{table:ApplicationResults} show a remarkable difference between the proposed estimator and the standard partial likelihood estimator. Specifically, the estimated values for $\beta_{z,1}$ and $\beta_{z,2}$, using the proposed estimator, are both positive and significant at the 95\% confidence level. This provides empirical evidence that both treatments increase the hazard rate of finding a job, that is they reduce unemployment duration. Conversely, the values for these coefficients estimated using the standard method are close to zero and not significant at the same confidence level, leading to conclude that there is no evidence of any effects of the treatments on the duration.

Additionally, the estimated value for $\beta_{x}$ using the proposed procedure is positive but not significantly different from zero at the 95\% confidence level. This could be interpreted as there being no evidence of the effect of the age of  non-white women on the duration of unemployment. In contrast, the estimated value for that coefficient is negative and significantly different from zero when the standard partial likelihood estimation procedure is used. This would mean that the older the woman, the more difficult it is to find a job.

In conclusion, the proposed estimator and the partial likelihood estimator yield different results leading to notably different cause-effect interpretations. It could be argued that our estimator provides a more reliable estimation of the regression coefficients compared to the partial likelihood estimator, especially when assessing the influence of the treatments on the duration outcome variable.

\section*{Acknowledgments}
The authors thank Gerda Claeskens, Elia Lapenta and Juan Carlos Pardo-Fern´andez for comments
that improved the paper. Jad Beyhum undertook most of this work while employed by CREST,
ENSAI. Ingrid Van Keilegom acknowledges support from the FWO and F.R.S.-FNRS under the
Excellence of Science (EOS) programme, project ASTeRISK (grant No. 40007517).

\newpage
\bibliographystyle{apalike}
\bibliography{main}
\appendix
\section*{Appendix}
The following Appendix includes the proofs of Theorems \ref{GID}, \ref{theo:F_results}, \ref{theo:consistency_beta} and \ref{theo:asymptotic_normality}, which are provided in Section \ref{appendix:GID}, \ref{appendix:F_results}, \ref{appendix:consistency} and \ref{appendix:asymptotic_normality}, respectively. The proofs make use of the technicalities and the notations discussed in Section \ref{appendix:technicalities_notation}.

\section{Technicalities and notation}\label{appendix:technicalities_notation}

\subsection{Technicalities}\label{appendix:technicalities}
The three steps of the estimation procedure discussed in the paper aim to clarify the generality of our method. However, in order to obtain asymptotic results on the estimator $\hat \beta$, it is necessary to properly formalize the fact that the estimator is based on  ``proxy'' observations. For that, we can argue as follows. 
Define the space 
$\mathcal{F}_{V}^{\bar U, \bar T} = \{\phi:\mathcal{V}\times [0,\bar U]\xrightarrow{} [0,\bar T]| \|\phi\|_{\infty}<\infty\}$, where $\|\phi\|_{\infty}  = \sup_{v\in\mathcal{V},u\in[0,\bar U]}|\phi(v,u)|$. Then, for any function $\phi\in\mathcal{F}_{V}^{\bar U, \bar T}$, and any continuous function $g:\mathcal{V}\mapsto \mathbb{R}^l$ for some integer $l\ge 1$, consider the quantities:
\begin{align*}
	Q^\phi(t) &= P(\phi(V,\tilde U)\ge t, \Delta = 1);\\
	E^\phi(g(v),t) &= E[g(V)I(\phi(V,\tilde U)\ge t)];\\
	E_1^\phi(g(v),t) &= E[g(V)I(\phi(V,\tilde U)\ge t,\Delta = 1)].
\end{align*}

Consider the operator $M$ which associates to a function $\phi\in \mathcal{F}_{V}^{\bar U, \bar T}$ a map from $\mathcal{B}$ to $\mathbb{R}^{d_V}$, denoted by $M(\cdot, \phi)$ and defined as 
\begin{align}\label{eq:M_operator}
	M(\beta,\phi) = E_1^\phi(v,0) + \int_0^{\bar T}\frac{E^\phi(v\exp(v^\top\beta),s)}{E^\phi(\exp(v^\top\beta),s)}\text{d}Q^\phi(s).
\end{align}
The identification result of Theorem \ref{BID} can be equivalently stated saying that $\beta_0$ is the unique point such that  
\begin{align}\label{eq:M_equal_zero}
	\|M(\beta_0,\varphi)\|= 0.
\end{align}
This is because $\|M(\beta,\varphi)\|=0$ is the first order condition of the maximization problem in $\beta$ of the expected value of the PH partial likelihood based on the variables $\varphi(Z,\tilde U), \Delta, Z$ with respect to $(\tilde U, \Delta, Z)$. For any function $\phi\in\mathcal{F}_{V}^{\bar U, \bar T}$, and any continuous function $v\mapsto g(v)$, we define the sample analogue of the quantities $ Q^\phi(t), E^\phi(g(v),t)$ and $E_1^\phi(g(v),t)$, respectively, as 
\begin{align*}
	\hat Q^\phi(t) &=  n^{-1}\sum_{i=1}^n I\big(\phi(V_i,\tilde U_i)\ge t, \Delta_i = 1\big);\\
	\hat E^\phi(g(v),t) &= n^{-1}\sum_{i=1}^ng(V_i)I\big(\phi(V_i,\tilde U_i)\ge t\big);\\
	\hat E_1^\phi(g(v),t) &= n^{-1}\sum_{i=1}^ng(V_i)I\big(\phi(V_i,\tilde U_i)\ge t,\Delta_i = 1\big).
\end{align*}

Then, we consider the sample analogue of the operator $M$, namely $M_n$, defined as 
\begin{align}
	M_n(\beta,\phi) = \hat E_1^\phi(v,0) + \int_0^{\bar T}\frac{\hat E^\phi\big(v\exp(v^\top\beta),s\big)}{\hat E^\phi\big(\exp(v^\top\beta),s\big)}\text{d}\hat Q^\phi(s),
\end{align}

Then, the estimator $\hat{\beta}$ can be rewritten in a convenient mathematical reformulation, which underlines its dependency on the estimator of $\hat \varphi$:
\begin{align}\label{eq:estimator_beta}
	\hat \beta \in \argmin_{\beta\in\mathcal{B}}\big\|M_n(\beta, \hat\varphi)\big\|.
\end{align}

The results we propose hold for any chosen distribution for $ U_g^c$, which, for a technical reason, is required to have a continuous density on the interior of its support, which is included in $[0,\bar U]$. This means that we can allow that the distribution of $U_g^c$ has support equal to $[\bar U-\tau,\bar U]$ for any $\tau>0$. As a consequence, it can be inferred that in practical applications and simulations, one could also employ a degenerate variable $U_g^c$ with a single point mass at $\bar U$. 

We now fix an arbitrarily small value $\tau>0$, and we suppose $U^c_g$ has a uniform distribution on the interval $[\bar U-\tau,\bar U]$. Since $\tilde U = \min(U_g,U_g^c)$ and $U_g$ has a uniform distribution on $[0,1]$, we have:
\begin{align*}
	P(\tilde U \le u) &= \begin{cases}
		0, & \text{if $u<0$}\\
		u, & \text{if $0\le u<\bar U-\tau$}\\ 
		(1-\frac{\bar U}{\tau}) + u(\frac{1}{\tau}+\frac{\bar U}{\tau})-\frac{u^2}{\tau}& \text{if $\bar U-\tau\le u< \bar U$}\\
		1, &\text{if $u\ge\bar U$}
	\end{cases};\\
	P(\tilde U \le u,\Delta =1) &= 
	\begin{cases}
		0, & \text{if $u<0$}\\
		u, & \text{if $0\le u<\bar U-\tau$}\\ 
		-\frac{u^2}{2\tau}+\frac{u\bar U}{\tau}-\frac{(\bar U-\tau)^2}{2\tau}, &  \text{if $\bar U-\tau\le u< \bar U$}\\
		\frac{\bar U^2-(\bar U-\tau)^2}{2\tau}, &\text{if $u\ge\bar U$}
	\end{cases}.
\end{align*}

Note also that the choice of the distribution for $U_g^c$ has an impact on the distribution of $\tilde U$ and therefore it affects the distribution of the proxies $(\hat \varphi(Z,X,\tilde U),Z,X,\Delta)$. While the estimator $\hat \beta$ is consistent with any choice of the distribution of $U_g^c$, depending on the resulting distribution of the proxies the asymptotic variance of $\sqrt{n}(\hat \beta-\beta_0)$ can vary.  We conjecture that the minimum of the variance is obtained by choosing a distribution for $U_g^c$ maximizing $P(\Delta =1)$, that is when the distribution of $U_g^c$ is as close as possible to the probability mass point concentrated in  $\bar U$, but it is outside the scope of this paper to prove such a result. 
\subsection{Notations}
In the sequel, we will use the following quantities. We denote by $\Gamma_1(\beta,\phi)$ the ordinary derivative of $M(\beta,\phi)$ with respect to $\beta$, so  
$\Gamma_1(\beta,\phi)= \frac{\partial}{\partial \beta} M(\beta,\phi)$, that is  
\begin{align*}
	\Gamma_1(\beta,\phi)& = \int_0^{\bar T}\Big[\frac{E^\phi(vv^\top\exp(v^\top\beta),s)}{E^\phi(\exp(v^\top\beta),s)} - \Big(\frac{E^\phi(v\exp(v^\top\beta),s)}{E^\phi(\exp(v^\top\beta),s)}\Big)^2 \Big]\text{d}Q^\phi(s).     
\end{align*}
We denote by $\Gamma_2(\beta,\phi)[\bar \phi - \phi]$ the pathwise derivative of $M(\beta,\phi)$ at $\phi$ in direction $[\bar\phi - \phi]$ when it exists, i.e.
\begin{align}
	\Gamma_2(\beta,\phi)[\bar \phi - \phi] = \lim_{\varepsilon\xrightarrow{}0} \frac{M(\beta,\phi+\varepsilon[\bar\phi-\phi]) - M(\beta,\phi)}{\varepsilon}.
\end{align}
Define also the function $\varrho(v,t)$ as $\varrho(v,t) = \varphi(v,\cdot)^{-1}(t)$, where $t\in[0,\bar T]$ and $v\in\mathcal{V}$. We denote by $1_{L}$ the vector in $\mathbb{R}^{L}$ with each component equal to $1$. 
Lastly, to simplify the notation, for a function $f(v,s)$ we denote by $f'(v,s)$ and  by $f''(v,s)$ the first and second derivative of $f(v,s)$ with respect to the last argument $s$, respectively. 

\section{Proof of Theorem \ref{GID}}\label{appendix:GID}
First, remark that, by the ``Kaplan-Meier type'' arguments mentioned in Section \ref{sec:estimation}, the left-hand-side of equation \eqref{eq:system} is identified for all $\left(\theta_\ell\right)_{\ell=1}^{L} \in \prod_{\ell=1}^{L} [0, \bar t \wedge c_{z_\ell,x}]$, where $c_{z_\ell,x} = \min_{k=1,...,L}  c_{z_\ell,x,w_k}$. 

Next, notice that
\begin{equation} \label{definition:baru}
\bar{u}^x = \sup \left\{ u \in [0, 1]\ :\ \text{the system of equation \eqref{eq:system} has a solution in }\prod_{\ell=1}^{L}[0, \bar t \wedge c_{z_\ell,x}]\right\}.
\end{equation}
Indeed, for $u \leq \bar{u}^x$, since $\bar{u}^x = \inf_{z\in Z,w\in W} \varphi(z, x, \cdot)^{-1}(\bar t \wedge c_{z,x,w})$, we have $\varphi^x(u) \in \prod_{\ell=1}^{L}[0, \bar t \wedge c_{z_\ell,x}]$, so that \eqref{eq:system} has a solution in $\prod_{\ell=1}^{L}[0, \bar t \wedge c_{z_\ell,x}]$ which is $\varphi^x(u)$. Next, for $u > \bar{u}^x$, $\varphi^x(u) \notin \prod_{\ell=1}^{L}[0, \bar t \wedge c_{z_\ell,x}]$, so that \eqref{eq:system} cannot have a solution in $\prod_{\ell=1}^{L}[0, \bar t \wedge c_{z_\ell,x}]$ (otherwise it would have two solutions in $\mathbb{R}^L$ which contradicts the assumption of the theorem). Equation \eqref{definition:baru}  shows that $\bar{u}^x$ is identified since its right-hand-side is identified. Then, for all $u \leq \bar{u}^x$, $\varphi^x(u)$ is identified as the unique solution to \eqref{definition:baru} in $\left(\theta_\ell\right)_{\ell=1}^{L} \in \prod_{\ell=1}^{L}[0, \bar t \wedge c_{z_\ell,x}]$.\hfill $\Box$\\

\section{Proof of Theorem \ref{theo:F_results}}\label{appendix:F_results}
\subsection{Lemmas}
\begin{lemma}\label{theo:uniform_convergence_Ft_cov}
	Under Assumption \ref{assumtpion:K}, we have  
	\begin{itemize}
		\item[(i)] $
	\sup\limits_{t\in[0,\bar T],z\in\mathcal{Z},x\in\mathcal{X},w\in\mathcal{W}} |\hat F(t|z,x,w) -F(t|z,x,w) | = O_p((\log n/(nh))^{1/2} + h^\nu + \epsilon^\pi ) .$
		\item[(ii)] $\sup\limits_{t\in[0,\bar T],z\in\mathcal{Z},x\in\mathcal{X},w\in\mathcal{W}} |\hat f(t|z,x,w) -f(t|z,x,w) | = O_p((\log n/(nh\epsilon))^{1/2} + h^\nu + \epsilon^\pi ) .$
		\item[(iii)] $\sup\limits_{t\in[0,\bar T],z\in\mathcal{Z},x\in\mathcal{X},w\in\mathcal{W}} |\frac{\partial}{\partial t}\hat f(t|z,x,w) -\frac{\partial}{\partial t}f(t|z,x,w) | = o_p(1).$
		\item[(iv)] $\sup\limits_{t\in[0,\bar T],z\in\mathcal{Z},x\in\mathcal{X},w\in\mathcal{W}} |\frac{\partial^2}{\partial t^2}\hat f(t|z,x,w) -\frac{\partial^2}{\partial t^2}f(t|z,x,w) | = o_p(1).$
		\item[(iv)] $\sup\limits_{t_1,t_2\in[0,\bar T],z\in\mathcal{Z},x\in\mathcal{X},w\in\mathcal{W}} \frac{|\frac{\partial^2}{\partial t^2}(\hat f(t_1|z,x,w)-\hat f(t_2|z,x,w) -f(t_1|z,x,w)+f(t_2|z,x,w) )|}{|t_1-t_2|^\alpha} = o_p(1).$
	\end{itemize}    
\end{lemma}
\textit{Proof} First, Proposition 1 in \cite{akritas2001non} implies
\begin{align}\label{eq:condF_van1997estimation}
	\sup_{t\in[0,\bar T]}\sup_{z\in\mathcal{Z}w\in\mathcal{W},x\in\mathcal{X}}|\tilde F(t|z,x,w) -F(t|z,x,w) | = O_p((\log n/(nh))^{1/2} +h^\nu).
\end{align}
The result in $(i)$ then follows by integration by parts and the following decomposition:
\begin{align*}
	\hat F(t|z,x,w) &-F(t|z,x,w) \\
	&= \int H\left(\frac{t-u}{\epsilon}\right)\text{d}\tilde F(u|z,x,w) - F(t|z,x,w) \\
	&= \int \tilde F(t-\epsilon u|z,x,w) \tilde K(u)\text{d}u  - F(t|z,x,w) \\
	&= \int \big( \tilde F(t-\epsilon u|z,x,w) - F(t-\epsilon u|z,x,w)\big)  \tilde K(u)\text{d}u  \\
	&\quad\quad+  \int \big(  F(t-\epsilon u|z,x,w) -F(t|z,x,w)\big)\tilde  K(u)\text{d}u.
\end{align*}
Specifically, we obtain that 

\begin{align*}
	\Big| \int \big( &\tilde F(t-\epsilon u|z,x,w) - F(t-\epsilon u|z,x,w)\big)  \tilde K(u)\text{d}u \Big| \\
	&\le    \int O_p((\log n/(nh))^{1/2} +h^\nu) \tilde K(u)\text{d}u\\
	&={\ifband\color{blue} \else \color{black} \fi O_p((\log n/(nh))^{1/2} + h^\nu).}
\end{align*}
{\ifband \color{blue} 
$O_p((\log n/(nh))^{1/2} + h^\nu)=o_p(n^{-1/4})\implies$
	\begin{itemize}
		\item $u<1/2$
            \item $u>1/(4\nu)$ 
	\end{itemize}	
	\else\fi
}
\noindent Using Assumption \ref{assumtpion:K}, the Taylor expansion $F(t-\epsilon u|z,x,w) -F(t|z,x,w) = -\epsilon u f(t|z,x,w) + \frac{1}{2}\epsilon^2 u^2 \frac{\partial}{\partial t}f(t|z,x,w) + \dots + O(\epsilon^\pi)$ and the boundedness of $f(t|z,x,w)$ and its derivatives in the argument $t$, we have
{\ifband\color{blue} \else \color{black}\fi
\begin{align*}
	\int \big(  F(t-\epsilon u|z,x,w)\big) -F(t|z,x,w)\big)\tilde  K(u)\text{d}u =  O(\epsilon^\pi).
\end{align*}
}
{\ifband \color{blue} 
$O(\epsilon^\pi) = o(n^{-1/4})\implies$
	\begin{itemize}
		\item$w>1/(4\pi)$
	\end{itemize}	
	\else\fi
}
The result in $(ii)$ corresponds to the rate of convergence of the term $(T_3)$ in the proof of Lemma 2 in the Supplementary Material of  \cite{de2020linear}.  More precisely, write $\hat f(t|z,x,w) - f(t|z,x,w) =I_1+I_2$, where
\begin{align*}
	I_1 &= \epsilon^{-1}\int \tilde K\left(\frac{t-s}{\epsilon}\right) \text{d}\big(\tilde F(s|z,x,w)-F(s|z,x,w)\big);\\
	I_2 &= \epsilon^{-1}\int \tilde K\left(\frac{t-s}{\epsilon}\right) \text{d}F(s|z,x,w) - f(t|z,x,w).
\end{align*}
Now, integrating by parts and using standard change of variables, we can write $I_1 = I_{11} + I_{12} + I_{13}$, where 
\begin{align*}
	I_{11}&=\epsilon^{-1}\int \big( \tilde F(t-u\epsilon|z,x,w) \\
	&\quad\quad- E \tilde F(t-u\epsilon|z,x,w)  - \tilde F(t|z,x,w) + E \tilde F(t|z,x,w)  \big)\tilde K'(u)\text{d}u;\\
	I_{12}&=\epsilon^{-1}\int \big( E \tilde F(t-u\epsilon|z,x,w)   \\
	&\quad\quad-  F(t-u\epsilon|z,x,w) - E \tilde F(t|z,x,w)  +F(t|z,x,w)  \big)\tilde K'(u)\text{d}u;\\
	I_{13}&=\epsilon^{-1}\big( \tilde F(t|z,x,w)-F(t|z,x,w)\big)\int \tilde K'(u)\text{d}u.
\end{align*}
Now, $I_{13} $ is straightforwardly equal to 0, because of Assumption \ref{assumtpion:K}. Following the proof of Theorem 3(b) of \cite{van1996uniform} it can be shown that there exists a finite constant $M_1$ such that 
\begin{align*}
	I_{11}&\le M_1\epsilon^{-1}\sup_{t,s\in[0,\bar T], |t-s|\le \epsilon,z\in\mathcal{Z},x\in\mathcal{X},w\in\mathcal{W}} \big| \tilde F(t|z,x,w) - E \tilde F(t|z,x,w)   \\
	&\quad\quad\quad\quad\quad\quad\quad\quad\quad\quad\quad\quad\quad\quad\quad\quad- \tilde F(s|z,x,w) + E \tilde F(s|z,x,w)  \big| \\
	&= O_p((\log n/(nh\epsilon))^{1/2}).
\end{align*}
For the term $I_{12}$, it is possible to follow the work of \cite{van1997estimation} with consideration here of higher-order kernels. We have that
\begin{align*}
	I_{12} \le \epsilon^{-1}\int \big(b(t-u\epsilon|z,x,w)h^\nu +o(h^{\nu}) +O(n^{-1}) -b(t|z,x,w)h^\nu \big)\tilde K'(u)\text{d}u,
\end{align*}
where the function $b$ is in equation (3.3) in \cite{van1997estimation}, and where their Taylor expansion term $o(h^\nu)$ can be replaced by $O(h^{\nu+ 1})$.  Hence, under Lipschitz continuous requirements for the function $b(t|z,x,w)$ with respect to $t$ induced by Assumption \ref{assumtpion:K}, we obtain {\ifband \color{blue}\else\fi$I_{12} = \epsilon^{-1}O(\epsilon h^\nu + h^{\nu+1}+n^{-1}) = O(h^\nu)$}, due to the assumptions on the bandwidths. 
{\ifband \color{blue} 
$O(\epsilon^{-1}h^{\nu+1}) = O(h^\nu)\implies$
\begin{itemize}
	\item$w>u$
\end{itemize}	
\else\fi
}
This argument shows that $I_{1} = O_p((\log n/(nh\epsilon))^{1/2}+h^\nu)$. For the term $I_2$, we can write 
\begin{align*}
	I_2 = \epsilon^{-1}\int \tilde K\left(\frac{t-s}{\epsilon}\right)\big(f(s|z,x,w)-f(t|z,x,w)\big)\text{d}s, 
\end{align*}
and apply standard Taylor expansion and change of variables to show that $I_2 = O(\epsilon^\pi)$.

For the result in $(iii)$, we can use a similar argument, where we will have $\epsilon^{-2}\tilde K''(u)$ in place of $\epsilon^{-1}\tilde K'(u)$, obtaining a rate of convergence for $\frac{\partial}{\partial t}\hat f(t|z,x,w) -\frac{\partial}{\partial t}f(t|z,x,w)$ equal to $O_p((\log n)^{1/2}n^{-1/2}h^{-1/2}\epsilon^{-3/2}) + \epsilon^{-2}O_p( h^{\nu+1})+O_p(h^\nu) +\epsilon^{-2}O_p(n^{-1})+ O(\epsilon^\pi)$. 

{\ifband \color{blue} In details, we have 
\begin{align*}
	I_1' &= \epsilon^{-2}\int \tilde K'\left(\frac{t-s}{\epsilon}\right) \text{d}\big(\tilde F(s|z,x,w)-F(s|z,x,w)\big);\\
	I_2' &= \epsilon^{-2}\int \tilde K'\left(\frac{t-s}{\epsilon}\right) \text{d}F(s|z,x,w) - f'(t|z,x,w).
\end{align*}
and, we the same argument as before, we have 
\begin{align*}
	I_{11}'&=\epsilon^{-2}\int \big( \tilde F(t-u\epsilon|z,x,w) \\
	&\quad\quad- E \tilde F(t-u\epsilon|z,x,w)  - \tilde F(t|z,x,w) + E \tilde F(t|z,x,w)  \big)\tilde K''(u)\text{d}u;\\
	I_{12}'&=\epsilon^{-2}\int \big( E \tilde F(t-u\epsilon|z,x,w)   \\
	&\quad\quad-  F(t-u\epsilon|z,x,w) - E \tilde F(t|z,x,w)  +F(t|z,x,w)  \big)\tilde K''(u)\text{d}u;\\
	I_{13}'&=\epsilon^{-2}\big( \tilde F(t|z,x,w)-F(t|z,x,w)\big)\int \tilde K''(u)\text{d}u.
\end{align*}
$I_{13}'$ is still zero by assumption, $I_{11}' = O_p((\log n)^{1/2}n^{-1/2}h^{-1/2}\epsilon^{-3/2})$ and 
\begin{align*}
	I_{12}' &\le \epsilon^{-2}\int \big(b(t-u\epsilon|z,x,w)h^\nu +O(h^{\nu+1}) +O(n^{-1}) -b(t|z,x,w)h^\nu \big)\tilde K''(u)\text{d}u\\
 &= \epsilon^{-2}h^\nu \int \big(\epsilon u b'(t|z,x,w)+O(\epsilon^2)\big)\tilde K''(u)\text{d}u  +\epsilon^{-2}O(h^{\nu+1}) +\epsilon^{-2}O(n^{-1}) \\
 &= O(h^\nu) +\epsilon^{-2}O(h^{\nu+1}) +\epsilon^{-2}O(n^{-1})
\end{align*}
For $I_2'$ we can argue as before to obtain $I_2' = O(\epsilon^\pi)$.
\else\fi
}

{\ifband \color{blue} 
\newpage
For (iv) we will similarly have 
\begin{align*}
	I_{12}'' &\le \epsilon^{-3}\int \big(b(t-u\epsilon|z,x,w)h^\nu +O(h^{\nu+1}) +O(n^{-1}) -b(t|z,x,w)h^\nu \big)\tilde K'''(u)\text{d}u\\
 &= \epsilon^{-3}h^\nu \int \big(\epsilon u b'(t|z,x,w)-\frac{1}{2}\epsilon^2 u^2 b''(t|z,x,w) +O(\epsilon^3)\big)\tilde K'''(u)\text{d}u  +\epsilon^{-3}O(h^{\nu+1}) +\epsilon^{-3}O(n^{-1}) \\
 &= O(h^\nu) +\epsilon^{-3}O(h^{\nu+1}) +\epsilon^{-3}O(n^{-1}),
\end{align*}
where in the last equality we used integration by parts and the fact that $\tilde K''(\pm 1)=0$.
\newpage
\else\fi}
The same applies for the result in $(iv)$ with rate of convergence 
$O_p((\log n)^{1/2}n^{-1/2}h^{-1/2}\epsilon^{-5/2}) + \epsilon^{-3}O_p( h^{\nu+1})+O_p(h^\nu) +\epsilon^{-3}O_p(n^{-1})+ O(\epsilon^\pi)$.

The result in $(v)$ follows by the almost sure behavior of the modulus of continuity of $\frac{\text{d}^2}{\text{d}t^2}\hat f(t|z,x,w) -\frac{\text{d}^2}{\text{d}t^2}f(t|z,x,w)$, with a similar proof as the one given in Lemma 2.11 in \cite{van1998nonparametric}. Thus
\begin{align*}
&\sup\limits_{t_1,t_2\in[0,\bar T],z\in\mathcal{Z},x\in\mathcal{X},w\in\mathcal{W}} \frac{|\frac{\partial^2}{\partial t^2}(\hat f(t_1|z,x,w)-\hat f(t_2|z,x,w) -f(t_1|z,x,w)+f(t_2|z,x,w) )|}{|t_1-t_2|^\alpha} \\
&= {\ifband\color{blue} \else \color{black} \fi O_p((\log n)^{1/2}n^{-1/2}h^{-1/2}\epsilon^{-5/2-\alpha}) + \epsilon^{-3-\alpha}O_p(h^{\nu+1})+\epsilon^{-\alpha}O_p(h^\nu)+\epsilon^{-3-\alpha}O_p(n^{-1})+O(\epsilon^{\pi})}.
\end{align*}\hfill $\Box$\\
{\ifband \color{blue} 
\begin{itemize}
	\item$w<(1-u)/(5+2\alpha)$
	\item $w<u(\nu+1)/(3+\alpha)$
        \item $w<u\nu/\alpha$ 
	\item $w<1/(3+\alpha)$
\end{itemize}	
\else\fi
}

\begin{lemma} \label{theo:p_convergence}
	Under Assumption \ref{assumtpion:K}, we have  
	\begin{itemize}
		\item[(i)]$\sup\limits_{z\in\mathcal{Z},x\in\mathcal{X},w\in\mathcal{W}}|\hat p_{z,x,w} - p_{z,x,w}| =  O_p((\log n/(nh))^{1/2} + h^\nu ).$
		\item[(ii)]$\sup\limits_{z\in\mathcal{Z},x\in\mathcal{X},w\in\mathcal{W}}|\frac{\partial}{\partial x}\hat p_{z,x,w} -\frac{\partial}{\partial x} p_{z,x,w}| =  o_p(1).$
		\item[(iii)]$\sup\limits_{z\in\mathcal{Z},x\in\mathcal{X},w\in\mathcal{W}}|\frac{\partial^2}{\partial x^2}\hat p_{z,x,w} -\frac{\partial^2}{\partial x^2} p_{z,x,w}| =  o_p(1).$
		\item[(iv)]$\sup\limits_{z\in\mathcal{Z},x_1,x_2\in\mathcal{X},w\in\mathcal{W}}\frac{|\frac{\partial^2}{\partial x^2}(\hat p_{z,x_1,w} -\hat p_{z,x_2,w}  -p_{z,x_1,w} +p_{z,x_2,w})|}{|x_1-x_2|^\alpha}= o_p(1).$	
	\end{itemize}
\end{lemma}
\textit{Proof} The quantity $\hat p_{z,x,w}$ corresponds to an adjusted Nadaraya-Watson estimator designed for estimating a conditional probability of a discrete outcome variable.  The outcome variable, in this case, is the variable $Z$,  the conditioning variable is $X$ and the weight for the $i$th observation is equal to the indicator of whether $W_i$ equals $w$ or not. The result in $(i)$ then follows from \cite{hall1999methods}. The results in $(ii)-(iii)$ can be proved similarly as in Proposition 5.3 of \cite{van1998nonparametric}, obtaining 
\begin{align*}
	\sup\limits_{z\in\mathcal{Z},x\in\mathcal{X},w\in\mathcal{W}}|\frac{\partial}{\partial x}\hat p_{z,x,w} -\frac{\partial}{\partial x} p_{z,x,w}| =  O_p((\log n/(nh^3))^{1/2} + h^{\nu} ),
\end{align*}
and
\begin{align*}
	\sup\limits_{z\in\mathcal{Z},x\in\mathcal{X},w\in\mathcal{W}}|\frac{\partial^2}{\partial x^2}\hat p_{z,x,w} -\frac{\partial^2}{\partial x^2} p_{z,x,w}| =  O_p((\log n/(nh^5))^{1/2} + h^{\nu} ).
\end{align*}
The result in $(iv)$ follows from Lemma 2.11 in \cite{van1998nonparametric}, and we obtain 
{\ifband \color{blue}\else\fi
$$\sup\limits_{z\in\mathcal{Z},x_1,x_2\in\mathcal{X},w\in\mathcal{W}}\frac{|\frac{\partial^2}{\partial x^2}(\hat p_{z,x_1,w} -\hat p_{z,x_2,w}  -p_{z,x_1,w} +p_{z,x_2,w})|}{|x_1-x_2|^\alpha} = O_p((\log n)^{1/2} (nh^{5+2\alpha})^{-1/2}+h^\nu)$$
}
{\ifband \color{blue} 
	\begin{itemize}
		\item$u<1/(5+2\alpha)$
	\end{itemize}	
	\else\fi
}

\hfill $\Box$\\

\begin{lemma}\label{theo:xderivF}
		Under Assumption \ref{assumption:C}, we have 
		\begin{itemize}
			\item[(i)]$	\sup\limits_{t\in[0,\bar T], x\in\mathcal{X},z\in\mathcal{Z},w\in\mathcal{W}} |\frac{\partial}{\partial x}\hat F(t|z,x,w)-\frac{\partial}{\partial x} F(t|z,x,w)|= o_p(1). $
			\item[(ii)]$	\sup\limits_{t\in[0,\bar T], x\in\mathcal{X},z\in\mathcal{Z},w\in\mathcal{W}} |\frac{\partial^2}{\partial x^2}\hat F(t|z,x,w)-\frac{\partial^2}{\partial x^2} F(t|z,x,w)|= o_p(1). $
			\item[(iii)] $\sup\limits_{t\in[0,\bar T], x_1,x_2\in\mathcal{X},z\in\mathcal{Z},w\in\mathcal{W}} \frac{|\frac{\partial^2}{\partial x^2}(\hat F(t|z,x_1,w)-\hat F(t|z,x_2,w)-F(t|z,x_1,w)+F(t|z,x_2,w))|}{|x_1-x_2|^\alpha}= o_p(1).$
		\end{itemize}
\end{lemma}
\textit{Proof} The results in $(i)$ and $(iii)$ follow from Proposition 5.5 in \cite{van1998nonparametric}:
\begin{align*}
	\sup\limits_{t\in[0,\bar T], x\in\mathcal{X},z\in\mathcal{Z},w\in\mathcal{W}}& |\frac{\partial}{\partial x}\hat F(t|z,x,w)-\frac{\partial}{\partial x} F(t|z,x,w)|\\
	&=  O_p((\log n/(nh^3))^{1/2} + h^{\nu} + \epsilon^\pi ),
\end{align*}
and 
\begin{align*}
	\sup\limits_{t\in[0,\bar T], x\in\mathcal{X},z\in\mathcal{Z},w\in\mathcal{W}}& |\frac{\partial^2}{\partial x^2}\hat F(t|z,x,w)-\frac{\partial^2}{\partial x^2} F(t|z,x,w)|\\
	&=  O_p((\log n/(nh^5))^{1/2} + h^{\nu} + \epsilon^\pi h^{-1} ).
\end{align*}
The result in $(iii)$ follows from Proposition 5.6 in \cite{van1998nonparametric},
{\ifband \color{blue}\else\fi
\begin{align*}
	\sup\limits_{t\in[0,\bar T], x_1,x_2\in\mathcal{X},z\in\mathcal{Z},w\in\mathcal{W}} & \frac{|\frac{\partial^2}{\partial x^2}(\hat F(t|z,x_1,w)-\hat F(t|z,x_2,w)-F(t|z,x_1,w)+F(t|z,x_2,w))|}{|x_1-x_2|^\alpha}\\
	& =  O_p((\log n/(nh^{5+2\alpha}))^{1/2} + h^{\nu} + \epsilon^\pi h^{-1-\alpha} ).
\end{align*}
}
\hfill $\Box$\\
 {\ifband \color{blue} 
 	\begin{itemize}
 		\item$u<1/(5+2\alpha)$
 		\item $w>u(1+\alpha)/\pi$
 	\end{itemize}	
 	\else\fi
 }
\begin{lemma}\label{theo:iid_representation1}
	Under Assumption \ref{assumtpion:K}, we have 
	\begin{itemize}
	\item[(i)] 
		\begin{align*}
			\hat F&(t|z,x,w) -F(t|z,x,w) \\
			&=(nh)^{-1}\sum_{i=1}^n K\left(\frac{x - X_i}{h}\right)\xi^F(Y_i, \delta_i, Z_i, W_i, t, z, x, w)  \\
			&\quad\quad+ R_n(t,z,x,w),
		\end{align*}  
  where
   \begin{align}
      \sup_{ t\in [0,\bar T],z\in\mathcal{Z},x\in\mathcal{X},w\in\mathcal{W}}|R_n(t,z,x,w)| =  O_p\left((\log n/(nh))^{3/4} + h^\nu + \epsilon^\pi \right).
  \end{align}
	\item[(ii)] 
		\begin{align*}  
			\hat f&(t|z,x,w) -f(t|z,x,w) \\
			&= (nh\epsilon)^{-1}\sum_{i=1}^n K\left(\frac{x - X_i}{h}\right)\int \tilde K'(u)\xi^F(Y_i, \delta_i, Z_i,  W_i, t-u\epsilon, z, x, w)\text{d}u \\
			&\quad\quad+  r_n(t,z,x,w),
		\end{align*}
  where 
    \begin{align*}
       \sup_{ t\in [0,\bar T],z\in\mathcal{Z},x\in\mathcal{X},w\in\mathcal{W}}|r_n(t,z,x,w)| = O_p\big((\log n/(nh\epsilon))^{3/4} + h^\nu + \epsilon^\pi \big).
  \end{align*}
  
\end{itemize}
\end{lemma}
\textit{Proof}
The result in $(i)$ is an easy adaptation of Lemma 1 in the Supplementary Material of \cite{de2020linear}. The result in $(ii)$ follows from $(i)$ and the fact that 
\begin{align*}
	\hat f(t|z,x,w) -f(t|z,x,w)  &= \epsilon^{-1}\int \tilde K\left(\frac{t-s}{\epsilon}\right)\text{d}\big( \tilde F(s|z,x,w) -F(s|z,x,w) \big) \\
	&\quad + \epsilon^{-1}\int \tilde K\left(\frac{t-s}{\epsilon}\right)\text{d} F(s|z,x,w) - f(t|z,x,w).
\end{align*}

More precisely, rewrite $\hat f(t|z,x,w) -f(t|z,x,w) $ as 
\begin{align}\label{eq:iidf}
	&\hat f(t|z,x,w) -f(t|z,x,w)  \nonumber\\
	&=\epsilon^{-2}\int \tilde K^{'}\left(\frac{t-s}{\epsilon}\right)\big( \tilde F(s|z,x,w) -F(s|z,x,w) \big)\text{d}s \nonumber\\
	&\quad\quad+ \epsilon^{-1}\int \tilde K\left(\frac{t-s}{\epsilon}\right)\big(f(s|z,x,w) - f(t|z,x,w)\big)\text{d} s\nonumber\\
	&=\epsilon^{-2}\int \tilde K^{'}\left(\frac{t-s}{\epsilon}\right)\big( \hat F(s|z,x,w) -F(s|z,x,w) \big)\text{d}s \nonumber\\
	&\quad\quad+ \epsilon^{-1}\int \tilde K\left(\frac{t-s}{\epsilon}\right)\big(f(s|z,x,w) - f(t|z,x,w)\big)\text{d} s \nonumber\\
	&\quad\quad+  O_p\left((\log n/(nh))^{3/4} + h^\nu + \epsilon^3 \right)\nonumber\\
	& = (nh\epsilon)^{-1} \sum_{i=1}^n \epsilon^{-1}\int \tilde K^{'}\left(\frac{t-s}{\epsilon}\right) K\left(\frac{x - X_i}{h}\right)\xi^F(Y_i, \delta_i, Z_i, W_i, s, z, x, w)\text{d}s  \nonumber\\
	&\quad\quad+  O_p\left((\log n/(nh))^{3/4} + h^\nu + \epsilon^3 \right)\nonumber\\
	&\quad\quad+ \int \tilde K(u)\big(f(t-\epsilon u|z,x,w) - f(t|z,x,w)\big)\text{d} u. 
\end{align}

Since 
\begin{align*}
	f(t-\epsilon u|z,x,w)  = f(t|z,x,w) - u\epsilon \frac{\partial  f(t|z,x,w)}{\partial t} + \frac{1}{2}u^2\epsilon^2 \frac{\partial^2  f(t|z,x,w)}{\partial t^2} +\dots+ O(\epsilon^\pi),
\end{align*}
and because of Assumption \ref{assumtpion:K}, we have
\begin{align*}
	\int \tilde K(u)\big(f(t-\epsilon u|z,x,w) - f(t|z,x,w)\big)\text{d} u = O(\epsilon^\pi).
\end{align*}
Therefore, by standard change of variable, equation \eqref{eq:iidf} leads to the result in $(ii)$. \hfill $\Box$\\

\subsection{Proof of Theorem  \ref{theo:F_results} }
The result follows from Lemma \ref{theo:uniform_convergence_Ft_cov} and Lemma \ref{theo:p_convergence}. More precisely, for $(i)$ we can write
\begin{align}\label{eq:scomposion_Fp}
	|\hat F(t,z|x,w) &-F(t,z|x,w)| \nonumber\\
	&\le  |\hat F(t|z,x,w)-F(t|z,x,w)|\, |\hat p_{z,x,w} - p_{z,x,w}| \nonumber\\
	&\quad + F(t|z,x,w)\, |\hat p_{z,x,w} - p_{z,x,w}| +  p_{z,x,w}\, |\hat F(t|z,x,w)-F(t|z,x,w)|\nonumber \\
	&= O_p\left((\log n/(nh))^{1/2} + h^\nu + \epsilon^\pi\right) \, O_p\left((\log n/(nh))^{1/2} + h^\nu \right) \nonumber\\
	&\quad + O_p\left((\log n/(nh))^{1/2} + h^\nu\right) + O_p\left((\log n/(nh))^{1/2} + h^\nu + \epsilon^\pi\right) \nonumber\\
	&= O_p\left((\log n/(nh))^{1/2} + h^\nu + \epsilon^\pi\right).
\end{align}

The same argument can be used for $(ii)-(viii)$ using Lemmas \ref{theo:uniform_convergence_Ft_cov}, \ref{theo:p_convergence} and \ref{theo:xderivF}. The result in $(ix)$ and $(x)$ are corollaries of Lemma \ref{theo:iid_representation1}. In detail, observing that 
\begin{align}
    \sup_{x\in\mathcal{X},w\in\mathcal{W}}|(nh)^{-1}\sum_{i=1}^nI(W_i=w)K(\frac{x-X_i}{h}) - f_{X,W}(x,w)| = O_p((\log n/(nh))^{1/2} + h^\nu),
\end{align}
we can write 
\begin{align*}
	\hat p_{z,x,w} - p_{z,x,w} &=   (nh)^{-1}\sum_{i=1}^n \frac{I(W_i=w)K(\frac{x-X_i}{h})[I(Z_i=z) - p_{z,x,w}]}{(nh)^{-1}\sum_{i=1}^n I(W_i=w)K(\frac{x-X_i}{h})}\\
 &=  
 (nh)^{-1}\sum_{i=1}^nK\left(\frac{x-X_i}{h}\right)I(W_i=w)\frac{ I(Z_i=z) - p_{z,x,w}}{f_{X,W}(x,w)} +O(h^\nu + \log n/(nh)).
\end{align*}
Therefore, with a similar decomposition as in \eqref{eq:scomposion_Fp}, we have   
\begin{align*}
	\hat F(t,&z|x,w) -F(t,z|x,w)\\
	& =  O_p\left((\log n/(nh))^{1/2} + h^\nu + \epsilon^\pi \right)O_p\left((\log n/(nh))^{1/2} + h^\nu \right) \\
	&\quad+  F(t|z,x,w)\big( \hat p_{z,x,w}-p_{z,x,w} \big)+ p_{z,x,w} \big(\hat  F(t|z,x,w)-F(t|z,x,w) \big).
\end{align*}
From this and Lemma \ref{theo:iid_representation1} the result in $(ix)$ follows. With a similar argument, we also obtain $(x)$.\hfill $\Box$\\

\section{Proof of Theorem \ref{theo:consistency_beta}}\label{appendix:consistency}
\subsection{Lemmas}\label{appendix:consistency_lemmas}

\begin{lemma}\label{theo:consistency_phi}
	Under Assumptions \ref{assumtpion:K} and \ref{assumption:C}, it holds $\|\hat\varphi-\varphi\|_{\infty} = o_p(1)$.
\end{lemma}
\textit{Proof}  Theorem \ref{theo:F_results} implies that  $\sup_{x\in\mathcal{X}}\|A(\hat \varphi^x,\hat F^x)-A(\hat \varphi^x,F^x)\|_{\infty} = o_p(n^{-1/4})$. Now, the proof of the result we are interested in is similar to the one of Theorem 4.1 in \cite{beyhum2022nonparametric}. More precisely,
\begin{align}\label{A1}
	\|A(\hat \varphi^x,F^x)\|_{\infty} &\le   \|A(\hat \varphi^x,\hat F^x)\|_{\infty} + \|A(\hat \varphi^x,\hat F^x)-A(\hat \varphi^x,F^x)\|_{\infty} \nonumber\\
	&\le  \|A(\hat \varphi^x,\hat F^x)\|_{\infty}+ \sup_{x\in\mathcal{X}}\|A(\hat \varphi^x,\hat F^x)-A(\hat \varphi^x,F^x)\|_{\infty}\nonumber\\
	&=\|A(\hat \varphi^x,\hat F^x)\|_{\infty} + o_p(n^{-1/4}).
\end{align}
By definition, 
\begin{align*}
	\|A(\hat\varphi^x,\hat F^x)\|_{\infty} 
	&= \inf_{\theta\in\mathcal{F}_{Z}^{\bar U,\bar T}} \|A(\theta,\hat F^x)\|_{\infty}\\
	&\le \inf_{\theta\in\mathcal{F}_{Z}^{\bar U,\bar T}} \|A(\theta, F^x)\|_{\infty} +\sup_{\theta\in\mathcal{F}_{Z}^{\bar U,\bar T}} \|A(\theta,\hat F^x)- A(\theta, F^x)\|_{\infty} \\
	&\le\sup_{x\in\mathcal{X}}\inf_{\theta\in\mathcal{F}_{Z}^{\bar U,\bar T}} \|A(\theta, F^x)\|_{\infty} +\sup_{x\in\mathcal{X}}\sup_{\theta\in\mathcal{F}_{Z}^{\bar U,\bar T}} \|A(\theta,\hat F^x)- A(\theta, F^x)\|_{\infty} \\
	&=0 + o_p(n^{-1/4}),
\end{align*}
where in the last equality we used Assumption \ref{assumption:C}. 

This argument and \eqref{A1} imply that $\sup_{x\in\mathcal{X}}\|A(\hat\varphi^x,F^x)\|_{\infty}=o_p(n^{-1/4})$, and so $\|\hat\varphi^x-\varphi^x\|_{\infty}=o_p(1)$ uniformly in $x\in\mathcal{X}$ by Assumption \ref{assumption:C}. \hfill $\Box$\\

\subsection{Proof of Theorem \ref{theo:consistency_beta}}
The proof is based on Lemma \ref{theo:consistency_phi} and on semiparametric Z-estimator-related theory. In particular, we refer to Theorem 1 in \cite{chen2003estimation}, and in the following, we show the fulfillment of conditions (1.1)-(1.5) of that theorem.


\begin{itemize}
	\item[(1.1)] The result follows from the definition of $\hat\beta$.
	\item[(1.2)] The result is a consequence of the identification of $\beta_0$, given in Theorem \ref{BID} and the continuity of the map $\beta\mapsto M(\beta,\varphi)$.
	\item[(1.3)] Note that $P(\tilde U\le u)$ and $P(\tilde U\le u, \Delta = 1)$ are continuous in $u$. In addition, the functions $g:\mathcal{V}\times\mathcal{B}\xrightarrow{}\mathbb{R}$ and $g_1:\mathcal{V}\times\mathcal{B}\xrightarrow{}\mathbb{R}^{d_V}$ defined as $g(v,\beta) = \exp(v^\top\beta)$ and $g_1(v,\beta) = v\exp(v^\top\beta)$ are Lipschitz in $\beta$ uniformly in $v$, since $\mathcal{V}$ and $\mathcal{B}$ are compact. Thus, the result easily follows.
	\item[(1.4)] The result follows from Lemma \ref{theo:consistency_phi}.
	\item[(1.5)] The result is a consequence of the proof of Theorem 3 in \cite{chen2003estimation}, which is shown to be valid in point (2.5) under Theorem \ref{theo:asymptotic_normality} below.
\end{itemize} \hfill $\Box$\\

\section{Proof of Theorem \ref{theo:asymptotic_normality}}\label{appendix:asymptotic_normality}

\subsection{Lemmas}\label{appendix:asympotic_normality_lemmas}

\begin{lemma}\label{theo:ex_conditition_Cii} Under Assumption \ref{assumtpion:K} and \ref{assumption:N}, 
	there exist $\varsigma,c>0$ such that for any $\theta \in \mathcal{F}_{Z}^{\bar U, \bar T}$ such that $\|\theta-\varphi^x\|_{\infty}\le \varsigma$, it holds $\|\theta-\varphi^x\|_{\infty}\le c  \|A(\theta,F^x)\|_{\infty} $, where $\varsigma, c$ do not depend on $x\in\mathcal{X}$.
\end{lemma}
\textit{Proof} Since $F(\cdot,z|x,w)$ is twice differentiable, also $A(\cdot, F^x)$ is twice differentiable in the first argument. Since $F(\cdot,z|x,w)$ has a bounded second derivative, uniformly in $z,w,x$, also the second derivative of $A(\cdot, F^x)(u)$ is bounded in $[0,\bar T]$ by a constant $M_1$ which does not depend on $x$. Using the Taylor expansion, for  $\theta\in\mathcal{F}_Z^{\bar U, \bar T}$, it holds
\begin{align*}
	A(\theta,F^x)(u) \le A(\varphi^x,F^x)(u) + \Gamma(\varphi^x,F^x)(u)(\theta(u)-\varphi^x(u)) + M_1\|\theta(u)-\varphi^x(u)\|^2.
\end{align*}
This implies that
\begin{align*}
    \|A(\theta,F^x)(u) - A(\varphi^x,F^x)(u) - \Gamma &(\varphi^x,F^x)(u)(\theta(u)-\varphi^x(u))\| \\
    &\le  M_1 \|\theta(u)-\varphi^x(u)\|^2\\
    &\le   M_1 \|\theta-\varphi^x\|_\infty^2.
\end{align*}
The previous inequality yields 
\begin{align*}
	\|A(\theta,F^x) -A(\varphi^x,F^x) - \Gamma(\varphi^x,F^x)(\theta-\varphi^x)\|_{\infty}\le M_1\|\theta-\varphi^x\|_{\infty}^2.
\end{align*}
Since $A(\varphi^x,F^x)=0$, we have 
\begin{align*}
	\| \Gamma(\varphi^x,F^x)(\theta-\varphi^x)\|_{\infty}\le     \|A(\theta,F^x)\|_\infty +  M_1\|\theta-\varphi^x\|_{\infty}^2.
\end{align*}
By Assumption \ref{assumption:N}, there exists a constant $M_2>0$, which does not depend on $x$ or $u\in[0,\bar U]$ such that
\begin{align*}
	M_2\|\theta-\varphi^x\|_\infty - M_1\|\theta-\varphi^x\|_\infty^2 \le \|A(\theta,F^x)\|_{\infty}.
\end{align*}
Choosing $\varsigma>0$ small enough, we obtain that when $\|\theta-\varphi^x\|_\infty\le \varsigma$, then 
\begin{align*}
	\frac{M_2}{2}\|\theta-\varphi^x\|_\infty \le \|A(\theta,F^x)\|_{\infty},
\end{align*}
which finishes the proof.\hfill $\Box$\\

\begin{lemma}\label{theo:rate_phi}
	Under Assumptions \ref{assumtpion:K}, \ref{assumption:C} and \ref{assumption:N}, we have $\|\hat\varphi-\varphi\|_{\infty} = o_p(n^{-1/4})$.
\end{lemma}
\textit{Proof} Similarly as in the proof of Lemma \ref{theo:consistency_phi}, we obtain  $\sup_{x\in\mathcal{X}}\|A(\hat\varphi^x,F^x)\|_{\infty}=o_p(n^{-1/4})$, and $\|\hat\varphi^x-\varphi^x\|_{\infty}=o_p(1)$ uniformly in $x\in\mathcal{X}$. Hence, by Lemma \ref{theo:ex_conditition_Cii}, we have $\|\hat\varphi^x-\varphi^x\|_{\infty}\le c\|A(\hat \varphi^x,F^x)\|_{\infty}$ with probability approaching to $1$, where $c$ does not depend on $x$. We obtain, 
\begin{align*}
	\sup_{x\in\mathcal{X}} \|\hat \varphi^x - \varphi^x\|_\infty\le c\sup_{x\in\mathcal{X}}\|A(\hat \varphi^x,F^x)\|_{\infty} = o_p(n^{-1/4}),
\end{align*}
which is the assertion.\hfill $\Box$\\

\begin{lemma} \label{theo:gateaux_inverse}
	Given a function $f(x)$, denote by $f^{-1}(y)$  its inverse, such that $f^{-1}(f(x)) = x$.
	Denote by $T$ the operator that associates to a function $f(x)$ its inverse $T(f)(y) = f^{-1}(y)$. Denote by $\Gamma(f)[h]$ the Gâteaux derivative of the operator $T$ at $f$ in direction $h$, defined as 
	\begin{align*}
		\Gamma(f)[h](y)&= \lim_{t \rightarrow 0} \frac{T(f + th)(y) - T(f)(y)}{t}.
	\end{align*}
	Then, the Gâteaux derivative of the operator $T$ takes the following form:
	\begin{align*}
		\Gamma(f)[h](y) &= - \frac{ h(f^{-1}(y))}{f'(f^{-1}(y))},
	\end{align*}
	where $f'(x) = \frac{d}{dx} f(x)$.
\end{lemma}
\textit{Proof} 
Solving for $x$ the equation $y = f(x)+th(x)$, we have $x = f^{-1}(y-th(x)) =f^{-1}(y)-th(f^{-1}(y))\frac{d}{dy}f^{-1}(y)+O(t^2)$, where $O(t^2)$ denotes terms of order $t^2$ or higher. 
Therefore,
\begin{align*}
	T(f + th)(y) &= f^{-1}(y) - th(f^{-1}(y)) \frac{ 1}{f'(f^{-1}(y))}+ O(t^2).
\end{align*}
The assertion then follows.\hfill $\Box$\\ 

\begin{lemma}\label{theo:gateaux_derivative_inverse}
	Given a function $f(x)$, denote by $f^{-1}(y)$  its inverse, as in Lemma \ref{theo:gateaux_inverse}.  Denote by $T$ the operator that associates to a function $f(x)$ the derivative of its inverse $T(f)(y) = \frac{d}{dy}f^{-1}(y)$, and denote by $\Gamma(f)[h]$ the Gâteaux derivative of $T$, defined as in Lemma \ref{theo:gateaux_inverse}. Then, $\Gamma(f)[h]$ takes the following form:
	\begin{align*}
		\Gamma(f)[h](y) &= \frac{h\big(f^{-1}(y)\big)\frac{d}{dy} f^{-1}(y) f''\big( f^{-1}(y)\big)-h'\big( f^{-1}(y)\big)}{f'\big( f^{-1}(y)\big)^2},
	\end{align*}
	where $f'(x) = \frac{d}{dx} f(x)$ and $f''(x)=\frac{d^2}{dx^2}f(x)$.
\end{lemma}
\textit{Proof} 
In the proof of Lemma \ref{theo:gateaux_inverse} we have shown that 
$$
(f+th)^{-1}(y) = f^{-1}(y) - t h(f^{-1}(y))\frac{d}{dy} f^{-1}(y)+O(t^2).
$$
Therefore, 
\begin{align*}
	&T(f + th)\\
	&= \frac{d}{dy}(f+th)^{-1}(y) \\
	&=\frac{1}{(f'+th')((f+th)^{-1}(y))}\\
	&= \big\{f'\big( f^{-1}(y) - t h(f^{-1}(y))\frac{d}{dy} f^{-1}(y)+O(t^2) \big) \\
	&\quad\quad+ th'\big( f^{-1}(y) - t h(f^{-1}(y))\frac{d}{dy} f^{-1}(y)+O(t^2)\big)\big\}^{-1}\\
	&=\frac{1}{f'\big( f^{-1}(y)\big) - t h\big(f^{-1}(y)\big)\frac{d}{dy} f^{-1}(y) f''\big( f^{-1}(y)\big) + th'\big( f^{-1}(y)\big) +O(t^2) }\\
	&=\frac{1}{f'\big( f^{-1}(y)\big)}\Big\{1 - \frac{t}{f'\big( f^{-1}(y)\big)}\big[h'\big( f^{-1}(y)\big) -h\big(f^{-1}(y)\big)\frac{d}{dy} f^{-1}(y) f''\big( f^{-1}(y)\big)  \big] + O(t^2)\Big\}.
\end{align*}
The assertion then follows. \hfill $\Box$\\ 

\begin{lemma}\label{theo:iid_representation_phi} Under Assumption \ref{assumption:N}, we have 
	\begin{itemize}
		\item[(i)]\begin{align*}
	\hat{\varphi}^x(u)-\varphi^x(u) &= (nh)^{-1}\sum_{i=1}^n K\left(\frac{x-X_i}{h}\right)\eta^\varphi(Y_i,\delta_i,Z_i,W_i,x,u) +S_n(u,x);
\end{align*}

		where 
		\begin{align*}
			\eta^\varphi(Y_i,\delta_i,Z_i,W_i,x,u)&=\Gamma(\varphi^x,F^x)^{-1}\left(\sum_{l=1}^L\eta^F(Y_i, \delta_i, Z_i,  W_i, \varphi(z_l,x,u), z_l, x, w_k)\right)_{k=1}^L.
		\end{align*}
  and 
  \begin{align}
      \sup_{u\in[0,\bar U],x\in\mathcal{X}}|S_n(u,x)| = o_p(n^{-1/2}).
  \end{align}
		\item[(ii)] Denoting by $\eta_j^\varphi$ the $j$th component of $\eta^\varphi$, we have
		\begin{align*}
			&\frac{\partial}{\partial u}\hat{\varphi}^x(u)- \frac{\partial}{\partial u}\varphi^x(u) \\
			&=  (nh)^{-1}\sum_{i=1}^n K\left(\frac{x - X_i}{h}\right)\left(\sum_{j=1}^L \kappa_j^l(x,u)\eta_j(Y_i,\delta_i,Z_i,W_i,x,u)\right)_{l=1}^L\\
			&+  (nh)^{-1}\sum_{i=1}^n  K\left(\frac{x - X_i}{h}\right)\\
			&\quad\times \left(\sum_{j,k=1}^L \zeta_{jk}^l(x,u)
     I(W_i=w_k)\frac{I(Z_i=z_l)-p_{z_l,x,w_k}}{f_{X,W}(x,w_k)} 
   f\big( \varphi(z_l,x,u)|z_l,x,w_k\big)\right)_{l=1}^L\\
			&+ (nh\epsilon)^{-1}\sum_{i=1}^n  K\left(\frac{x - X_i}{h}\right) \\
			&\quad\times\int \tilde K'(s)\left(\sum_{j,k=1}^L \zeta_{jk}^l(x,u)\xi^F(Y_i, \delta_i, Z_i,  W_i,  \varphi(z_l,x,u)-s\epsilon, z_l, x, w_k) \right)_{l=1}^L\text{d}s \\
			&+s_n(u,x)
		\end{align*}
		for suitable functions $\kappa^l_j(x,u)$ and  $\zeta^{l}_{jk}(x,u)$, varying $j,k,l$ in $1,\dots,L$, and where 
   \begin{align}
      \sup_{u\in[0,\bar U],x\in\mathcal{X}}|s_n(u,x)| = o_p(n^{-1/2}).
  \end{align}
	\end{itemize}
\end{lemma}
\textit{Proof} 
Theorem \ref{theo:F_results} provides a representation of the quantity $\hat F(t,z|x,w) - F(t,z|x,w)$, and similar arguments to those used in the proof of Theorem 4.2 in \cite{beyhum2022nonparametric} yield the first equality. Specifically, 
the first-order conditions of the optimization program \eqref{eq:estimator_phi} are
\begin{align*}
	\Gamma(\hat\varphi^x,\hat F^x)^\top A(\hat\varphi^x,\hat F^x) = 0,
\end{align*}
which leads to
\begin{align*}
	\Gamma(\hat\varphi^x,\hat F^x)^\top A(\varphi^x,\hat F^x) + \Gamma(\hat \varphi^x,\hat F^x)^\top\Gamma(\varphi^x,\hat F^x)(\hat \varphi^x-\varphi^x) + \Gamma(\hat \varphi^x,\hat F^x)^\top R^x = 0,
\end{align*}
where $R^x = A(\hat \varphi^x,\hat F^x) - A(\varphi^x,\hat F^x) - \Gamma(\varphi^x,\hat F^x)(\hat \varphi^x-\varphi^x)$. Thus, we have
\begin{align*}
	\hat \varphi^x-\varphi^x &= -[\Gamma(\hat \varphi^x,\hat F^x)^\top\Gamma(\varphi^x,\hat F^x)]^{-1}
 \Gamma(\hat\varphi^x,\hat F^x)^\top A(\varphi^x,\hat F^x)- [\Gamma(\hat \varphi^x,\hat F^x)^\top\Gamma(\varphi^x,\hat F^x)]^{-1}\Gamma(\hat \varphi^x,\hat F^x)^\top R^x\\
 &=-\Gamma(\varphi^x,\hat F^x)^{-1}A(\varphi^x,\hat F^x)-\Gamma(\varphi^x,\hat F^x)^{-1} R^x.
\end{align*}
Now, Theorem \ref{theo:F_results} implies that $\|A(\varphi^x, \hat F^x)\|_\infty =  O_p((\log n/(nh))^{1/2} + h^\nu + \epsilon^\pi ) $ and that $  \Gamma(\varphi^x,\hat F^x)=  \Gamma(\varphi^x, F^x) + O_p((\log n/(nh\epsilon))^{1/2} + h^\nu + \epsilon^\pi )$. Under Assumption \ref{assumption:N}, we have  
\begin{align}\label{eq:inverse_matrix_argument}
	\Gamma(\varphi^x,\hat F^x)^{-1}&=  (\Gamma(\varphi^x, F^x) + O_p((\log n/(nh\epsilon))^{1/2} + h^\nu + \epsilon^\pi )\big)^{-1}\nonumber\\
	&=\big(\Gamma(\varphi^x, F^x)(I + \Gamma(\varphi^x, F^x)^{-1}O_p((\log n/(nh\epsilon))^{1/2} + h^\nu + \epsilon^\pi ))\big)^{-1}\nonumber\\
	&=\big(I + \Gamma(\varphi^x, F^x)^{-1}O_p((\log n/(nh\epsilon))^{1/2} + h^\nu + \epsilon^\pi )\big)^{-1}\Gamma(\varphi^x, F^x)^{-1}\nonumber\\
	&= \big(I - \Gamma(\varphi^x, F^x)^{-1}O_p((\log n/(nh\epsilon))^{1/2} + h^\nu + \epsilon^\pi )\big)\Gamma(\varphi^x, F^x)^{-1}\nonumber\\
	&= \Gamma(\varphi^x, F^x)^{-1} -  \big(\Gamma(\varphi^x, F^x)^{-1}\big)^2 O_p((\log n/(nh\epsilon))^{1/2} + h^\nu + \epsilon^\pi)\nonumber\\
	&= \Gamma(\varphi^x, F^x)^{-1} + O_p((\log n/(nh\epsilon))^{1/2} + h^\nu + \epsilon^\pi)
\end{align}
where the last equality is due to the boundedness of the involved quantities. Thus, we obtain 
\begin{align*}
	\Gamma(\varphi^x,\hat F^x)^{-1}&A(\varphi^x,\hat F^x) \\
	&= [\Gamma(\varphi^x,\hat F^x)^{-1} - \Gamma(\varphi^x,F^x)^{-1}]A(\varphi^x,\hat F^x) +\Gamma(\varphi^x,F^x)^{-1}A(\varphi^x,\hat F^x)\\
	&=  O_p((\log n/(nh\epsilon))^{1/2} + h^\nu + \epsilon^\pi) O_p((\log n/(nh))^{1/2} + h^\nu + \epsilon^\pi )  \\
	&\quad\quad+\Gamma(\varphi^x,F^x)^{-1}A(\varphi^x,\hat F^x)\\
	&= \Gamma(\varphi^x,F^x)^{-1}A(\varphi^x,\hat F^x) + o_p(n^{-1/2})
\end{align*}
The proof of Theorem \ref{theo:consistency_beta} implies that $R^x = o_p(n^{-1/2})$. Since, for every $k=1,\dots,L$, it holds $u = \sum_{l=1}^L F^x(\varphi^x_{l}(u), z_l|w_k)$, we obtain
\begin{align*}
	&\hat\varphi^x(u) - \varphi^x(u) \\
	&= \Gamma(\varphi^x,F^x)^{-1}A(\varphi^x,\hat F^x) + o_p(n^{-1/2})\\
	&=\Gamma(\varphi^x,F^x)^{-1}\left(\sum_{l=1}^L \hat F^x(\varphi^x_{l}(u), z_l|w_k) -F^x(\varphi^x_{l}(u), z_l|w_k) \right)_{k=1}^L + o_p(n^{-1/2})\\
	&=(nh)^{-1}\sum_{i=1}^n K\left(\frac{x-X_i}{h}\right)\Gamma(\varphi^x,F^x)^{-1}\left(\sum_{l=1}^L\eta^F(Y_i, \delta_i, Z_i,  W_i, \varphi(z_l,x,u), z_l, x, w_k)\right)_{k=1}^L\\
	&\quad\quad+o_p(n^{-1/2}).
\end{align*}

We now show the result in $(ii)$. To simplify the argument we assume $L=2$, but it will be clear that the result holds for any integer $L\ge 1$. Denote by $1_L$ the vector in $\R^L$ with all entrances equals to 1. We have shown in equation \eqref{eq:derivative_phi_u} that
\begin{align*}
	&\frac{\partial}{\partial u}\hat\varphi^x -   \frac{\partial}{\partial u}\varphi^x 
	= \Big(\Gamma(\hat\varphi^x,\hat F^x)^{-1}-\Gamma(\varphi^x,F^x)^{-1}\Big)1_{L} \\
	&=\Bigg(\frac{1}{\hat f^x(\hat \varphi^x_1,z_1|w_1)\hat f^x(\hat \varphi^x_2,z_2|w_2)-\hat f^x(\hat \varphi^x_2,z_2|w_1)\hat f^x(\hat \varphi^x_1,z_1|w_2)}\\
	&\quad\quad\quad\quad\times \begin{bmatrix}
		\hat f^x(\hat \varphi^x_2,z_2|w_2) & -\hat f^x(\hat \varphi^x_2,z_2|w_1)\\
		-\hat f^x(\hat \varphi^x_1,z_1|w_2) & \hat f^x(\hat \varphi^x_1,z_1|w_1)
	\end{bmatrix}\\
	&\quad- \frac{1}{ f^x( \varphi^x_1,z_1|w_1) f^x( \varphi^x_2,z_2|w_2)- f^x( \varphi^x_2,z_2|w_1) f^x( \varphi^x_1,z_1|w_2)}\\
	&\quad\quad\quad\quad\times \begin{bmatrix}
		f^x( \varphi^x_2,z_2|w_2) & - f^x( \varphi^x_2,z_2|w_1)\\
		- f^x( \varphi^x_1,z_1|w_2) &  f^x( \varphi^x_1,z_1|w_1)
	\end{bmatrix}\Bigg)1_{L}
\end{align*}
We now show that the aimed iid representation holds for the first component of $\frac{\partial}{\partial u}\hat\varphi^x -   \frac{\partial}{\partial u}\varphi^x$, that is $\frac{\partial}{\partial u}\hat\varphi^x_1 -   \frac{\partial}{\partial u}\varphi^x_1$, and clearly the same argument can be used for the second component. So, write
\begin{align*}
	\frac{\partial}{\partial u}\hat\varphi^x_1 -   \frac{\partial}{\partial u}\varphi^x_1&=\frac{\hat f^x(\hat \varphi^x_2,z_2|w_2) -\hat f^x(\hat \varphi^x_2,z_2|w_1)}{\hat f^x(\hat \varphi^x_1,z_1|w_1)\hat f^x(\hat \varphi^x_2,z_2|w_2)-\hat f^x(\hat \varphi^x_2,z_2|w_1)\hat f^x(\hat \varphi^x_1,z_1|w_2)} \\
	&\quad\quad-  \frac{ f^x( \varphi^x_2,z_2|w_2) - f^x( \varphi^x_2,z_2|w_1)}{ f^x( \varphi^x_1,z_1|w_1) f^x( \varphi^x_2,z_2|w_2)- f^x( \varphi^x_2,z_2|w_1) f^x( \varphi^x_1,z_1|w_2)}\\
	&= I_1 + I_2,
\end{align*}
where 
\begin{align*}
	I_1&=\frac{\hat f^x(\hat \varphi^x_2,z_2|w_2) -\hat f^x(\hat \varphi^x_2,z_2|w_1)}{\hat f^x(\hat \varphi^x_1,z_1|w_1)\hat f^x(\hat \varphi^x_2,z_2|w_2)-\hat f^x(\hat \varphi^x_2,z_2|w_1)\hat f^x(\hat \varphi^x_1,z_1|w_2)} \\
	&\quad\quad- \frac{\hat f^x( \varphi^x_2,z_2|w_2) -\hat f^x( \varphi^x_2,z_2|w_1)}{\hat f^x( \varphi^x_1,z_1|w_1)\hat f^x( \varphi^x_2,z_2|w_2)-\hat f^x(\varphi^x_2,z_2|w_1)\hat f^x(\varphi^x_1,z_1|w_2)};\\
	I_2&=\frac{\hat f^x( \varphi^x_2,z_2|w_2) -\hat f^x( \varphi^x_2,z_2|w_1)}{\hat f^x( \varphi^x_1,z_1|w_1)\hat f^x( \varphi^x_2,z_2|w_2)-\hat f^x(\varphi^x_2,z_2|w_1)\hat f^x(\varphi^x_1,z_1|w_2)}\\
	&\quad\quad - \frac{ f^x( \varphi^x_2,z_2|w_2) - f^x( \varphi^x_2,z_2|w_1)}{ f^x( \varphi^x_1,z_1|w_1) f^x( \varphi^x_2,z_2|w_2)- f^x( \varphi^x_2,z_2|w_1) f^x( \varphi^x_1,z_1|w_2)}.
\end{align*}
We first start with $I_1$, and using standard Taylor expansion, we can write $ I_1 =  I_{11}+ I_{12}$, where 
\begin{align*}
	I_{11}& = \Big(\big(\hat f^x( \varphi^x_1,z_1|w_1)\hat f^x( \varphi^x_2,z_2|w_2)-\hat f^x(\varphi^x_2,z_2|w_1)\hat f^x(\varphi^x_1,z_1|w_2)\big)^{-2}\\
	&\quad \quad\times \big(\frac{d}{dt}\hat f^x( \varphi^x_2,z_2|w_2) -\frac{d}{dt}\hat f^x( \varphi^x_2,z_2|w_1)\big)\\	
	&\quad\quad\times (\hat f^x( \varphi^x_1,z_1|w_1)\hat f^x( \varphi^x_2,z_2|w_2)
	-\hat f^x(\varphi^x_2,z_2|w_1)\hat f^x(\varphi^x_1,z_1|w_2))\\
	&\quad- \big(\hat f^x( \varphi^x_1,z_1|w_1)\hat f^x( \varphi^x_2,z_2|w_2)-\hat f^x(\varphi^x_2,z_2|w_1)\hat f^x(\varphi^x_1,z_1|w_2)\big)^{-2}\\
	&\quad\quad\times\big(\hat f^x( \varphi^x_2,z_2|w_2) -\hat f^x( \varphi^x_2,z_2|w_1)\big)\\
	&\quad\quad\times(\hat f^x( \varphi^x_1,z_1|w_1)\frac{d}{dt}\hat f^x( \varphi^x_2,z_2|w_2)-\frac{d}{dt}\hat f^x(\varphi^x_2,z_2|w_1)\hat f^x(\varphi^x_1,z_1|w_2))\Big)\\
	&\times (\hat\varphi_2^x-\varphi_2^x) + O_p(\|\hat\varphi_2^x-\varphi_2^x\|_\infty^2);\\
	I_{12}& = -\big(\hat f^x( \varphi^x_1,z_1|w_1)\hat f^x( \varphi^x_2,z_2|w_2)-\hat f^x(\varphi^x_2,z_2|w_1)\hat f^x(\varphi^x_1,z_1|w_2)\big)^{-2}\\
	&\quad\quad\times\big(\hat f^x( \varphi^x_2,z_2|w_2) -\hat f^x( \varphi^x_2,z_2|w_1)\big)\\
	&\quad\quad\times\big(\frac{d}{dt}\hat f^x( \varphi^x_1,z_1|w_1)\hat f^x( \varphi^x_2,z_2|w_2)-\hat f^x(\varphi^x_2,z_2|w_1)\frac{d}{dt}\hat f^x(\varphi^x_1,z_1|w_2)\big)\\
	&\quad\quad\times (\hat\varphi_1^x-\varphi_1^x) + O_p(\|\hat\varphi_1^x-\varphi_1^x\|_\infty^2)
\end{align*}
Now, $ I_{12} = I_{121} + I_{122}$, with 
\begin{align*}
	I_{121} &=  \Big(-\big(\hat f^x( \varphi^x_1,z_1|w_1)\hat f^x( \varphi^x_2,z_2|w_2)-\hat f^x(\varphi^x_2,z_2|w_1)\hat f^x(\varphi^x_1,z_1|w_2)\big)^{-2}\\
	&\quad\quad\times\big(\hat f^x( \varphi^x_2,z_2|w_2) -\hat f^x( \varphi^x_2,z_2|w_1)\big)\\
	&\quad\quad\times\big(\frac{d}{dt}\hat f^x( \varphi^x_1,z_1|w_1)\hat f^x( \varphi^x_2,z_2|w_2)-\hat f^x(\varphi^x_2,z_2|w_1)\frac{d}{dt}\hat f^x(\varphi^x_1,z_1|w_2)\big)\\
	&\quad+\big( f^x( \varphi^x_1,z_1|w_1) f^x( \varphi^x_2,z_2|w_2)- f^x(\varphi^x_2,z_2|w_1) f^x(\varphi^x_1,z_1|w_2)\big)^{-2}\\
	&\quad\quad\times\big( f^x( \varphi^x_2,z_2|w_2) - f^x( \varphi^x_2,z_2|w_1)\big)\\
	&\quad\quad\times(\frac{d}{dt} f^x( \varphi^x_1,z_1|w_1) f^x( \varphi^x_2,z_2|w_2)- f^x(\varphi^x_2,z_2|w_1)\frac{d}{dt} f^x(\varphi^x_1,z_1|w_2))\Big) \\
	&\times (\hat\varphi_1^x-\varphi_1^x);\\
	I_{122}&=  -\big( f^x( \varphi^x_1,z_1|w_1) f^x( \varphi^x_2,z_2|w_2)- f^x(\varphi^x_2,z_2|w_1) f^x(\varphi^x_1,z_1|w_2)\big)^{-2}\\
	&\quad\quad\times\big( f^x( \varphi^x_2,z_2|w_2) - f^x( \varphi^x_2,z_2|w_1)\big)\\
	&\quad\quad\times(\frac{d}{dt} f^x( \varphi^x_1,z_1|w_1) f^x( \varphi^x_2,z_2|w_2)- f^x(\varphi^x_2,z_2|w_1)\frac{d}{dt} f^x(\varphi^x_1,z_1|w_2))\\
	&\quad\quad\times (\hat\varphi_1^x-\varphi_1^x) + O_p(\|\hat\varphi_1^x-\varphi_1^x\|_\infty^2).
\end{align*}
It is easy to show that  $I_{121}=o_p(n^{-1/2})$ because $\frac{\partial}{\partial t}\hat f^x$ is bounded with probability tending to 1, $\hat f^x-f^x = O_p(n^{-1/4})$ and $\hat \varphi_1^x - \varphi_1^x = o_p(n^{-1/4})$. Denote by $\eta^\varphi_l$ the $l$th functional component  of $\eta^\varphi$ for $l=1,\dots,L$. It follows that 
\begin{align*}
	I_{12} &= o_p(n^{-1/2}) + I_{122}\\
	&=-\big( f^x( \varphi^x_1,z_1|w_1) f^x( \varphi^x_2,z_2|w_2)- f^x(\varphi^x_2,z_2|w_1) f^x(\varphi^x_1,z_1|w_2)\big)^{-2}\\
	&\quad\quad\times\big( f^x( \varphi^x_2,z_2|w_2) - f^x( \varphi^x_2,z_2|w_1)\big)\\
	&\quad\quad\times(\frac{d}{dt} f^x( \varphi^x_1,z_1|w_1) f^x( \varphi^x_2,z_2|w_2)- f^x(\varphi^x_2,z_2|w_1)\frac{d}{dt} f^x(\varphi^x_1,z_1|w_2))\\
	&\quad\quad\times (nh)^{-1}\sum_{i=1}^n K\left(\frac{x-X_i}{h}\right)\eta_1^\varphi(Y_i,\delta_i,Z_i,W_i,x,u) + o_p(n^{-1/2}).
\end{align*}
A similar argument can be used for $I_{11}$. 

We now focus on $I_2$, which, by standard decomposition argument
, can be written as $I_2 = I_{21} +I_{22} +o_p(n^{-1/2})$, where
\begin{align*}
	I_{21} &= \frac{ f^x( \varphi^x_2,z_2|w_2) - f^x( \varphi^x_2,z_2|w_1)-\hat f^x( \varphi^x_2,z_2|w_2) -\hat f^x( \varphi^x_2,z_2|w_1)}{ f^x( \varphi^x_1,z_1|w_1) f^x( \varphi^x_2,z_2|w_2)- f^x( \varphi^x_2,z_2|w_1) f^x( \varphi^x_1,z_1|w_2)}\\
	I_{22} &= \frac{ \big(f^x( \varphi^x_2,z_2|w_2) - f^x( \varphi^x_2,z_2|w_1)\big)}{\big( f^x( \varphi^x_1,z_1|w_1) f^x( \varphi^x_2,z_2|w_2)- f^x( \varphi^x_2,z_2|w_1) f^x( \varphi^x_1,z_1|w_2)\big)^2}\\
	&\quad\times \big(f^x( \varphi^x_1,z_1|w_1) f^x( \varphi^x_2,z_2|w_2)- f^x( \varphi^x_2,z_2|w_1) f^x( \varphi^x_1,z_1|w_2) \\
	&\quad\quad\quad\quad-\hat f^x( \varphi^x_1,z_1|w_1)\hat f^x( \varphi^x_2,z_2|w_2)+\hat f^x(\varphi^x_2,z_2|w_1)\hat f^x(\varphi^x_1,z_1|w_2)\big)
\end{align*}

This equality and standard decomposition arguments similar to the ones used before, imply that it is possible to use the iid representation of $\hat f^x - f^x$ to obtain a representation for $I_{21}$ and $I_{22}$. This means that there exist four functions $\zeta_{lk}^0(x,u)$ for $l,k=1,2$ such that
\begin{align*}
	I_{2} &=\sum_{l,k=1}^2\zeta_{lk}^0(x,u) \times \Big((nh)^{-1}\sum_{i=1}^n K\left(\frac{x - X_i}{h}\right)
   I(W_i=w_k)\frac{I(Z_i=z_l)-p_{z_l,x,w_k}}{f_{X,W}(x,w_k)} 
 f( \varphi(z_l,x,u)|z_l,x,w_k) \\
	&+ (nh\epsilon)^{-1}\sum_{i=1}^n p_{z_l,x,w_k} K\left(\frac{x - X_i}{h}\right) \int \tilde K(s)\xi^F(Y_i, \delta_i, Z_i,  W_i, \varphi(z_l,x,u)-s\epsilon, z_l, x, w_k)\text{d}s\Big)\\
	&+o_p(n^{-1/2}).
\end{align*}
To summarize, there exist some differentiable functions $\kappa^l_j(x,u)$ and  $\zeta^{l}_{jk}(x,u)$, for $j,k,l = 1,\dots,L$ such that  
\begin{align*}
	&\frac{\partial}{\partial u}\hat{\varphi}^x(u)- \frac{\partial}{\partial u}\varphi^x(u) \\
	&=  (nh)^{-1}\sum_{i=1}^n K\left(\frac{x - X_i}{h}\right) \left(\sum_{j=1}^L \kappa_j^l(x,u)\eta_j(Y_i,\delta_i,Z_i,W_i,x,u)\right)_{l=1}^L\\
	& +  (nh)^{-1}\sum_{i=1}^n  K\left(\frac{x - X_i}{h}\right) \left(\sum_{j,k=1}^L \zeta_{jk}^l(x,u)
   I(W_i=w_k)\frac{I(Z_i=z_l)-p_{z_l,x,w_k}}{f_{X,W}(x,w_k)} 
 f( \varphi(z_l,x,u)|z_l,x,w_k)\right)_{l=1}^L\\
	& + (nh\epsilon)^{-1}\sum_{i=1}^n  K\left(\frac{x - X_i}{h}\right) \\
	&\quad\quad\times\int \tilde K'(s)\left(\sum_{j,k=1}^L \zeta_{jk}^l(x,u)\xi^F(Y_i, \delta_i, Z_i,  W_i,  \varphi(z_l,x,u)-s\epsilon, z_l, x, w_k) \right)_{l=1}^L\text{d}s\\
	&\,+o_p(n^{-1/2}).
\end{align*}
This corresponds to the result in $(ii)$. \hfill $\Box$\\     

\subsection{Proof of Theorem \ref{theo:asymptotic_normality}}\label{appendix:asymptotic_normality_proof}
Fix a value $\alpha \in(0,1)$. We are interested in showing that $\hat f(t,z|x,w)$ is Lipschitz continuous up to order $2+\alpha$. Any value of $\alpha\in(0,1)$ is suitable, taking into account that a higher value of $\alpha$ restricts the conditions on the bandwidths $h$ and $\epsilon$.  Now, for a function $f:\mathcal{Z}\times \mathcal{X}\times[0,\bar U]\xrightarrow{} \mathbb{R}$ define the norm  $\|\cdot \|_{\Phi}$ as 
\begin{align*}
	\|f\|_{\Phi}= \sum_{z\in\mathcal{Z}}&\Big\{\sup_{u\in[0,\bar U], x\in\mathcal{X}}|f(z,x,u)| \\
	&+  \sup_{u\in[0,\bar U], x\in\mathcal{X}}|\frac{\partial}{\partial u }f(z,x,u)| \\
	&+  \sup_{u\in[0,\bar U], x\in\mathcal{X}}|\frac{\partial^2}{\partial u^2 }f(z,x,u)| \\
	&+ \sup_{u\in[0,\bar U], x\in\mathcal{X}}|\frac{\partial}{\partial x }f(z,x,u)|\\
	&+ \sup_{u\in[0,\bar U], x\in\mathcal{X}}|\frac{\partial^2}{\partial x^2 }f(z,x,u)|\\
	&+\sup_{u_1,u_2\in[0,\bar U],x\in\mathcal{X}}\frac{|\frac{\partial^2}{\partial u^2 }f(z,x,u_1)-\frac{\partial^2}{\partial u^2 }f(z,x,u_2)}{|u_1-u_2|^\alpha}\\
		&+\sup_{x_1,x_2\in\mathcal{X},u\in[0,\bar U]}\frac{|\frac{\partial^2}{\partial x^2 } f(z,x_1,u)-\frac{\partial^2}{\partial x^2 }f(z,x_2,u)}{|x_1-x_2|^\alpha}\Big\}.
\end{align*}
With the abuse of notation of denoting the restriction of $\varphi(v,u)$ on  $\mathcal{V}\times[0,\bar U]$ again as $\varphi$, under our regularity assumptions, for any $\alpha\in(0,1)$, there exists a constant $m_0$ such that $\|\varphi\|_{\Phi}<m_0$. This result follows from the implicit function theorem and our regularity assumptions \ref{assumption:C} and \ref{assumption:N}. Therefore, $\varphi\in\Phi$, where $\Phi = \{f:\mathcal{Z}\times \mathcal{X}\times[0,\bar U]\xrightarrow{} \mathbb{R}| f(z,x,\cdot)$ and $f(z,\cdot,u)$ are twice differentiable and such that 
$\|f\|_{\Phi}\le m_0\}$. The norm $\|\cdot\|_{\Phi}$ is defined specifically to establish the set $\Phi$. It ensures that the necessary level of regularity for the functions involved in the theorem is met. This is crucial for achieving the Donsker property results of the given class of functions. The set $\Phi$ is endowed with the norm $\|\phi\|_{1}  = \sup_{v\in\mathcal{V},u\in[0,\bar U]}|\phi(v,u)| + \sup_{v\in\mathcal{V},u\in[0,\bar U]}|\phi'(v,u)|$ for any $\phi\in\Phi$. The  operator $M$ and its sample analog $M_n$ are considered as operators on the space  $(\Phi,\|\cdot\|_1)$. 

The proof utilizes Lemmas provided in Appendix \ref{appendix:asympotic_normality_lemmas} and Z-estimator related theory. In particular, we refer to Theorem 2 of \cite{chen2003estimation} and, in the following, we show the fulfillment of conditions (2.1)-(2.6) of that theorem.
\begin{itemize}
	\item[(2.1)] The result follows by definition of $\hat\beta$.
	\item[(2.2)] It is easy to see that $\Gamma_1(\beta,\varphi)$ is continuous in $\beta$ and that $\Gamma_1(\beta_0,\varphi)$ is negative definite, and so it has full rank. 
	\item[(2.3)] First, note that, thanks to our regularity conditions, the functions $\varphi'(v,u)$, $\varphi''(v,u)$, $\varrho(v,t)$ and $\varrho'(v,t)$ exist. Assumption \ref{RI} $(ii)$ also implies that, on $\mathcal{V}\times[0,\bar U]$, the function $\varphi'(v,u)$ admits a minimum strictly greater than zero. This, together with the definition of $\Phi$ which ensures that for any $\phi\in\Phi$, the functions $\phi(v,u)$ and $\phi'(v,u)$ are uniformly bounded by the constant $m_0$, we obtain that there exists  $\epsilon= \epsilon(m_0)>0$ such that for any $\phi\in\Phi$ the function $\varphi + t\phi$ is invertible for every $|t|<\epsilon$. 
	
	Denoting the measure induced on $\mathcal{V}$ by the probability of the variable $V$ by $\mu(v)$, we can use Lemma \ref{theo:gateaux_inverse} and the boundedness of the involved quantities to justify the following equalities:
	\begin{align*}
		&\lim_{\varepsilon\xrightarrow{} 0 }\frac{E^{\varphi +\varepsilon\phi}( g(v),t)-E^{\varphi}( g(v),t)}{\varepsilon} \\ 
		&=\lim_{\varepsilon\xrightarrow{} 0 } \frac{1}{\varepsilon}\int_{\mathcal{V}} g(v) \Big[ P\big((\varphi +\varepsilon\phi)(V,\tilde U)\ge t|V=v\big)-P\big(\varphi(V,\tilde U)\ge t|V=v\big)\Big]\text{d}\mu(v)\\
		&=\lim_{\varepsilon\xrightarrow{} 0 } \frac{1}{\varepsilon}\int_{\mathcal{V}} g(v) \Big[ P\big((\varphi +\varepsilon\phi)(v,\tilde U)\ge t|V=v\big)-P\big(\varphi(v,\tilde U)\ge t|V=v\big)\Big]\text{d}\mu(v)\\
		&=\lim_{\varepsilon\xrightarrow{} 0 } \frac{1}{\varepsilon}\int_{\mathcal{V}} g(v) \Big[ P\big((\varphi +\varepsilon\phi)(v,\tilde U)\ge t\big)-P\big(\varphi(v,\tilde U)\ge t\big)\Big]\text{d}\mu(v)\\
		&=\lim_{\varepsilon\xrightarrow{} 0 } \frac{1}{\varepsilon}\int_{\mathcal{V}} g(v) \Big[ P\big(\tilde U\ge (\varphi +\varepsilon\phi)(v,\cdot)^{-1}(t)\big)-P\big(\tilde U\ge \varphi(v,\cdot)^{-1}(t)\big)\Big]\text{d}\mu(v)\\
		&=-\int_\mathcal{V}g(v)\Big\{I(\varrho(v,t)<\bar U-\delta) + I(\bar U-\delta\le \varrho(v,t)\le \bar U)\frac{1+\bar U-2\varrho(v,t)}{\delta}\Big\}\\
		&\quad\quad\quad\quad\times\frac{\phi(v,\varrho(v,t))}{\varphi'(v,\varrho(v,t))} \text{d}\mu(v),
	\end{align*}
 where we used the expression of the distribution of $\tilde U$.	With a similar argument, we also obtain 
	\begin{align*}
		&\lim_{\varepsilon\xrightarrow{} 0 }\frac{E_1^{\varphi +\varepsilon\phi}( g(v),t)-E_1^{\varphi}( g(v),t)}{\varepsilon} \\ 
		&=-\int_{\mathcal{V}} g(v) \Big\{I(\varrho(v,t)<\bar U-\delta) + I(\bar U-\delta\le \varrho(v,t)\le \bar U)\frac{\bar U-\varrho(v,t)}{\delta}\Big\}\\
		&\quad\quad\times\frac{\phi(v,\varrho(v,t))}{\varphi'(v,\varrho(v,t))} \text{d}\mu(v).
	\end{align*}
	The quantity $\frac{\text{d}Q^\phi(t)}{\text{d}t}$ can be written as 
	\begin{align*}
		&\frac{\text{d}Q^\phi(t)}{\text{d}t}= \frac{\partial}{\partial t} \int_\mathcal{V} P(\tilde U \ge \phi(v,\cdot)^{-1}(t),\Delta =1)\text{d}\mu(v)\\
		&= \int_{\mathcal{V}}\Big\{I(\phi(v,\cdot)^{-1}(t)<\bar U-\delta) + I(\bar U-\delta\le \phi(v,\cdot)^{-1}(t)\le \bar U)\frac{\bar U-\phi(v,\cdot)^{-1}(t)}{\delta}\Big\}\\
		&\quad\quad\times\frac{\partial}{\partial t} \phi(v,\cdot)^{-1}(t)\text{d}\mu(v).
	\end{align*}
	Therefore, Lemma \ref{theo:gateaux_derivative_inverse} and again Lemma \ref{theo:gateaux_inverse} yield
	\begin{align*}
		&\lim_{\varepsilon\xrightarrow{} 0 }\frac{ \frac{\text{d}Q^{\varphi +\varepsilon\phi}(t)}{\text{d}t}- \frac{\text{d}Q^\phi(t)}{\text{d}t}}{\varepsilon} \\
		&=\int_{\mathcal{V}} \Big\{I(\varrho(v,t)<\bar U-\tau) + I(\bar U-\tau\le \varrho(v,t)\le \bar U)\frac{\bar U}{\tau}\Big\}\\
		&\quad\quad\times \frac{\phi(v,\varrho(v,t))\varrho'(v,t) \varphi''(v,\varrho(v,t))-\phi'(v,\varrho(v,t))}{(\varphi'(v,\varrho(v,t)))^2} \text{d}\mu(v)\\
		&\quad+\int_{\mathcal{V}} \Big\{ I(\bar U-\tau\le \varrho(v,t)\le \bar U)\frac{\varrho(v,t)'}{\tau}\Big\} \frac{\phi(v,\varrho(v,t))}{\varphi'(v,\varrho(v,t))}\text{d}\mu(v).
	\end{align*}
	Then, we can define 
	\begin{align*}
		&e^\varphi(g,t)[\phi] \\
		&= -\int_{\mathcal{V}} g(v)\Big\{I(\varrho(v,t)<\bar U-\tau) + I(\bar U-\tau\le \varrho(v,t)\le \bar U)\frac{1+\bar U-2\varrho(v,t)}{\tau}\Big\}\\
		&\quad\quad\quad\quad\times\frac{\phi(v,\varrho(v,t))}{\varphi'(v,\varrho(v,t))} \text{d}\mu(v);\\
		&e_1^\varphi(g,t)[\phi]  \\
		&=  -\int_{\mathcal{V}} g(v) \Big\{I(\varrho(v,t)<\bar U-\tau) + I(\bar U-\tau\le \varrho(v,t)\le \bar U)\frac{\bar U-\varrho(v,t)}{\tau}\Big\}\\
		&\quad\quad\quad\quad\times
		\frac{\phi(v,\varrho(v,t))}{\varphi'(v,\varrho(v,t))} \text{d}\mu(v);\\
		&q^\varphi(t)[\phi] = \int_{\mathcal{V}}\Big\{I(\varrho(v,t)<\bar U-\tau) + I(\bar U-\tau\le \varrho(v,t)\le \bar U)\frac{\bar U}{\tau}\Big\}\\
		&\quad\quad\quad\quad\times
		\frac{\phi(v,\varrho(v,t))\varrho'(v,t) \varphi''(v,\varrho(v,t))-\phi'(v,\varrho(v,t))}{(\varphi'(v,\varrho(v,t)))^2}  \text{d}\mu(v)\\
		&\quad\quad+\int_{\mathcal{V}}\Big\{ I(\bar U-\tau\le \varrho(v,t)\le \bar U)\frac{\varrho(v,t)'}{\tau}\Big\} \frac{\phi(v,\varrho(v,t))}{\varphi'(v,\varrho(v,t))} \text{d}\mu(v);\\
		&q^\varphi(t) = \int_{\mathcal{V}}\Big\{I(\varrho(v,t)<\bar U-\tau) + I(\bar U-\tau\le \varrho(v,t)\le \bar U)\frac{\bar U-\varrho(v,t)}{\tau}\Big\}  \\&\quad\quad\quad\quad\times\frac{\varrho(v,t)}{\frac{\partial}{\partial u}\varphi(v,\varrho(v,t))}\text{d}\mu(v).
	\end{align*}
	After some computation, it can be shown that
	\begin{align}\label{eq:gamma2}
		&\Gamma_2(\beta,\varphi)[\phi]= e_1(v,0)[\phi] \nonumber\\
		& - \int_0^{\bar T}\frac{\big\{e^\varphi(v\exp(v^\top\beta),t)[\phi]q(t)+E^\varphi(v\exp(v^\top\beta),t)q^\varphi(t)[\phi]\big\}E^\varphi(\exp(v^\top\beta),t)}{E^\varphi(\exp(v^\top\beta),t)^2}\text{d}t
		\nonumber\\
		&+ \int_0^{\bar T}\frac{E^\varphi(v\exp(v^\top\beta),t)q^\varphi(t)e^\varphi(\exp(v^\top\beta),t)[\phi]}{E^\varphi(\exp(v^\top\beta),t)^2}\text{d}t.
	\end{align}
	Therefore, by performing a routine Taylor expansion of $M(\beta, \phi)$ around $\varphi$, we can verify the desired condition due to the fact that the quantities $e^\varphi(g,t)[\phi]$, $ e_1^\varphi(g,t)[\phi]$ and $q^\varphi(t)[\phi]$ are linear with respect to $(\phi(v,u),\phi'(v,u))$ and the fact that $\Gamma_2(\beta,\varphi)[\phi]$ is linear with respect to $(e^\varphi(g,t)[\phi], e_1^\varphi(g,t)[\phi],q^\varphi(t)[\phi])$. This shows that condition $(2.3)(i)$ of \cite{chen2003estimation} is satisfied. Condition (2.3)$(ii)$ is instead implied by the differentiability in $\beta$ of $\Gamma_2(\beta,\varphi)[\phi]$.
	\item[(2.4)] The result follows from the Implicit Function Theorem. We now explain the argument in detail.
	Consider the map $\hat g:\mathbb{R}\times \mathbb{R}\times \mathbb{R}^{L}\xrightarrow{}  \mathbb{R}^{L}$, defined as 
	\begin{align}
		\hat g: (u,x,\theta)\mapsto (\hat g_1(u,x,\theta),\dots,\hat g_L(u,x,\theta))^\top,
	\end{align}
	where $\hat g_{k}(u,x,\theta) = \sum_{l=1}^{L}\hat F^x(\theta_l,z_l|w_k) - u$. By definition, $\hat\varphi$ satisfies $\hat g(u,x,\hat\varphi^x(u)) = 0$, for each $u\in[0,\bar U],x\in\mathcal{X}$.
	Theorem \ref{theo:F_results} implies $\sup_{u\in[0,\bar U], x\in\mathcal{X}}\|\Gamma(\hat\varphi^x,\hat F^x)(u)-\Gamma(\varphi^x, F^x)\| = o_p(1)$. Therefore, with probability tending to 1, $\Gamma(\hat\varphi^x,\hat F^x)(u)$ is invertible, because of Assumption \ref{assumption:N}. Thus, the Implicit Function Theorem implies
	\begin{align}\label{eq:derivative_phi_u}
		\frac{\partial}{\partial u}\hat\varphi^x(u) =& - \Gamma(\hat\varphi^x,\hat F^x)(u)^{-1}\big(\frac{\partial \hat g_0}{\partial u}( u,x,\hat\varphi^x(u)),\dots,\frac{\partial \hat g_L}{\partial u} ( u,x,\hat\varphi^x(u))\big)^\top\nonumber\\
		&=\Gamma(\hat\varphi^x,\hat F^x)(u)^{-1} 1_{L}.
	\end{align}
	Using the same argument as in Equation \eqref{eq:inverse_matrix_argument}, we obtain 
	\begin{align}\label{eq:decomposition_derivative_phi}
		&\frac{\partial}{\partial u}\hat\varphi^x -   \frac{\partial}{\partial u}\varphi^x \nonumber\\
		&= \Big(\Gamma(\hat\varphi^x,\hat F^x)^{-1}-\Gamma(\varphi^x,F^x)^{-1}\Big)1_{L}\nonumber\\
		&= \Big(\Gamma(\hat\varphi^x, F^x)^{-1} -\Gamma(\varphi^x,F^x)^{-1} + O_p((\log n/(nh\epsilon))^{1/2} + h^\nu + \epsilon^\pi)\Big)1_{L}
	\end{align}
	Using the boundedness of the second derivative in $t$ of $F(t,z|x,w)$ uniformly in $x\in\mathcal{X}, z\in\mathcal{Z},w\in\mathcal{W},t\in[0,\bar T]$,   Lemma \ref{theo:rate_phi}, and the differentiability of the operator that associates to a matrix $M$ its inverse $M^{-1}$, it holds $\sup_{u\in[0,\bar U], x\in\mathcal{X}}\|\Gamma(\hat\varphi^x, F^x)^{-1}-\Gamma(\varphi^x, F^x)^{-1} \|= o_p(n^{-1/4})$. This results, together with Lemma \ref{theo:rate_phi} implies $\|\hat\varphi - \varphi\|_{1} = o_p(n^{-1/4})$.
	The regularity of $\hat F(t,z|x,w)$ in both the arguments $t,x$ implies, with a similar argument, that also $\hat\varphi$ has the same regularity in $u,x$ and so $\hat\varphi\in\Phi$ with probability tending to 1.
 
\item[(2.5)]For $\gamma>0$, define $\Phi_\gamma = \{\phi\in\Phi: \|\phi-\varphi\|_{1}<\gamma\}$. 
We first show that the family 
$$\mathcal{F}=\{(v,u)\mapsto I(\phi(v,u)\ge t)|\phi\in\Phi_{\gamma},t\in[0,\bar T]\},$$ 
is Donsker. Using Theorem 2.5.6 of \cite{Vaart1996}, it is sufficient to show that 
\begin{align*}
	\int_0^\infty\sqrt{\log N_{[]}(\bar \epsilon,\mathcal{F},L_2(P))}\text{d}\bar \epsilon<\infty,
\end{align*}
where $N_{[]}$ is the bracketing number, $P$ is the probability measure corresponding to the joint distribution of $(V,\tilde U)$, and $L_2(P)$ is the $L_2$-norm. Fix $\bar \epsilon>0$. In Corollary 2.7.2 of \cite{Vaart1996}, it is stated that
\begin{align*}
	m = N_{[]}(\bar \epsilon^2, \Phi_{\gamma},L_2(P))\le \exp(K\bar \epsilon^{-4/(2+\alpha)}).
\end{align*}
Let $\phi_1^L\le\phi_1^T,\dots,\phi_{m}^L\le\phi_m^T$ be the functions defining the $m$ brackets for $\Phi_\gamma$. Thus, for each fixed $t$ and $\phi\in\Phi_\gamma$, we have 
\begin{align*}
	I(\phi_i^L(v,u)\ge t )\le I(\phi(v,u)\ge t )\le I(\phi_i^T(v,u)\ge t ),
\end{align*}
for some $i=1,\dots,m$. Define $F_i^L(t)=P(\phi_i^L(V,\tilde U)\ge t)$, and let $t^L_{ij}$ for $j=1,\dots,O(\bar \epsilon^{-2})$, partition the line in segments having $F^L_i$-probability less than or equal to a fraction of $\bar \epsilon^2$. Similarly, define $F_i^T(t)=P(\phi_i^T(V,\tilde U)\ge t)$, and let $t^T_{ij}$ for $j=1,\dots,O(\bar \epsilon^{-2})$, partition the line in segments having $F^T_i$-probability less than or equal to a fraction of $\bar \epsilon^2$. Now define the following bracket for $t$:
$$
t^L_{ij_1}\le t\le t^T_{ij_2},
$$
where $t^L_{ij_1}$ is the largest of the $t^L_{ij}$ with the property of being less than or equal to $t$ and $t^T_{ij2}$ is the smallest of the $t^T_{ij}$ with the property of being greater than or equal to $t$. We will now show that the brackets for $\mathcal{F}$ are given by
\begin{align*}
	I(\phi_i^L(v,u)\ge t_{ij_2}^T )\le I(\phi(v,u)\ge t )\le I(\phi_i^T(v,u)\ge t_{ij_1}^L).
\end{align*}
Note that, taking eventually a smaller $\gamma>0$, for $\phi_1,\phi_2\in\Phi_{\gamma}$, there exist $\phi_1(v,\cdot)^{-1}$ and $\phi_2(v,\cdot)^{-1}$, which we can denote by $\rho_1(v,t)$ and $\rho_2(v,t)$, respectively. In addition, by the regularity assumptions on $\varphi$, it is easy to show that there exists a constant $\Theta$, such that, for any integer $k\ge 1$, we have
\begin{align}
	\|\rho_1-\rho_2\|_{P,k}\le \|\rho_1-\rho_2\|_{\infty}\le \Theta \|\phi_1-\phi_2\|_{1},
\end{align}
where $\|\cdot\|_{P,k}$ is the $L_k(P)$  norm. This also implies that there exists a constant $K$ such that
\begin{align*}
	\|	I&(\phi_i^L(V,\tilde U)\ge t_{ij_2}^T )- I(\phi_i^T(V,\tilde U)\ge t_{ij_1}^L)\|_{P,2}^2\\
	&=P(\phi_i^T(V,\tilde U)\ge t_{ij_1}^L) - P(\phi_i^L(V,\tilde U)\ge t_{ij_2}^T ) \\
	&=F_i^T(t_{ij1}^L) - F_i^L(t_{ij2}^T)\\
	&\le F_i^T(t)-F_i^L(t) +K \epsilon^2.
\end{align*}
Now, denote by $f_{\tilde U}(u)$ the quantity $f_{\tilde U}(u)=\text{d}P(\tilde U\le u)/\text{d}u$, which exists a.e., and by $\rho_i^L(v,t)$ and $\rho_i^T(v,t)$ the functions $\phi_i^L(v,\cdot)^{-1}$ and $\phi_i^T(v,\cdot)^{-1}$, respectively. Then, using a Taylor expansion, we obtain 
\begin{align}\label{eq:_ingrid_paper_21}
	F_i^T(t)-F_i^L(t) &= \int_{\mathcal{V}}\Big(P(\tilde U\ge \rho_i^T(v,t)) - P(\tilde U\ge \rho_i^L(v,t))\Big)\text{d}\mu(v)\nonumber\\
	&=\int_{\mathcal{V}}f_{\tilde U}(\xi(v,t)) \big(\rho_i^T(v,t) - \rho_i^L(v,t)\big)\text{d}\mu(v),
\end{align}
where $\xi(v,t)$ is a point between $\rho_i^T(v,t)$ and $\rho_i^L(v,t)$. Therefore, for a suitable constant $J$ bounding $f_U$, we obtain 
\begin{align*}
		F_i^T(t)-F_i^L(t) &\le J \|\rho_i^T(v,t) - \rho_i^L(v,t)\|_{P,1}\\
		&\le J  \|\rho_i^T(v,t) - \rho_i^L(v,t)\|_{P,2} \\
		&\le J\Theta \bar \epsilon^2.
\end{align*}
 Hence for the class $\mathcal{F}$ and for each $\bar \epsilon>0$, we have at most $O(\bar \epsilon^{-2}\exp(K\bar \epsilon^{-4/(2+\alpha)}))$ brackets in total. However, for $\bar \epsilon>1$, one bracket suffices. So, it holds 
\begin{align*}
	\int_{0}^\infty\sqrt{\log N_{[]}(\bar \epsilon,\mathcal{F},L_2(P))}\text{d}\bar \epsilon<\infty.
\end{align*} 
This shows that the class $\mathcal{F}$ is Donsker. Together with Theorem 2.10.6 in \cite{Vaart1996}, this result also implies that the following families of functions are Donsker:
\begin{align*}
		\mathcal{F}_1 &=\{\mathcal{V}\times[0,\bar U]\ni (v,u)\mapsto \exp(v^\top\beta)I(\phi(v,u)\ge t)| t\in[0,\bar T],\beta\in\mathcal{B},\phi\in\Phi_\gamma\};\\
	\mathcal{F}_2 &=\{\mathcal{V}\times[0,\bar U]\times\{0,1\}\ni (v,u,\Delta)\mapsto I(\phi(v,u)\ge t)I(\Delta=1)| t\in[0,\bar T],\phi\in\Phi_\gamma\};\\
		\mathcal{F}_3 &=\{\mathcal{V}\times[0,\bar U]\times\{0,1\}\ni (v,u,\Delta)\mapsto v\exp(v^\top\beta)I(\phi(v,u)\ge t)I(\Delta=1)\\
	&\quad\quad\quad\quad\quad\quad\quad\quad\quad\quad\quad\quad\quad\quad\quad\quad| t\in[0,\bar T],\beta\in\mathcal{B},\phi\in\Phi_\gamma\}.
\end{align*}
Now, consider $\mathcal{F}_1$. What remains to show is that if $\beta_n$ is a sequence in $n$ converging to $\beta_0$ (with respect to the Euclidean norm) and $\phi_n$ is a sequence in $n$ converging to  $\varphi$  (with respect to $\|\cdot\|_1$) then
\begin{align*}
	E\Bigl\{&\Bigl[\Bigl(\exp\bigl(V^\top\beta_n\bigr)I\bigl(\phi_n(V,\tilde U)\ge t\bigr) - \exp\bigl(V^\top\beta\bigr)I\bigl(\varphi(V,\tilde U)\ge t\bigr)\Bigr) \\
	&- 	E\Bigl(\exp\bigl(V^\top\beta_n\bigr)I\bigl(\phi_n(V,\tilde U)\ge t\bigr) - \exp\bigl(V^\top\beta\bigr)I\bigl(\varphi(V,\tilde U)\ge t\bigr)\Bigr)\Bigr]^2\Bigr\} \xrightarrow{}0
\end{align*}
This result easily follows from the regularity of the distribution of $\tilde U$, the compactness of the support of $V$, the independence between $V$ and $\tilde U$, the definition of the norm $\|\cdot\|_1$ and the Lipschitz continuity of $v\mapsto \exp(v^\top\beta)$ in $\beta$ uniformly in $v$. More precisely, write 
\begin{align*}
    E\Bigl\{&\Bigl[\Bigl(\exp\bigl(V^\top\beta_n\bigr)I\bigl(\phi_n(V,\tilde U)\ge t\bigr) - \exp\bigl(V^\top\beta\bigr)I\bigl(\varphi(V,\tilde U)\ge t\bigr)\Bigr) \\
	&- 	E\Bigl(\exp\bigl(V^\top\beta_n\bigr)I\bigl(\phi_n(V,\tilde U)\ge t\bigr) - \exp\bigl(V^\top\beta\bigr)I\bigl(\varphi(V,\tilde U)\ge t\bigr)\Bigr)\Bigr]^2\Bigr\} \\
 &= I_1+I_2,
\end{align*}
where
\begin{align*}
	I_1 &= E[\exp(2V^\top\beta_n)I(\phi_n(V,\tilde U)\ge t)] \\
	&\quad+ E[\exp(2V^\top\beta_0)I(\varphi(V,\tilde U)\ge t)]\\ &\quad-2E[\exp(V^\top\beta_n)I(\phi_n(V,\tilde U)\ge t)\exp(V^\top\beta_n)I(\phi_n(V,\tilde U)\ge t)];
\end{align*}
and 
\begin{align*}
	I_2 &= \Big(E[(V^\top\beta_n)P(\phi_n(V,\tilde U))]-E[(V^\top\beta_0)P(\varphi(V,\tilde U))]\Big)^2.
\end{align*}
Now, using similar arguments as before, we have 
\begin{align*}
	\sup_{v\in\mathcal{V}}|P(\phi_n&(v,\tilde U)\ge t)-P(\varphi(v,\tilde U)\ge t)| \\
	&=	\sup_{v\in\mathcal{V}}|	P(\tilde U\ge \rho_n(v,t))-P(\tilde U \ge\varrho(v,t)|\\
	&\le K\|\phi_n-\varphi\|_1
\end{align*}
for a suitable constant $K$, where $\rho_n(v,t) = \phi_n(v,\cdot)^{-1}(t)$. Also, 
\begin{align*}
	\sup_{v\in\mathcal{V}}|\exp(v^\top\beta_n) -\exp(v^\top\beta_0) | \le K\|\beta_n-\beta_0\|,
\end{align*}
for a further constant $K$. Therefore, it is easy to see that $I_2\xrightarrow{}0$. Using the independence between $V$ and $\tilde U$, similar arguments also show that $I_1\xrightarrow{}0$.
Similar results can be shown for the families $\mathcal{F}_2$ and $\mathcal{F}_3$. Thus, from Corollary 2.3.12 in \cite{Vaart1996}, we have that, for any sequence $\varsigma_n=o(1)$, it holds
\begin{align}\label{eq:final_donsker}
	&\sup_{s\in[0,\bar T],\|\beta-\beta_0\|\le \varsigma_n,\|\phi-\varphi\|_{1}\le \varsigma_n} |\hat Q^\phi(s) - Q^\phi(s) - \hat Q^\varphi(s) +Q^\varphi(s)|= o_p(n^{-1/2});\nonumber\\
	&\sup_{s\in[0,\bar T],\|\beta-\beta_0\|\le \varsigma_n,\|\phi-\varphi\|_{1}\le \varsigma_n} \|\hat E^\phi(g(\beta,v),s) \nonumber\\
	&\quad\quad- E^\phi(g(\beta,v),s) -\hat E^\varphi(g(\beta_0,v),s) + E^\varphi(g(\beta_0,v),s) \|= o_p(n^{-1/2});\\
	&\sup_{s\in[0,\bar T],\|\beta-\beta_0\|\le \varsigma_n,\|\phi-\varphi\|_{1}\le \varsigma_n} \|\hat E_1^\phi(g(\beta,v),s)) \nonumber\\
	&\quad\quad- E_1^\phi(g(\beta,v),s) -\hat E_1^\varphi(g(\beta_0,v),s)+ E_1^\varphi(g(\beta_0,v),s) \|= o_p(n^{-1/2}).\nonumber
\end{align}
Now, with a similar decomposition as the one in \cite{tsiatis1981large} (page 97), we can write
\begin{align*}
	M_n(\beta,\phi)-&M(\beta,\varphi) - (M_n(\beta,\varphi)-M(\beta_0,\varphi))\\
	&=\sum_{j=1}^6D_{jn}(\beta,\phi)+\sum_{j=1}^8R_{jn} (\beta,\phi) ,
\end{align*}
where $D_{jn}(\beta,\phi) = C_{jn}(\beta,\phi) - C_{jn}(\beta_0,\varphi)$,  
\begin{align*}
	C_{1n}(\beta,\phi) &= \hat E_1^\phi(v,0)-E_1^\phi(v,0);\\
	C_{2n}(\beta,\phi)&=-[\hat Q^\phi(0)-Q^\phi(0)]E^\phi(v\exp(v^\top\beta),0)/E^\phi(\exp(v^\top\beta),0);\\
	C_{3n}(\beta,\phi)&=\int_0^{\bar T}[(\hat Q^\phi(s)-Q^\phi(s))/E^\phi(\exp(v^\top\beta),s)]\text{d}E^\phi(v\exp(v^\top\beta),s);\\
	C_{4n}(\beta,\phi)&=\int_0^{\bar T}[(\hat Q^\phi(s)-Q^\phi(s))E^\phi(v\exp(v^\top\beta),s)/E^\phi(\exp(v^\top\beta),s)^2]\\
	&\quad\quad\text{d}E^\phi(\exp(v^\top\beta),s);\\
	C_{5n}(\beta,\phi)&=\int_0^{\bar T}[(\hat E^\phi(v\exp(v^\top\beta),s) -E^\phi(v\exp(v^\top\beta),s))/E^\phi(\exp(v^\top\beta),s) ]\text{d}Q^\phi(s);\\
	C_{6n}(\beta,\phi)&=\int_0^{\bar T}[(\hat E^\phi(\exp(v^\top\beta),s) -E^\phi(\exp(v^\top\beta),s))\\
	&\quad\quad\times E^\phi(v\exp(v^\top\beta),s)/E^\phi(\exp(v^\top\beta),s)^2 ]\text{d}Q^\phi(s);
\end{align*}
and 
\begin{align*}
	R_{1n}(\beta,\phi) &= \int_0^{\bar T}[(\hat E^\phi(v\exp(v^\top\beta),s)-E^\phi(v\exp(v^\top\beta),s))/\hat E^\phi(\exp(v^\top\beta),s)]\\
	&\quad\quad\quad\quad\text{d}\big(\hat Q^\phi(s) - Q^\phi(s) - \hat Q^\varphi(s) +Q^\varphi(s)\big);\\
	R_{2n}(\beta,\phi) &= \int_0^{\bar T}\Big\{[(\hat E^\phi(v\exp(v^\top\beta),s)-E^\phi(v\exp(v^\top\beta_0),s))/\hat E^\phi(\exp(v^\top\beta),s)] \\
	&\quad\quad-[(\hat E^\varphi(v\exp(v^\top\beta_0),s)-E^\varphi(v\exp(v^\top\beta_0),s))/\hat E^\varphi(\exp(v^\top\beta),s)]\Big\}\\
	&\quad\quad\quad\quad\text{d}\big(\hat Q^\varphi(s) - Q^\varphi(s)\big);\\
	R_{3n}(\beta,\phi) &= \int_0^{\bar T}[(\hat E^\phi(v\exp(v^\top\beta),s)-E^\phi(v\exp(v^\top\beta),s))\\
	&\quad\quad\times(\hat E^\phi(\exp(v^\top\beta),s)-E^\phi(\exp(v^\top\beta),s))\\
	&\quad\quad \times (\hat E^\phi(\exp(v^\top\beta),s)E^\phi(\exp(v^\top\beta),s))^{-1}]\text{d} \big( Q^\phi(s) - Q^\varphi(s)\big);\\
	R_{4n}(\beta,\phi) &= \int_0^{\bar T}\Big\{[(\hat E^\phi(v\exp(v^\top\beta),s)-E^\phi(v\exp(v^\top\beta),s))\\
	&\quad\quad\times(\hat E^\phi(\exp(v^\top\beta),s)-E^\phi(\exp(v^\top\beta),s))\\
	&\quad\quad \times (\hat E^\phi(\exp(v^\top\beta),s)E^\phi(\exp(v^\top\beta),s))^{-1}]\\
	&\quad\quad-[(\hat E^\varphi(v\exp(v^\top\beta_0),s)-E^\varphi(v\exp(v^\top\beta_0),s))\\
	&\quad\quad\times(\hat E^\varphi(\exp(v^\top\beta_0),s)-E^\varphi(\exp(v^\top\beta_0),s))\\
	&\quad\quad \times (\hat E^\varphi(\exp(v^\top\beta_0),s)E^\varphi(\exp(v^\top\beta_0),s))^{-1}]\Big\}\text{d} Q^\varphi(s);\\
	R_{5n}(\beta,\phi) &= -\int_0^{\bar T}[(\hat E^\phi(\exp(v^\top\beta),s)-E^\phi(\exp(v^\top\beta),s))E^\phi(v\exp(v^\top\beta),s)\\
	&\quad\quad \times (\hat E^\phi(\exp(v^\top\beta),s)E^\phi(\exp(v^\top\beta),s))^{-1}]\\
	&\quad\quad\quad\quad\text{d} \big(\hat Q^\phi(s) - Q^\phi(s) - \hat Q^\varphi(s) +Q^\varphi(s)\big);\\
	R_{6n}(\beta,\phi) &= -\int_0^{\bar T}\Big\{[(\hat E^\phi(\exp(v^\top\beta),s)-E^\phi(\exp(v^\top\beta),s))E^\phi(v\exp(v^\top\beta),s)\\
	&\quad\quad \times (\hat E^\phi(\exp(v^\top\beta),s)E^\phi(\exp(v^\top\beta),s))^{-1}]\\
	&\quad\quad -[(\hat E^\varphi(\exp(v^\top\beta_0),s)-E^\varphi(\exp(v^\top\beta_0),s))E^\varphi(v\exp(v^\top\beta_0),s)\\
	&\quad\quad \times (\hat E^\varphi(\exp(v^\top\beta_0),s)E^\varphi(\exp(v^\top\beta_0),s))^{-1}]\Big\}\text{d} \big(\hat Q^\varphi(s) -Q^\varphi(s)\big);\\
	R_{7n}(\beta,\phi) &= \int_0^{\bar T}[(\hat E^\phi(\exp(v^\top\beta),s)-E^\phi(\exp(v^\top\beta),s))^2E^\phi(v\exp(v^\top\beta),s)\\
	&\quad\quad \times (\hat E^\phi(\exp(v^\top\beta),s)E^\phi(\exp(v^\top\beta_0),s)^2)^{-1}]\text{d} \big(Q^\phi(s) -Q^\varphi(s)\big);\\
	R_{8n}(\beta,\phi) &= \int_0^{\bar T}\Big\{[(\hat E^\phi(\exp(v^\top\beta),s)-E^\phi(\exp(v^\top\beta),s))^2E^\phi(v\exp(v^\top\beta),s)\\
	&\quad\quad \times (\hat E^\phi(\exp(v^\top\beta),s)E^\phi(\exp(v^\top\beta),s)^2)^{-1}]\\
	&\quad\quad -[(\hat E^\varphi(\exp(v^\top\beta_0),s)-E^\varphi(\exp(v^\top\beta_0),s))^2E^\varphi(v\exp(v^\top\beta_0),s)\\
	&\quad\quad \times (\hat E^\varphi(\exp(v^\top\beta_0),s)E^\varphi(\exp(v^\top\beta_0),s)^2)^{-1}]\Big\}\text{d} Q^\varphi(s).
\end{align*}
This decomposition, and the equalities in \eqref{eq:final_donsker} imply  the assertion. 
	\item[(2.6)] The statement can be reduced to showing that $n^{1/2}\{M_n(\beta_0,\varphi) + \Gamma_2(\beta_0,\varphi)[\hat\varphi-\varphi]\}$ can be expressed as a sum of iid terms of the form
	\begin{align*}
		&n^{1/2}\{M_n(\beta_0,\varphi) + \Gamma_2(\beta_0,\varphi)[\hat\varphi-\varphi]\\
		&\quad\quad = n^{-1/2}\sum_{i=1}^n\eta(Y_i,\delta_i,Z_i,W_i,X_i,\Delta_i,\tilde U_i) + o_p(1).
	\end{align*}
	The fact that $ n^{1/2} M_n(\beta_0,\varphi)$ admits such representation is thanks to the decomposition given in (2.5). We now show that the same holds for the term $ n^{1/2}\Gamma_2(\beta_0,\varphi)[\hat\varphi-\varphi]$, whose expression is given in \eqref{eq:gamma2}. 
	
	Define $p(z,x)= P(Z=z|X=x)$ and denote by $\nu(x)$ the density of $X$. First, consider $e_1(v,0)[\hat\varphi-\varphi]$, that is 
	\begin{align*}
		e_1(v,0)[\hat\varphi-\varphi] =- \int_{\mathcal{X}}\sum_{l=1}^{L} \frac{(z_l^\top,x^\top)^\top}{\frac{\partial}{\partial u}\varphi(z_l,x,0)}\big(\hat\varphi(z_l,x,0)-\varphi(z_l,x,0)\big)p(z_l,x)\nu(x)\text{d}x.
	\end{align*}
	Denote $\Gamma(\varphi^x,F^x)^{-1}_l$ the $l$th row the matrix $\Gamma(\varphi^x,F^x)^{-1}$. Given two vectors $v$ and $w$ denote by $vw$ the inner product, defined by $v^\top w$. We can write 
	\begin{align*}
		&e_1(v,0)[\hat\varphi-\varphi] \\
		&=- \int_{\mathcal{X}}\sum_{l=1}^{L} \frac{(z_l^\top,x^\top)^\top}{\frac{\partial}{\partial u}\varphi(z_l,x,0)}\Bigg((nh)^{-1}\sum_{i=1}^n K\left(\frac{x-X_i}{h}\right)\\
		&\times (\Gamma(\varphi^x,F^x)^{-1})_l\left(\sum_{l=1}^L\eta^F(Y_i, \delta_i, Z_i,  W_i, \varphi(z_l,x,0), z_l, x, w_k)\right)_{k=1}^L \Bigg)p(z_l,x)\nu(x)\text{d}x \\
		&+o_p(n^{-1/2})\\
		&=(nh)^{-1}\int_{\mathcal{X}}\sum_{i=1}^n K\left(\frac{x-X_i}{h}\right) b(Y_i,\delta_i,Z_i,W_i,x)\text{d}x+o_p(n^{-1/2}),
	\end{align*}
	where 
	\begin{align*}
		&b(Y_i,\delta_i,Z_i,W_i,x) = -\sum_{l=1}^{L} \frac{(z_l^\top,x^\top)^\top}{\frac{\partial}{\partial u}\varphi(z_l,x,0)}\Bigg((\Gamma(\varphi^x,F^x)^{-1})_l\\
		&\times\Big(\sum_{l=1}^L\eta^F(Y_i, \delta_i, Z_i,  W_i, \varphi(z_l,x,0), z_l, x, w_k)\Big)_{k=1}^L \Bigg)p(z_l,x)\nu(x).
	\end{align*}
	Thus, by standard change of variable and Taylor expansion, we obtain 
	\begin{align*}
		e_1&(v,0)[\hat\varphi-\varphi] \\
		&= n^{-1} \sum_{i=1}^n\int_{\mathcal{X}} b(Y_i,\delta_i,Z_i,W_i,hx+X_i)K(x)\text{d}x\\
		&= n^{-1} \sum_{i=1}^n\int_{\mathcal{X}} [b(Y_i,\delta_i,Z_i,W_i,X_i) \\
		&\quad\quad+ \sum_{j=1}^{\nu-1}(hx)^j\frac{\partial^j}{\partial x^j}b(Y_i,\delta_i,Z_i,W_i,X_i) +O_p(h^\nu)]K(x)\text{d}x \\
		&= n^{-1} \sum_{i=1}^n b(Y_i,\delta_i,Z_i,W_i,X_i) + o_p(n^{-1/2}),
	\end{align*}
	where in the last equality we used that $O_p(h^\nu) = o_p(n^{-1/2})$. 
	Now, write  $q^\varphi(t)[\hat\varphi-\varphi] = I_1 + I_2$, where 
	\begin{align*}
		I_1 &=\int_{\mathcal{V}}  g_0(v,t) \big(\hat\varphi(v,\rho(v,t))-\varphi(v,\rho(v,t))\big)\text{d}\mu(v);\\
		I_2 &= \int_{\mathcal{V}} g_1(v,t)\big(\frac{\partial}{\partial u}\hat\varphi(v,\rho(v,t))-\frac{\partial}{\partial u}\varphi(v,\rho(v,t))\big)\text{d}\mu(v);
	\end{align*}
	and 
	\begin{align*}
		g_0(v,t)&= \Big\{I(\varrho(v,t)<\bar U-\delta) + I(\bar U-\delta\le \varrho(v,t)\le \bar U)\frac{\bar U}{\delta}\Big\}\frac{\varrho'(v,t) \varphi''(v,\varrho(v,t))}{\varphi'(v,\varrho(v,t))} \\
		&\quad+ \Big\{ I(\bar U-\delta\le \varrho(v,t)\le \bar U)\frac{\varrho(v,t)'}{\delta}\Big\} \frac{1}{(\varphi'(v,\varrho(v,t)))^2} \\
		g_1(v,t)&=  \Big\{I(\varrho(v,t)<\bar U-\delta) + I(\bar U-\delta\le \varrho(v,t)\le \bar U)\frac{\bar U}{\delta}\Big\}\frac{1}{(\varphi'(v,\varrho(v,t)))^2}
	\end{align*}
	Using a similar argument as for $ e_1(v,0)[\hat\varphi-\varphi]$ we again can show that there exists a function $\tilde b(y,\delta, z, w, x)$ such that 
	\begin{align*}
		I_1 = n^{-1} \sum_{i=1}^n\tilde b(Y_i,\delta_i,Z_i,W_i,X_i) + o_p(n^{-1/2}).
	\end{align*}
	We now focus on $I_2$, and we write it as $I_2 = I_{21}+ I_{22}+o_p(n^{-1/2})$, where 
	\begin{align*}   
		&I_{21}=\int_{\mathcal{X}}\sum_{l=1}^{L}g_0(z_l,x,t)\Big[(nh)^{-1}\sum_{i=1}^n K\left(\frac{x - X_i}{h}\right)\\
		&\times\sum_{j=1}^L \kappa_j^l(x,\varrho(z_l,x,t))\eta_j(Y_i,\delta_i,Z_i,W_i,x,\varrho(z_l,x,t)) \\
		& + (nh)^{-1}\sum_{i=1}^n  K\left(\frac{x - X_i}{h}\right) \sum_{j,k=1}^L \zeta_{jk}^l(x,\varrho(z_l,x,t))
     I(W_i=w_k)\frac{I(Z_i=z_l)-p_{z_l,x,w_k}}{f_{X,W}(x,w_k)} 
  f(t|z_l,x,w_k)\Big]\\
		&\times p(z_l,x)\nu(x)\text{d}x\\
		&I_{22} = \int_{\mathcal{X}}\sum_{l=1}^{L}g_1(z_l,x,t)\Big[(nh\epsilon)^{-1}\sum_{i=1}^n  K\left(\frac{x - X_i}{h}\right)\\
		&\times \int \tilde K'(s)\sum_{j,k=1}^L \zeta_{jk}^l(x,\varrho(z_l,x,t))\xi^F(Y_i, \delta_i, Z_i,  W_i,  t-s\epsilon, z_l, x, w_k)\text{d}s \Big]\\
		&\times p(z_l,x)\nu(x)\text{d}x.
	\end{align*}
	So, again with a similar argument as before, $I_{21}$ admits such representation, and we now discuss $I_{22}$. Theorem 3 (b) of \cite{van1996uniform} implies that 
	\begin{align*}
		&(nh)^{-1}\sum_{i=1}^n  K\left(\frac{x - X_i}{h}\right) \xi^F(Y_i, \delta_i, Z_i,  W_i,  t-s\epsilon, z_l, x, w) \\
		&= (nh)^{-1}\sum_{i=1}^n  K\left(\frac{x - X_i}{h}\right) \xi^F(Y_i, \delta_i, Z_i,  W_i,  t, z, x, w) +  O_p\big((\epsilon\log n/(nh))^{1/2} +\epsilon^{1/2}h^\nu\big).
	\end{align*}
	Since $\int\tilde  K'(s)\text{d}s=0$, $\int s\tilde K'(s)\text{d}s=-1$ and $\int s^j\tilde K'(s)\text{d}s=0$ for $j=2,\dots,\pi$, we obtain 
	\begin{align*}
		&I_{22} = \int_{\mathcal{X}}\sum_{l=1}^{L}g_1(z_l,x,t)\Big[(nh\epsilon)^{-1}\sum_{i=1}^n  K\left(\frac{x - X_i}{h}\right)\\
		&\quad\times \int \tilde K'(s)\sum_{j,k=1}^L \zeta_{jk}^l(x,\varrho(z_l,x,t))F(t-\epsilon s,z|x,w)\text{d}s \Big]\\
		&\quad\times p(z_l,x)\nu(x)\text{d}x + O_p\big((\log n/( nh\epsilon))^{1/2} +\epsilon^{-1/2}h^\nu\big)\\
		&=\int_{\mathcal{X}}\sum_{l=1}^{L}g_1(z_l,x,t)\\
		&\quad\times\Big[(nh\epsilon)^{-1}\sum_{i=1}^n  K\left(\frac{x - X_i}{h}\right) \int \tilde K'(s)\sum_{j,k=1}^L \zeta_{jk}^l(x,\varrho(z_l,x,t))\big(-s\epsilon f(t,z_l|x,w_k) \\
		&\quad\quad\quad\quad+\frac{1}{2}s^2\epsilon^2 f'(t,z_l|x,w_k) -\frac{1}{6} s^3\epsilon^3 f''(t,z_l|x,w_k)+\dots+ O(\epsilon^{\pi+1})\big)\text{d}s \Big]\\
		&\quad\times p(z_l,x)\nu(x)\text{d}x + O_p\big((\log n/( nh\epsilon))^{1/2} +\epsilon^{-1/2}h^\nu\big))\\
		&= \int_{\mathcal{X}}\sum_{l=1}^{L}g_1(z_l,x,t)\\
		&\quad\times\Big[(nh)^{-1}\sum_{i=1}^n  K\left(\frac{x - X_i}{h}\right)\sum_{j,k=1}^L \zeta_{jk}^l(x,\varrho(z_l,x,t))(f(t,z_l|x,w_k) + O(\epsilon^\pi))\Big]\\
		&\quad\times p(z_l,x)\nu(x)\text{d}x + O_p\big((\log n/( nh\epsilon))^{1/2} +\epsilon^{-1/2}h^\nu\big)
	\end{align*}
	Note that $O_p\big((\log n/(nh\epsilon ))^{1/2} +\epsilon^{-1/2}h^\nu\big) = o_p(n^{-1/2})$ since {\ifband \color{blue}\else\color{black}\fi$w<2\nu u-1$}. All the other terms in $ \Gamma_2(\beta_0,\varphi)[\hat\varphi-\varphi]$ can follow the same arguments, from which the assertion follows.
\end{itemize}

{
\ifband\color{blue}
\begin{itemize}
    \item $u<1/2$
    \item $u>1/(4\nu)$ 
    \item$w>1/(4\pi)$
    \item $w>u$
    \item $w<(1-u)/(5+2\alpha)$
    \item $w<u(\nu+1)/(3+\alpha)$
    \item $w<u\nu/\alpha$ 
    \item $w<1/(3+\alpha)$
    \item $u<1/(5+2\alpha)$
    \item$u<1/(5+2\alpha)$
    \item $w>u(1+\alpha)/\pi$
    \item $w<2\nu u-1$
\end{itemize}
results in 
\begin{align*}
\frac{1}{4\nu}<u<\min(\frac{1}{2},\frac{1}{5+2\alpha})
\end{align*}
and
\begin{align*}
	\max(\frac{u(1+\alpha)}{\pi}, u,\frac{1}{4\pi})<w<\min(\frac{1}{3+\alpha},2\nu u-1,\frac{u\nu}{\alpha},\frac{u(\nu+1)}{3+\alpha},\frac{1-u}{5+2\alpha})
\end{align*}

\else\fi
}

\hfill $\Box$\\

\end{document}